%% file: main.tex
\documentclass[12pt]{article}
\usepackage{publication}
\usepackage{graphicx}
\usepackage{upgreek}
\usepackage{multicol,multirow}
\usepackage{amsmath,amssymb,amsfonts}
\usepackage{mathrsfs}
\usepackage{amsthm}
\usepackage[figuresright]{rotating}
\usepackage{appendix}
\usepackage[authoryear]{natbib}
\usepackage{ifpdf}
\usepackage{newtxtext}
\usepackage{newtxmath}
\usepackage{textcomp}
\usepackage{xcolor}
\usepackage{placeins}
\usepackage{subcaption}
\usepackage{siunitx}
\usepackage{floatrow}
\newfloatcommand{capbtabbox}{table}[][\FBwidth]

\usepackage{bigdelim} 
\usepackage{hhline} 

\usepackage[colorlinks,allcolors=blue]{hyperref}
\definecolor{jourcolor}{cmyk}{1,0.57,0.01,0.38}
\hypersetup{
	colorlinks,%
	citecolor=jourcolor,%
	filecolor=jourcolor,%
	linkcolor=jourcolor,%
	urlcolor=jourcolor
}

\theoremstyle{definition}

\usepackage[]{cleveref}
\crefrangelabelformat{subfigure}{#3#1#4\textendash#5#2#6}

\usepackage{pgfplots}
\usepackage{tikz}

\input{colourstuff.tex}

\newcommand{\inlineline}[1]{
	\tikz[baseline=-0.6ex,trim left=0.3em, trim right=0.7em]\draw[#1, very thick] (0em,0em) -- (1em,0em);
}

\begin{document}
	
	\title{A Low-Fidelity Method for Aerofoil Shape Optimisation for Curvilinear Blade Kinematics}
	
	\author{Benjamin Irwin$^{1*}$}
	\contact{$^*$B.Irwin@soton.ac.uk}
	
	\author{David Toal$^1$}
	
	\author{Swathi Krishna$^1$}
	
	\affil{$^1$Department of Aeronautics and Astronautics, University of Southampton, Southampton, SO17 1BJ, UK}
	
	\date{\today}
	
	\maketitle
	\begin{abstract}
		\footnotesize
		This study develops a low-fidelity framework for aerofoil shape optimisation under curvilinear blade kinematics, using a hovering cyclorotor as a representative case. Aerofoil optimisation can improve cyclorotor efficiency by suppressing leading-edge vortex separation during dynamic stall, but conventional approaches rely on computationally expensive CFD-based optimisation. The proposed method uses a single-streamtube model to estimate the rotor throughflow and optimises the aerofoil camberline using separate leading- and trailing-edge criteria. Assessed across configurations with varying blade counts and chord lengths, the framework consistently identifies aerofoils that improve hover efficiency, quantified by Figure of Merit. For the baseline four-bladed configuration, the low-fidelity optimum achieves 77\% of the Figure of Merit improvement obtained using high-fidelity optimisation at a fraction of the computational cost. The analysis also reveals an additional torque-minimising design family at increased chord lengths, highlighting the influence of trailing-edge loading. Aerofoil optimisation is also compared to blade-pitch kinematics optimisation, which improves the efficiency through similar control of the leading-edge vortex separation. While both approaches produce comparable improvements in efficiency, the optimised kinematics substantially reduces thrust. Aerofoil optimisation may therefore be more practical, as maintaining a target thrust with optimised pitch kinematics would require higher rotational speeds, potentially introducing structural and noise issues.
\end{abstract}

%
%
%
%

	\section{Introduction} \label{section:introduction}
	
	Operating aerofoils along circular trajectories exposes them to curvilinear flow conditions, producing a chordwise-varying angle-of-attack and an effect known as virtual camber, which alters the effective geometry of the aerofoil \citep{Migliore1980FlowAerodynamics}. A prominent application of this motion is the cyclorotor, where blades travel along a circular path while undergoing continuous, cyclic pitch variations. This cyclic pitching induces dynamic stall, unsteady flow separation and leading-edge vortex formation that further modify aerodynamic loads. Cyclorotors are primarily employed in the maritime industry as Voith-Schneider propellers, where their instantaneous thrust-vectoring capabilities are utilised to enhance ship manoeuvrability \citep{Voith2023VoithVSP}. In recent years, there has been growing interest in cyclorotors for the application of micro-air vehicles (MAVs), whose small size forces them to operate at typical Reynolds numbers of 10,000-40,000. In comparison to conventional rotors, cyclorotors have shown improved resilience to the detrimental effects present in this low-Reynolds regime \citep{Shrestha2017DevelopmentApplication}. 
	
	\begin{figure}[htbp]
		\begin{floatrow}
			\ffigbox[\FBwidth]
			{\includegraphics[width=\linewidth]{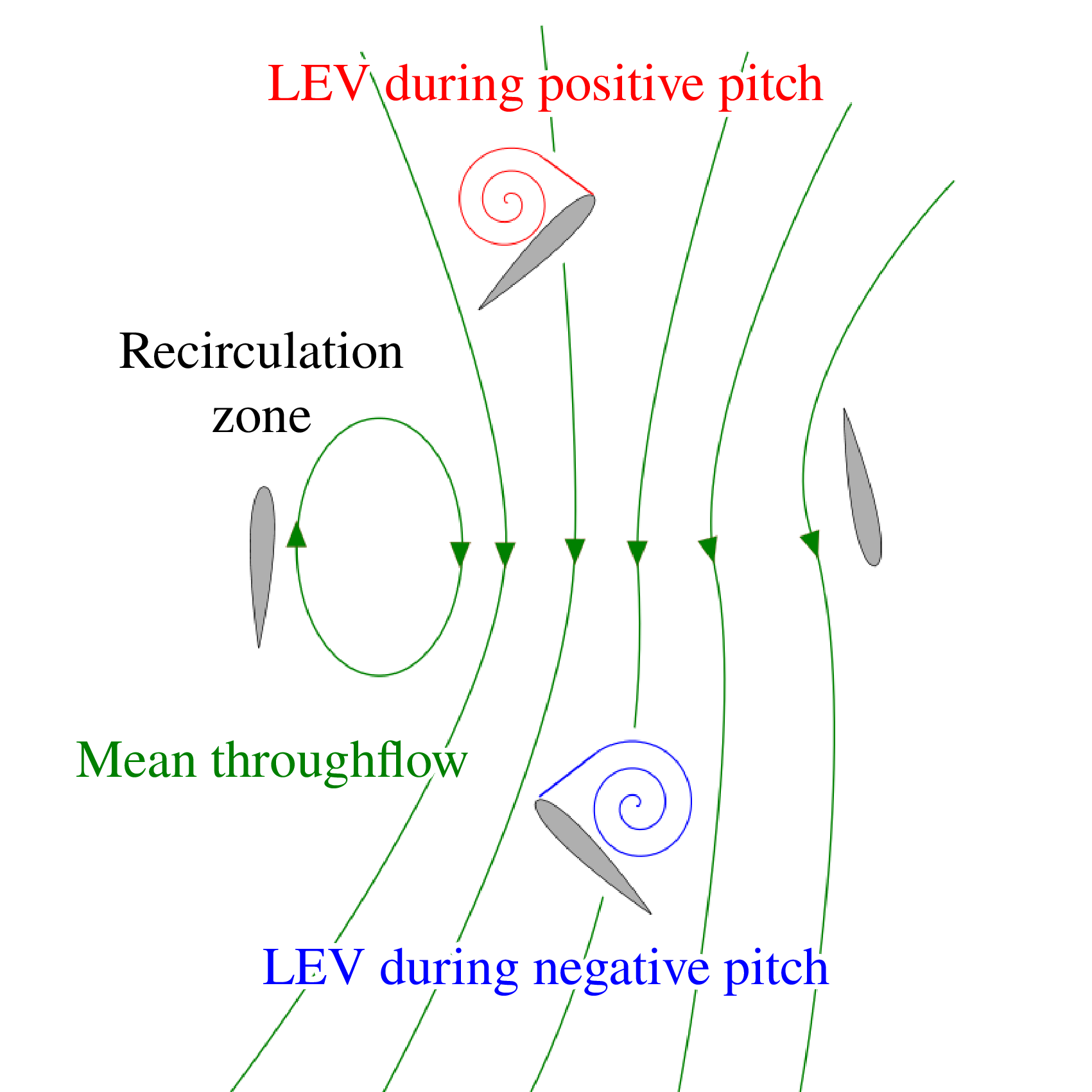}}
			{\caption{Main components of flow}\label{fig:cyclorotor_flow_diagram}}
			
			\ffigbox[\FBwidth]
			{\includegraphics[width=\linewidth]{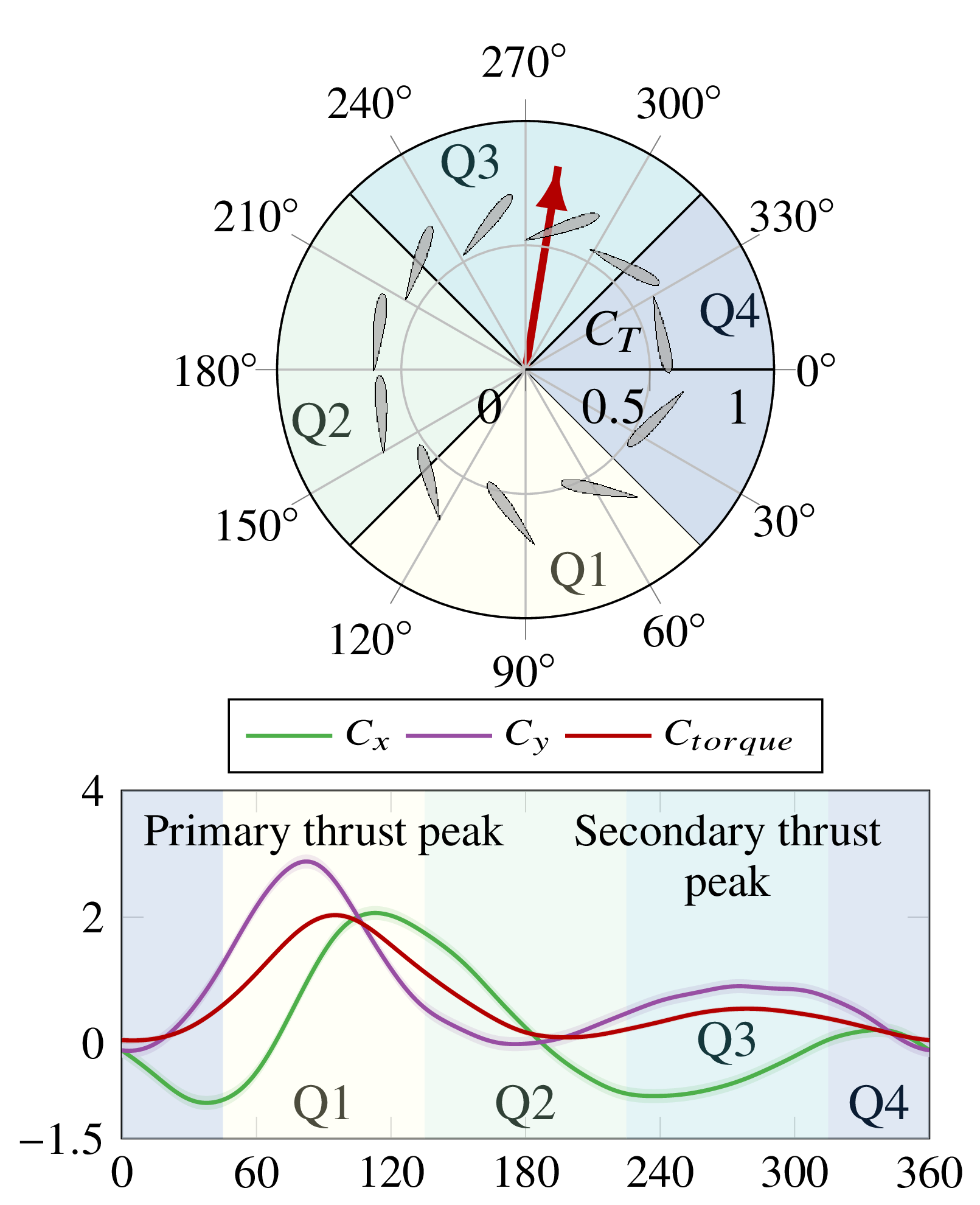}}
			{\caption{Coefficients of force and torque for a single bladed cyclorotor in hover \citep{irwin2026roledynamicstallaerofoil}}\label{fig:single_blade_loads}}
		\end{floatrow}
	\end{figure}

	A cyclorotor comprises multiple blades arranged around a rotating cylindrical structure, with each blade undergoing periodic pitch variation throughout the rotation (Fig. \ref{fig:cyclorotor_flow_diagram}). The blade pitch is considered negative over the lower half of the rotor and positive over the upper half. The minimum and maximum pitch angles correspond to the primary and secondary thrust peaks respectively  (Fig. \ref{fig:single_blade_loads}). The circular path of the blades subjects each aerofoil to curvilinear flow, producing a virtual camber effect, which modifies the effective shape of the aerofoil \citep{Migliore1980FlowAerodynamics}. This effect increases thrust during negative pitch, resulting in a larger primary thrust peak. In addition to curvilinear flow, the rotor generates a throughflow as the surrounding fluid is ingested through the upper half, passed through the rotor and ejected out the lower half (Fig. \ref{fig:cyclorotor_flow_diagram}). This throughflow acts as a downwash that  suppresses the effective angle-of-attack experienced by the blades\citep{Benedict2010FundamentalApplications}.

	The thrust and power required to drive the rotor are primarily governed by three parameters: rotational speed ($\omega$), blade-pitch amplitude ($\theta_{amp}$) and solidity ($\sigma$) \citep{Benedict2010FundamentalApplications}. Thrust scales with $\omega^2$ whereas power scales with $\omega^3$, making lower rotational speeds favourable for efficiency. This operational constraint incentivises higher blade-pitch amplitudes and solidities to maintain sufficient lift. At high blade-pitch amplitudes, cyclorotor blades experience dynamic stall: an aerodynamic phenomenon whereby unsteady blade kinematics enable transient lift generation on a blade well beyond its static stall angle \citep{Mulleners2012TheRevisited}. Dynamic stall is typically categorized into ``light" and ``deep" regimes \citep{McCroskey1981TheStall}. For a pitching blade, the flow will generally transition from light to deep dynamic stall as the pitching amplitude increases.
	In the light stall regime, the surface-normal thickness of the viscous zone around the aerofoil  is comparable to aerofoil thickness. In deep dynamic stall, this zone is on the order of the chord length and the flow is characterised by the formation and shedding of leading-edge vortices (LEVs), which are associated with a loss in lift and increase in drag.

\begin{figure}[h]
		\centering
		\includegraphics{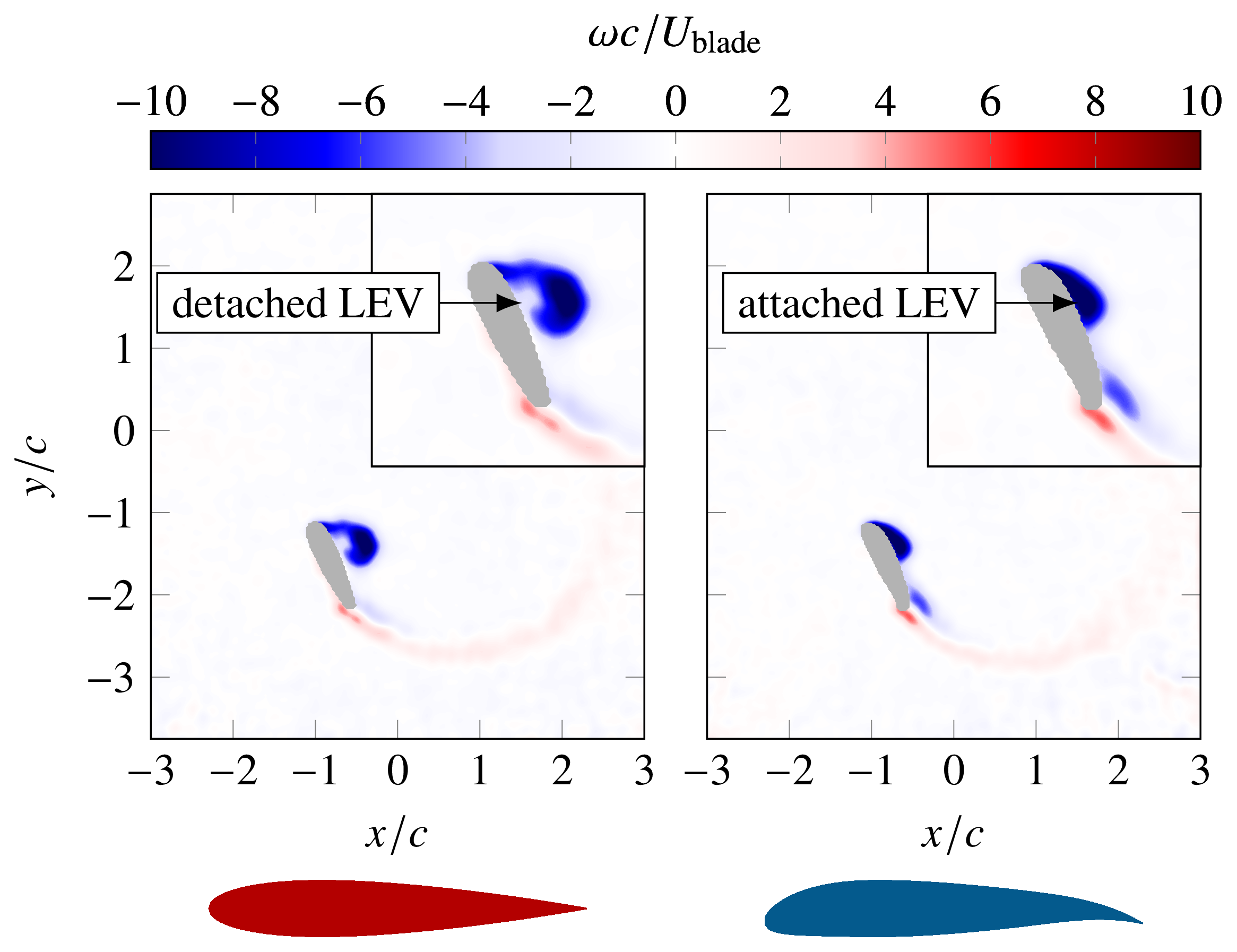}
		\caption{LEV characteristics of baseline (red) and optimised (blue) aerofoils during primary thrust peak \citep{irwin2026roledynamicstallaerofoil}}
		\label{fig:optimisation_summary}
	\end{figure}

	Aerofoil-shape optimisation can improve cyclorotor efficiency by suppressing LEV separation during the primary thrust peak (Fig. \ref{fig:optimisation_summary}) \citep{irwin2026roledynamicstallaerofoil,Tang2017UnsteadyModel, Zhang2018ThePropellers, Ferrier2020InvestigationMorphing}. However, the benefits of aerofoil optimisation are not unconditional. If the dynamic stall that the blades undergo is too deep, aerofoil geometry can prove insufficient for preventing flow separation. The impact of throughflow upon this dynamic stall behaviour was demonstrated in \cite{irwin2026roledynamicstallaerofoil}, using blade-count with constant blade-chord to control solidity and thereby the throughflow. The aerofoil shape optimisation was effective only at high blade-count/solidity, where the increased throughflow sufficiently suppressed the effective angles-of-attack. Although the optimal aerofoil is specific to a particular cyclorotor configuration, the mechanistic principle of dynamic stall control is more generalisable and has also been observed in vertical-axis wind turbines \citep{LeFouest2022TheTurbines}.

	Previous studies on optimising aerofoils for specific cyclorotor configurations have used 2D unsteady Reynolds-averaged Navier-Stokes simulation to evaluate candidate geometries \citep{irwin2026roledynamicstallaerofoil,Tang2017UnsteadyModel}. Although URANS already provides a simplified representation of the flow, its computational cost limits its use in extensive parametric optimisation. This study therefore aims to develop a low-cost, low-fidelity method for determining the optimum aerofoil shape for a given cyclorotor configuration. The model is evaluated across a range of configurations to assess its generalisability and to identify any additional constraints on optimisation. In particular, this study isolates the effects of blade-count and blade-chord from the effects of solidity by optimising aerofoils for different configurations with a fixed solidity but using different combinations of blade-count and blade-chord. The aerofoils predicted by the low-fidelity method are then compared with those obtained using the comparatively high-fidelity URANS methodology.
	
	Another approach to improve cyclorotor efficiency is through the optimisation of the blade-pitch kinematics \citep{Benedict2010FundamentalApplications, Walther2019SymmetricNumbers, Shi2022AnalysisRatio, Benedict2016DevelopmentVehicle}. The optimised pitching schedules generally have a more moderate blade-pitch amplitude and a slight positive bias, thus modifying the blade angle-of-attack in a similar way as the leading-edge droop that is characteristic of the optimised aerofoils. A direct comparison is therefore required to determine whether aerofoil optimisation offers performance or operational advantages beyond those achievable through kinematic control. This study applies both approaches to the same cyclorotor configuration to compare their effects on efficiency, thrust production and operating requirements.

	\section{Methodology}
	
\begin{figure}[h]
		\begin{floatrow}
			\capbtabbox{
				\begin{tabular}{|l|l|}
					\hline
					Rotor Radius ($R$)   & 150 \si{\milli\metre}        \\ \hline
					Blade Chord ($c$)     & 80 \si{\milli\metre}          \\ \hline
					Blade Span ($b$)       & 200 \si{\milli\metre}         \\ \hline
					Number of Blades ($N_b$) & 4           \\ \hline
					Solidity ($\sigma=N_b c/2 \pi R$) & 0.34 \\ \hline
					Max/Min Pitch    & $\pm$45 \si{\degree}            \\ \hline
					Linkage $L_1$	& 150 \si{\milli\metre}  \\ \hline
					Linkage $L_2$	& 17.21 \si{\milli\metre}  \\ \hline
					Linkage $L_3$	& 151.01 \si{\milli\metre}  \\ \hline
					Linkage $L_4$	& 24.5 \si{\milli\metre}  \\ \hline
					Rotational speed ($\omega$)        & 24 RPM     \\ \hline
					Reynolds Number   & $\sim$30,000 \\ \hline
					Working fluid & Water \\ \hline
					Default Aerofoil	& NACA 0015 \\ \hline
				\end{tabular}
			}{
				\caption{Baseline cyclorotor design parameters}
				\label{table:base_design}
			}\hfill
			\ffigbox{
				\includegraphics{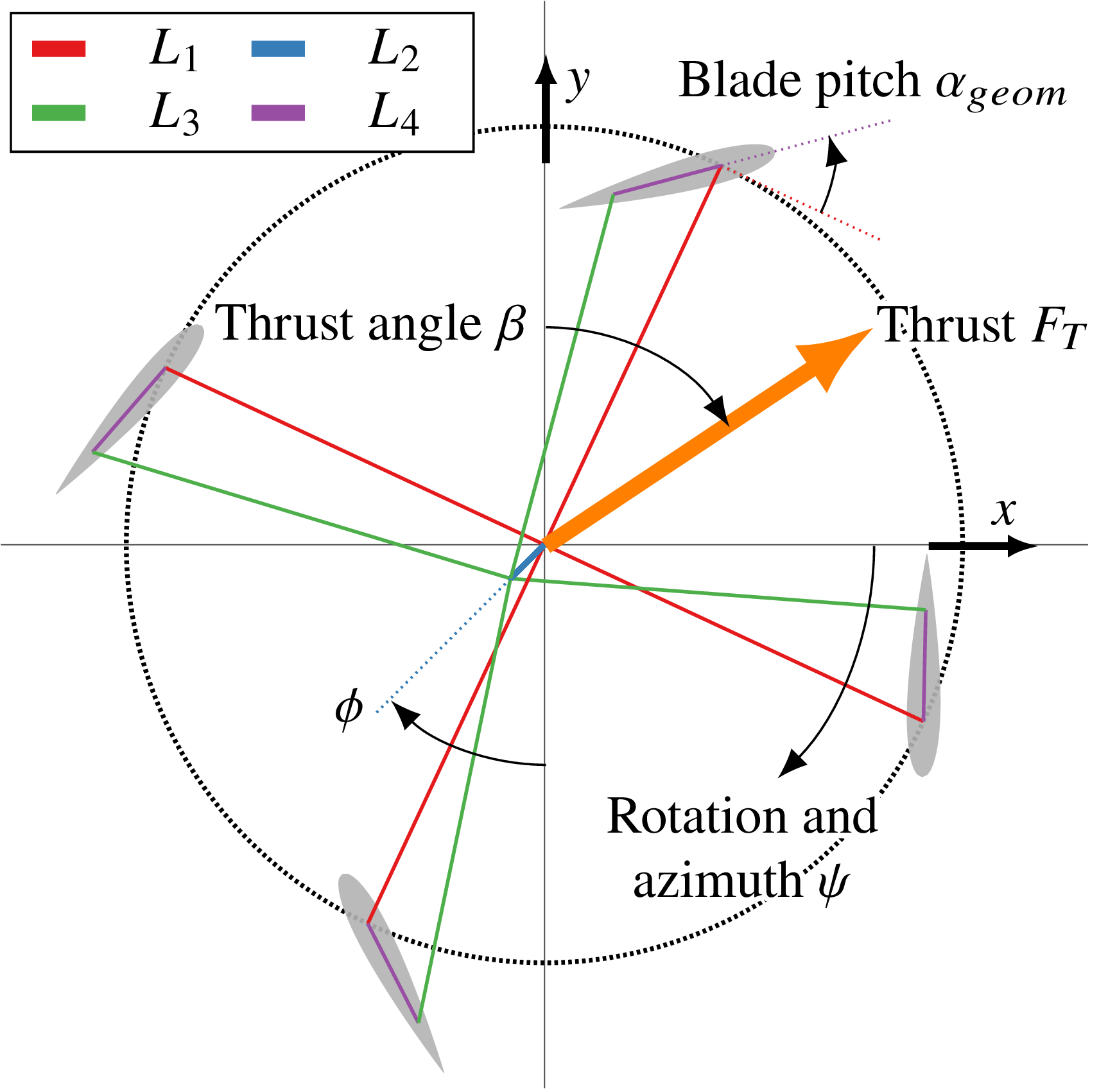}
			}{
				\caption{Cyclorotor 4-bar linkage system for blade pitching kinematics}
				\label{fig:basic_cyclorotor}
			}
		\end{floatrow}
	\end{figure}

	The current study builds upon the experimental and computational framework used in \cite{irwin2026roledynamicstallaerofoil}. A 4-bladed cyclorotor operating at a Reynolds number of 30,000 served as the primary baseline configuration (see Table \ref{table:base_design} for details). A 4-bar linkage system generated an approximately sinusoidal pitching profile $\alpha_{geom}$ between $\pm$45 \si{\degree} (Fig. \ref{fig:basic_cyclorotor}). The linkage equations used follow  \cite{Cogan2022NumericalState}.
	
	\subsection{Low-fidelity method for optimal aerofoil selection} \label{section:method_anal_optimisation}
	
	\begin{figure}[b]
		\centering
		\includegraphics{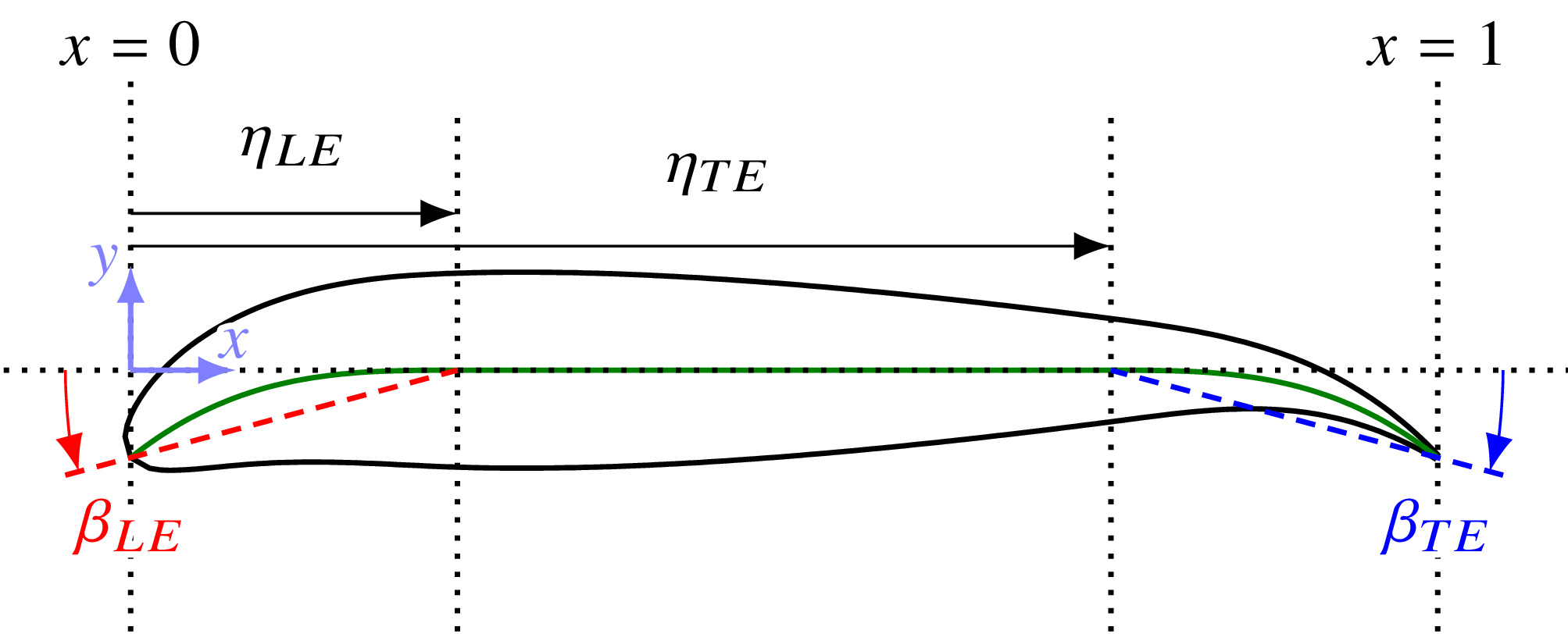}
		\caption{Blade-shape parametrization via chordwise hinge-point location ($\eta$) and droop angle ($\beta$) at LE and TE: $\eta_{\text{LE}}$, $\beta_{\text{LE}}$, $\eta_{\text{TE}}$ and $\beta_{\text{TE}}$ \citep{irwin2026roledynamicstallaerofoil}}
		\label{fig:parameters}
	\end{figure}

	A low-fidelity method is developed for selecting an optimal aerofoil for an initial cyclorotor design prior to conducting experiments or CFD. The method defines separate objective functions for the leading and trailing edge shape, which can be optimised with an arbitrary aerofoil-shape parametrization. The objective functions are described in Sections \ref{section:LE_obj_function}-\ref{section:TE_obj_function}. These functions do not predict physical quantities such as efficiency. Instead, they return dimensionless objective scalar values used solely to rank candidate geometries during optimisation. Their absolute values therefore have no independent physical interpretation.
	
	Optimised aerofoils were obtained by minimising these objective functions for the LE and TE using a simplex search method implemented in MATLAB's \textit{fminsearch}, using a symmetric NACA 0015 as the baseline geometry. In this study, aerofoil shape was parametrised with the same methodology used in \cite{irwin2026roledynamicstallaerofoil}, applying independent droops to the LE and TE by specifying a droop hinge-point ($\eta$) and droop-angle ($\beta$)  as illustrated in Fig. \ref{fig:parameters}. 
	The MATLAB code for this method has been made available as a GitHub repository\footnote{\url{https://github.com/BenIrwin95/Low-Fidelity-Cyclorotor-Aerofoil-Optimiser}}.

	\subsubsection{Leading-edge objective function} \label{section:LE_obj_function}
	
	Optimised aerofoils prevent boundary layer separation at the leading-edge (LE) as the blade undergoes maximum negative pitch during the primary thrust peak \citep{irwin2026roledynamicstallaerofoil}. Boundary layer separation occurs when a sufficiently strong adverse pressure gradient is present. On an aerofoil this will often occur near the LE where a suction peak forms. The magnitude of this suction peak and the surrounding gradients are generally proportional to the local angle-of-attack at the LE\citep{Abbott1959TheoryData,Carlson1987ApplicabilityEdges}. 
	
\begin{figure}[h]
		\centering
		\includegraphics{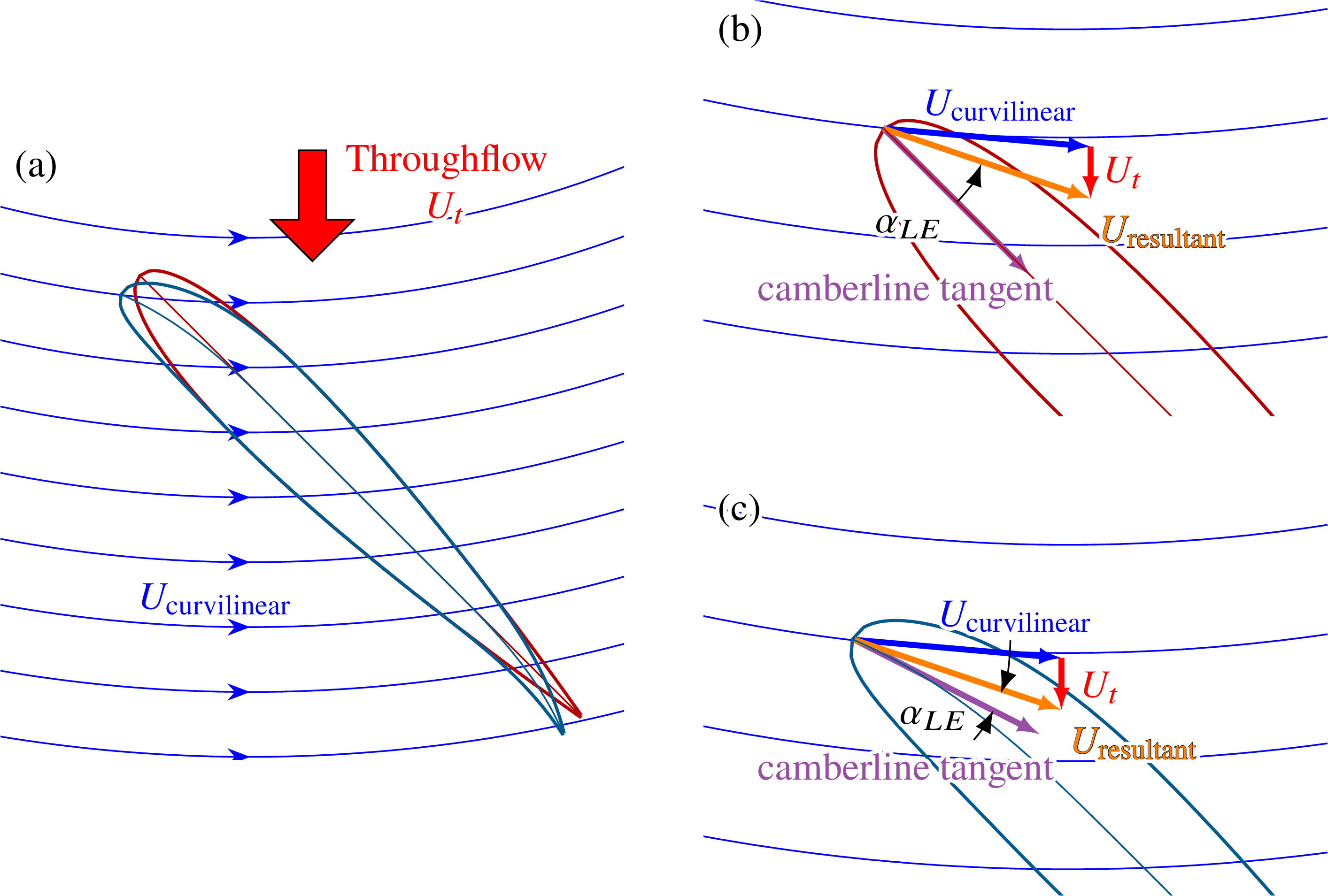}
		\caption{Primary flow components and angles around LE of cyclorotor blade at maximum negative pitch. (a) baseline (red) and optimised (light blue) aerofoil overlayed with curvilinear streamlines (blue) and throughflow (red). (b) $\alpha_{LE}$ for baseline aerofoil. (c) $\alpha_{LE}$ for optimised aerofoil.}
		\label{fig:shape_generalisation_diagram}
		\phantomsubcaption\label{fig:shape_generalisation_diagram_A}
		\phantomsubcaption\label{fig:shape_generalisation_diagram_B}
		\phantomsubcaption\label{fig:shape_generalisation_diagram_C}
	\end{figure}

	The baseline and optimised aerofoils for the 4-bladed cyclorotor configuration identified in previous work from \cite{irwin2026roledynamicstallaerofoil} are compared schematically in Fig. \ref{fig:shape_generalisation_diagram} with respect to the primary flow components. Both aerofoils are shown at their maximum negative pitch ($\alpha_{geom} = \ang{-45}$) corresponding to the primary thrust peak. The primary flow components comprises of the curvilinear flow generated by the blade motion and the throughflow exiting the rotor (Fig. \ref{fig:shape_generalisation_diagram_A}). These flow components determine the local angle-of-attack at the LE with respect to the camberline ($\alpha_{LE}$) (Fig. \ref{fig:shape_generalisation_diagram_B}). The LE droop of the optimised aerofoil reduces this angle and thus weakens the suction peak and suppresses separation.
	
\begin{figure}[t]
		\centering
		\includegraphics[]{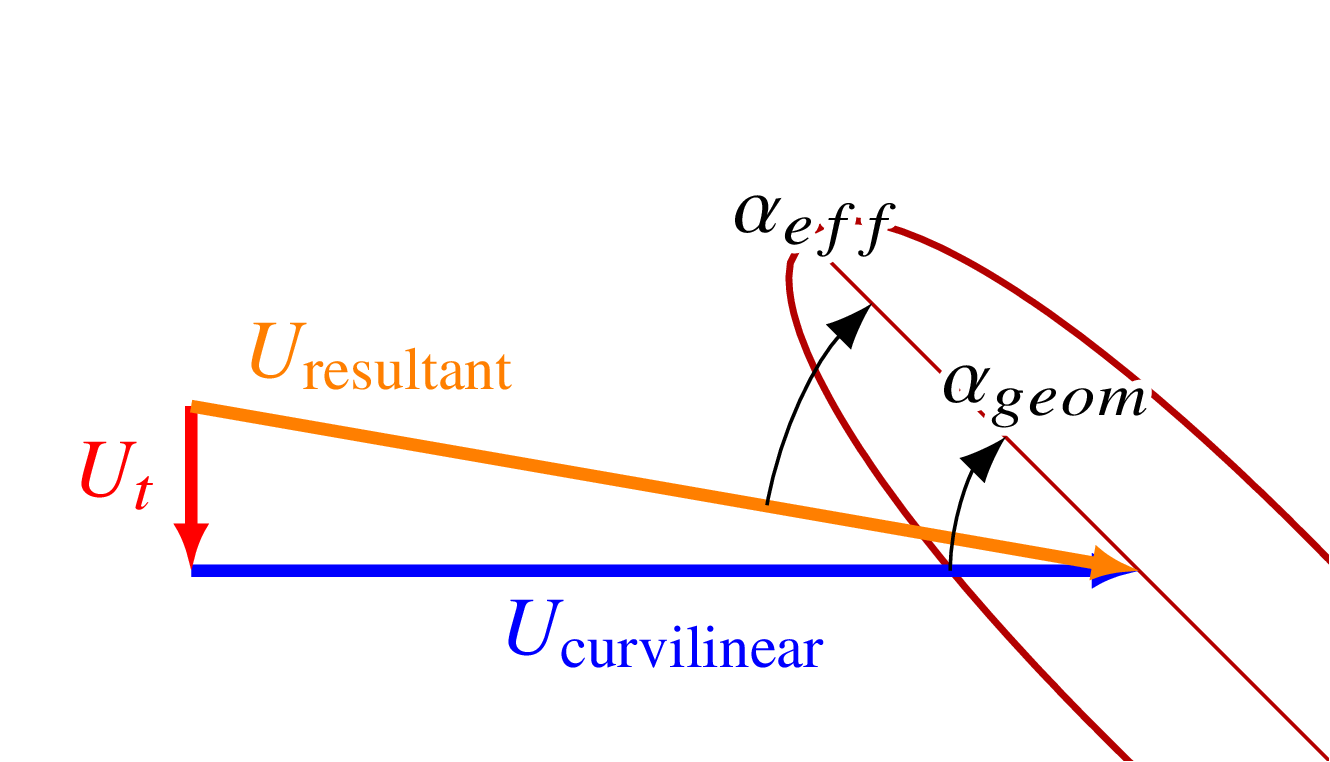}
		\caption{Diagram of geometric angle-of-attack and effective angle-of-attack after accounting for influence of throughflow $U_t$}
		\label{fig:AoA}
	\end{figure}

	This observation formed the logical basis for the LE objective function in the low-fidelity method. The LE geometry is optimised by minimising $\alpha_{LE}$ at the maximum negative blade pitch during the primary thrust peak ($\alpha_{geom} = \ang{-45}$). The throughflow velocity $U_t$ required to obtain $\alpha_{LE}$ was calculated using a single streamtube model \citep{Cogan2022NumericalState}. In this model, flow through a cyclorotor is approximated as a single streamtube with uniform flowspeed across its width. Within this streamtube, the cyclorotor is modelled as an actuator disk, allowing throughflow to be calculated for a given value of thrust. 
	The individual blade aerodynamic forces are calculated using the linear lift relation  $C_l = 2 \pi \alpha_{eff}$ (Fig. \ref{fig:AoA}). The throughflow modifies the effective blade angle-of-attack and consequently the aerodynamic forces. The model was therefore solved iteratively until the predicted thrust and induced throughflow converge.
	
	The single streamtube predictions for throughflow speed were compared with particle image velocimetry (PIV) data from \cite{irwin2026roledynamicstallaerofoil} for the 1- and 4-bladed configurations with a constant chord length using both the baseline NACA 0015 and optimised aerofoil (Table \ref{table:throughflow_comparison}). The two blade-counts were used to assess whether the model remained applicable beyond the baseline four-bladed configuration. The mean throughflow velocity is obtained from the PIV data by averaging the flowspeed within the rotor cage. For the 4-bladed configurations, the single streamtube model predicted  $U_t/U_{blade}$ with an error below 10\%. The 1-bladed configurations results exhibited slightly higher percentage errors (<20\%) due to the lower solidity and correspondingly weaker throughflow. Although the relative error increases at low blade count/solidity, aerofoil optimisation is primarily relevant at higher blade count/solidity, where the throughflow model is more accurate \citep{irwin2026roledynamicstallaerofoil}. To avoid aerofoils with sharp changes in curvature, the LE objective function was extended to minimise the average local angle-of-attack magnitude across the front 25\% of the LE.

	\begin{table}[h]
		\centering
		\begin{tabular}{lllll}
			& &\multicolumn{3}{c}{$U_t/U_{blade}$}                      \\ \hline
			\multicolumn{1}{|l|}{$N_b$} & \multicolumn{1}{l|}{Aerofoil}  & \multicolumn{1}{l|}{PIV}  & \multicolumn{1}{l|}{single streamtube model} & \multicolumn{1}{l|}{\% error} \\ \hline
			\multicolumn{1}{|l|}{1}    & \multicolumn{1}{l|}{Baseline}  & \multicolumn{1}{l|}{0.22} & \multicolumn{1}{l|}{0.26} & \multicolumn{1}{l|}{18\%}      \\ \hline
			\multicolumn{1}{|l|}{1}    & \multicolumn{1}{l|}{Optimised} & \multicolumn{1}{l|}{0.23} & \multicolumn{1}{l|}{0.26} & \multicolumn{1}{l|}{13\%}      \\ \hline
			\multicolumn{1}{|l|}{4}    & \multicolumn{1}{l|}{Baseline}  & \multicolumn{1}{l|}{0.44} & \multicolumn{1}{l|}{0.41} & \multicolumn{1}{l|}{-7\%}     \\ \hline
			\multicolumn{1}{|l|}{4}    & \multicolumn{1}{l|}{Optimised} & \multicolumn{1}{l|}{0.44} & \multicolumn{1}{l|}{0.42} & \multicolumn{1}{l|}{-5\%}     \\ \hline
		\end{tabular}
		\caption{Comparison of throughflow velocity within rotor cage of a cyclorotor using PIV and single streamtube model (PIV data from \citep{irwin2026roledynamicstallaerofoil})}
		\label{table:throughflow_comparison}
	\end{table}

	\subsubsection{Trailing-edge objective function} \label{section:TE_obj_function}
	
	\begin{table}[b]
		\centering
		\begin{tabular}{|l|l|l|l|}
			\hline
			Case & FM   & Thrust (\unit{\newton}) & Torque (\unit{\newton\metre}) \\ \hline
			Baseline     & 0.39 & 3.00   & 0.49   \\ \hline
			Optimised    & 0.65 & 3.52   & 0.37   \\ \hline
			Optimised LE & 0.57 & 3.40   & 0.40   \\ \hline
			Optimised TE & 0.39 & 3.03   & 0.50   \\ \hline
		\end{tabular}
		\caption{Isolated impact of LE and TE components of the optimised aerofoil for the 4-bladed cyclorotor configuration. All shown results are from URANS simulations.}
		\label{table:performance_shape_components}
	\end{table}
	
	To determine an optimisation criterion for trailing-edge (TE), URANS simulations were performed for the baseline 4-bladed configuration using the LE and TE portions of the optimised aerofoil independently as summarised in Table \ref{table:performance_shape_components}. The full optimised shape increased thrust and decreased torque. While the isolated LE portion of the optimised aerofoil also increased thrust and decreased torque, the isolated TE portion increased both thrust and torque. The TE camberline was observed to be approximately perpendicular to the local curvilinear flow, as shown in Fig. \ref{fig:shape_generalisation_diagram_A}. This orientation is expected to direct the aerofoil wake to leave the rotor radially and away from the path of the following blade. By encouraging the ejection of the flow from the rotor, thrust increases in accordance with momentum conservation. Consequently, the TE objective function was made to minimise the angle between the camberline normal at the TE and the corresponding curvilinear flow direction. In contrast to the LE objective function, the downwash velocity was not included. This was because the underlying principle to this objective function is to control the wake direction rather than control the angle-of-attack.

	\subsection{Computational methodology}
	
	The performance of candidate aerofoil geometries and the accuracy of the low-fidelity method optimisation method were assessed using results from the comparatively high-fidelity 2D URANS methodology established in \cite{irwin2026roledynamicstallaerofoil}. This approach has been successfully applied to cyclorotor simulations in previous studies \citep{Hu2016InvestigationSolver, Yun2007DesignCyclocopter, Xisto2017ParametricConditions, Zhang2018ThePropellers, Shi2022NumericalNumber, Yu2016Two-dimensionalHover, Hansen2021NumericalApproach, Ullah2022Two-DimensionalStall}.  The simulations are implemented with Ansys FLUENT using a $k-\omega$ SST turbulence model. The blade kinematics were implemented via overset meshing with 2000 timesteps per revolution. Each simulation was solved for 120 blade-passes to achieve statistically stationary conditions and results are averaged over the last 10 revolutions. A more detailed discussion of the limitations of this URANS methodology can be found in \cite{irwin2026roledynamicstallaerofoil}.
	
	The present study used URANS primarily to compare and rank candidate aerofoil geometries rather than determine absolute aerodynamic quantities. Comparisons with previously obtained PIV data showed that URANS predicted differences in the strength, coherence and persistence of vortical structures, but captured the presence/absence of blade-surface separation and reproduced the relative performance ranking of the baseline and optimised aerofoils (discussed further in Appendix \ref{section:URANS_comparison}). The absolute performance values and detailed vortical structures were therefore interpreted cautiously, while the relative differences between candidate designs were used to guide the optimisation. Accordingly, the objective of the present study is to assess whether the proposed low-fidelity method reproduces the aerofoil-design trends identified by the URANS optimisation framework. Further limitations of the URANS methodology are discussed in our previous work \citep{irwin2026roledynamicstallaerofoil}. Each URANS simulation for a given configuration took approximately 60 hours using 40 cores on the University of Southampton's Iridis HPC cluster.

	The URANS simulations were coupled with a Kriging surrogate model to determine the optimal aerofoil shape \citep{irwin2026roledynamicstallaerofoil}. An initial design of experiments comprising of 20 aerofoil geometries was generated via a space-filling Latin hypercube. Subsequent updated geometries were selected using the expected improvement criterion and the current global maximum of the surrogate model \citep{Jones1998EfficientFunctions}. Additional designs were iteratively added until no further improvement was observed in the best evaluated design after 6 consecutive updates. These methods were implemented in MATLAB using the toolbox from \cite{Forrester2008EngineeringGuide}. The objective was to maximise the Figure of Merit ($FM$), which measures hover efficiency, as defined in Eq. \ref{eq:fm} \citep{Benedict2015ExperimentalHover}):
	
	\begin{equation}
		FM = PL\sqrt{\frac{DL}{2 \rho}},
		\label{eq:fm}
	\end{equation}
	where, $PL$ and $DL$ are Power Loading and Disk Loading respectively and are given by-
	\begin{equation}
		PL= \frac{\text{thrust}}{\text{power}}= \frac{\text{thrust}}{\text{torque}\cdot\text{RPM}\cdot\frac{2\pi}{60}}, \quad DL= \frac{\text{thrust}}{\text{rotor projected area}} = \frac{\text{thrust}}{2Rb}.
	\end{equation}

	\subsection{Optimisation of cyclorotor kinematics}
	
	To compare blade-pitch kinematics optimisation with aerofoil-shape optimisation, the blade-pitch kinematics were parametrised using 
	pitch amplitude ($\theta_{amp}$) and mean blade pitch ($\theta_{mean}$). The new blade pitching kinematics were implemented with the same 4-bar linkage system as the baseline configuration showin in Fig. \ref{fig:basic_cyclorotor} following \citep{Cogan2022NumericalState}, keeping linkage lengths $L_1$ and $L_4$ identical to the baseline configuration (Table. \ref{table:base_design}). Linkage lengths $L_2$ and $L_3$ were optimised numerically to attain a specified value for $\theta_{amp}$ and $\theta_{mean}$.  This was done with \textit{fminsearch} in Matlab using a simplex search method and starting from an initial guess for $L_2$ and $L_3$ equal to their values for the original $\pm\ang{45}$ pitch kinematics. The parameters $\theta_{amp}$ and $\theta_{mean}$ were optimised to maximise $FM$ with the same high-fidelity method used in \cite{irwin2026roledynamicstallaerofoil}.
	
	\section{Results}
	
	\subsection{Optimal aerofoil for baseline 4-bladed configuration using low-fidelity method} \label{section:4blade_optimisation}
	
\begin{figure}[h]
		\centering
		\includegraphics{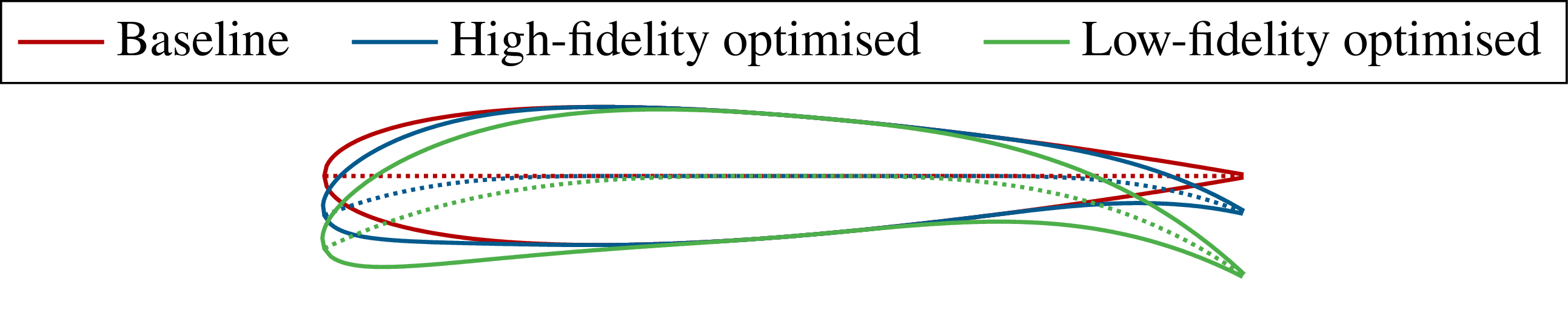}
		\caption{Comparison of aerofoil shapes optimised via different methods.}
		\label{fig:opt_shape_anal_baseline}
	\end{figure}

	The low-fidelity optimisation method was applied to the baseline four-bladed configuration, for which a high-fidelity optimal aerofoil had previously been established \citep{irwin2026roledynamicstallaerofoil}, to assess the accuracy of the method. This produced an aerofoil with similar shape but a larger droop to both the LE and TE (Fig. \ref{fig:opt_shape_anal_baseline}).

	The baseline, high-fidelity optimised and low-fidelity optimised aerofoils are compared using  URANS simulations (Table \ref{table:aerofoil_comparison_analytical}).
	The low-fidelity optimum increased the Figure of Merit by 51\% relative to the baseline, compared with a 67\% increase for the high-fidelity optimum. Both optimised aerofoils improved $FM$ by increasing thrust and decreasing torque. The low-fidelity optimum produced a greater increase in thrust but a smaller reduction in torque, resulting in a lower figure of merit than the high-fidelity optimum. The TE droop was previously shown to increase both thrust and torque (Section \ref{section:method_anal_optimisation}). The more pronounced TE droop of the low-fidelity optimal aerofoil is therefore consistent with the enhanced thrust and smaller torque reduction, limiting the improvement in overall efficiency relative to the high-fidelity optimum. However, the computational cost—and therefore runtime—of the low-fidelity method is significantly lower ($\approx$2 minutes) than that of the high-fidelity approach ($\approx$1 month).
	
	\begin{table}[h]
		\centering
		\begin{tabular}{|l|l|l|l|}
			\hline
			Case         & FM   & Thrust (\unit{\newton}) & Torque (\unit{\newton\metre}) \\ \hline
			Baseline                & 0.39 & 3.00   & 0.49   \\ \hline
			High-fidelity optimisation  & 0.65 & 3.52   & 0.37   \\ \hline
			Low-fidelity optimisation & 0.59 & 3.65   & 0.43   \\ \hline
		\end{tabular}
		\caption{URANS predicted performance of aerofoils optimised via different methods.}
		\label{table:aerofoil_comparison_analytical}
	\end{table}
	
	\begin{figure}[h]
		\centering
		\includegraphics{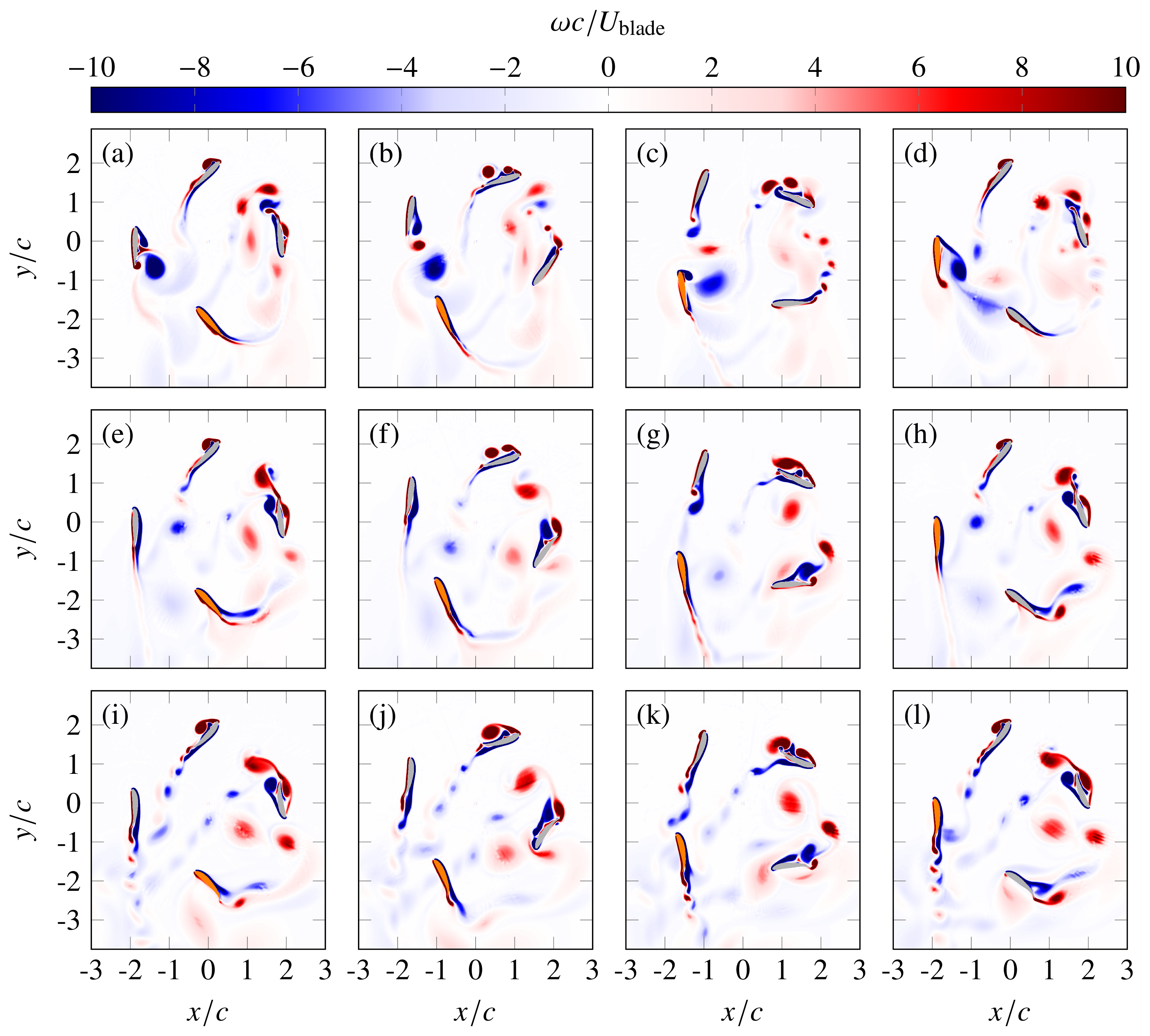}
		\caption{Comparison of URANS vorticity flow fields for 4-bladed cyclorotor using NACA 0015 (a-d),  high-fidelity optimised aerofoil (e-h) and low-fidelity optimised aerofoil (i-l). Columns from left-to-right show snapshots for $t/T=0, 0.07, 0.15, 0.25$. Blade 1 is marked in orange.}
		\label{fig:4blade_vorticity_baseline_opt_anal_opt}
		\phantomsubcaption\label{fig:4blade_vorticity_baseline_opt_anal_opt_a}
		\phantomsubcaption\label{fig:4blade_vorticity_baseline_opt_anal_opt_b}
		\phantomsubcaption\label{fig:4blade_vorticity_baseline_opt_anal_opt_c}
		\phantomsubcaption\label{fig:4blade_vorticity_baseline_opt_anal_opt_d}
		\phantomsubcaption\label{fig:4blade_vorticity_baseline_opt_anal_opt_e}
		\phantomsubcaption\label{fig:4blade_vorticity_baseline_opt_anal_opt_f}
		\phantomsubcaption\label{fig:4blade_vorticity_baseline_opt_anal_opt_g}
		\phantomsubcaption\label{fig:4blade_vorticity_baseline_opt_anal_opt_h}
		\phantomsubcaption\label{fig:4blade_vorticity_baseline_opt_anal_opt_i}
		\phantomsubcaption\label{fig:4blade_vorticity_baseline_opt_anal_opt_j}
		\phantomsubcaption\label{fig:4blade_vorticity_baseline_opt_anal_opt_k}
		\phantomsubcaption\label{fig:4blade_vorticity_baseline_opt_anal_opt_l}
	\end{figure}
	
	To investigate the aerodynamic mechanisms behind the performance differences between the baseline and optimised aerofoils, the corresponding URANS vorticity fields are examined in Fig. \ref{fig:4blade_vorticity_baseline_opt_anal_opt}. The selected time instants are chosen to highlight the formation and separation of the LEV during primary thrust peak. Both optimised aerofoils suppress LE flow separation on blade 1 during the primary thrust peak as expected (Fig. \ref{fig:4blade_vorticity_baseline_opt_anal_opt}). 
	The blade which undergoes maximum positive pitch in the upper-half of the rotor, sheds a positive LEV for all three cases as the orientation of the LE droop increases LE angle-of-attack (Figs. \ref{fig:4blade_vorticity_baseline_opt_anal_opt_a}, \ref{fig:4blade_vorticity_baseline_opt_anal_opt_e}, \ref{fig:4blade_vorticity_baseline_opt_anal_opt_i}).
	
	The primary difference in the flowfields for the two optimised aerofoils is distinguishable by the flow at the TE. The flow leaves the trailing edge of blade 1 cleanly in the high-fidelity case (Figs. \ref{fig:4blade_vorticity_baseline_opt_anal_opt_g} -\ref{fig:4blade_vorticity_baseline_opt_anal_opt_h}), whereas for the low-fidelity optimised aerofoil the flow leaves as a vortex street (Figs. \ref{fig:4blade_vorticity_baseline_opt_anal_opt_k}-\ref{fig:4blade_vorticity_baseline_opt_anal_opt_l}). This unsteady wake flow  at the TE is consistent with higher drag/torque and provides a plausible explanation for the smaller torque reduction achieved by the low-fidelity optimum. The flowfields support the interpretation that the TE droop is overly pronounced and adversely affects the flow behaviour.

	\begin{figure}[h]
		\includegraphics{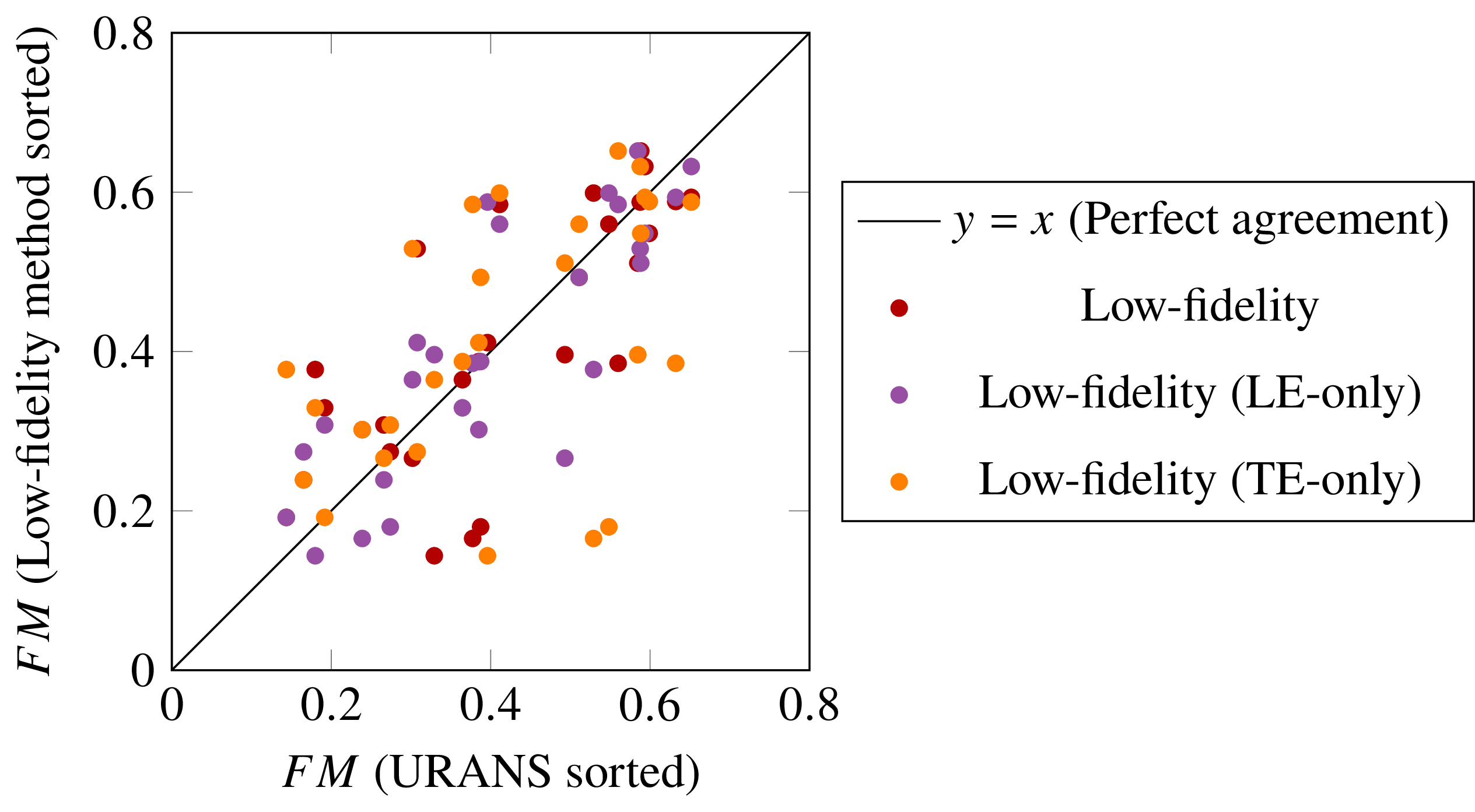}
		\caption{Comparison of low-fidelity method design ranking vs. URANS reference ranking} 
		\label{fig:ranking_comparison}
	\end{figure}
	
	To further assess the low-fidelity method's ability to rank aerofoil performance, it was applied to all aerofoil designs for the 4-bladed configuration that were evaluated with URANS. Each design therefore had a $FM$ value from the URANS simulation and an objective scalar value from the low-fidelity method. 
	The designs are sorted with respect to their $FM$ value and with respect to their low-fidelity objective scalar. The URANS-derived $FM$ values from each sorted list are be plotted against one-another to demonstrate their agreement in ranking (Fig. \ref{fig:ranking_comparison}). If the low-fidelity method were to rank the designs in the exact same order as simply comparing their URANS $FM$ values, then all points in this figure would be displayed along a $y=x$ line. Deviations from this line therefore indicate differences between rankings predicted by the low-fidelity method and URANS.

	The low-fidelity method reproduced the overall URANS ranking trend, with some deviations for individual designs. The agreement corresponded to a root-mean-square error (RMSE) of 0.11 and a cross-correlation coefficient of 0.75.
	This comparison was repeated using the LE and TE portions of the low-fidelity method in isolation. The LE objective produced a slightly improved ranking, with a RMSE of 0.09 and a cross-correlation coefficient of 0.83. In contrast, the TE objective produced a weaker ranking with a RMSE of 0.16 and a correlation coefficient of 0.44. Aerofoil performance was previously shown to depend more strongly on LE geometry than on TE geometry, which is consistent with the weaker performance of the TE-only ranking \citep{irwin2026roledynamicstallaerofoil}. The improved performance of the leading-edge-only objective suggests that the trailing-edge formulation contributes limited additional ranking capability in its current form and may benefit from further refinement.

	\subsection{Generalisability of the low-fidelity method for varying chord length} 
	\label{section:results_optimisation_double_chord}
	
	The low-fidelity optimisation method was developed using the results of the high-fidelity optimisation of the same 4-bladed configuration used in Section \ref{section:4blade_optimisation}. The present section therefore evaluates the low-fidelity method's generalisability for a different rotor configuration. Throughflow velocity was previously shown to influence the effectiveness of aerofoil optimisation \citep{irwin2026roledynamicstallaerofoil}. A two-bladed cyclorotor with twice the baseline chord length of $160\unit{\milli\metre}$, was investigated to isolate the effect of blade chord while minimising changes in throughflow. The rotor solidity was held constant to minimise corresponding changes in thrust and throughflow.

	\begin{table}[t]
		\begin{tabular}{|l|l|l|l|l|}
			\hline
			$N_b$ & Reduced frequency $k$    & Thrust (\unit{\newton}) & Torque(\unit{\newton\metre}) & FM   \\ \hline
			2    & 0.53 & 2.60   & 0.75   & 0.20 \\ \hline
			3    & 0.36 & 3.35   & 0.55   & 0.42 \\ \hline
			4    & 0.27 & 3.00   & 0.49   & 0.39 \\ \hline
			5    & 0.21 & 2.56   & 0.41   & 0.36 \\ \hline
			6    & 0.17 & 2.90   & 0.44   & 0.38 \\ \hline
			8    & 0.13 & 1.85   & 0.32   & 0.29 \\ \hline
		\end{tabular}
		\caption{Performance metrics for different blade-counts and fixed solidity ($\sigma=0.34$). Results from URANS.}
		\label{table:constant_solidity}
	\end{table}

	\begin{figure}[h]
		\centering
		\includegraphics{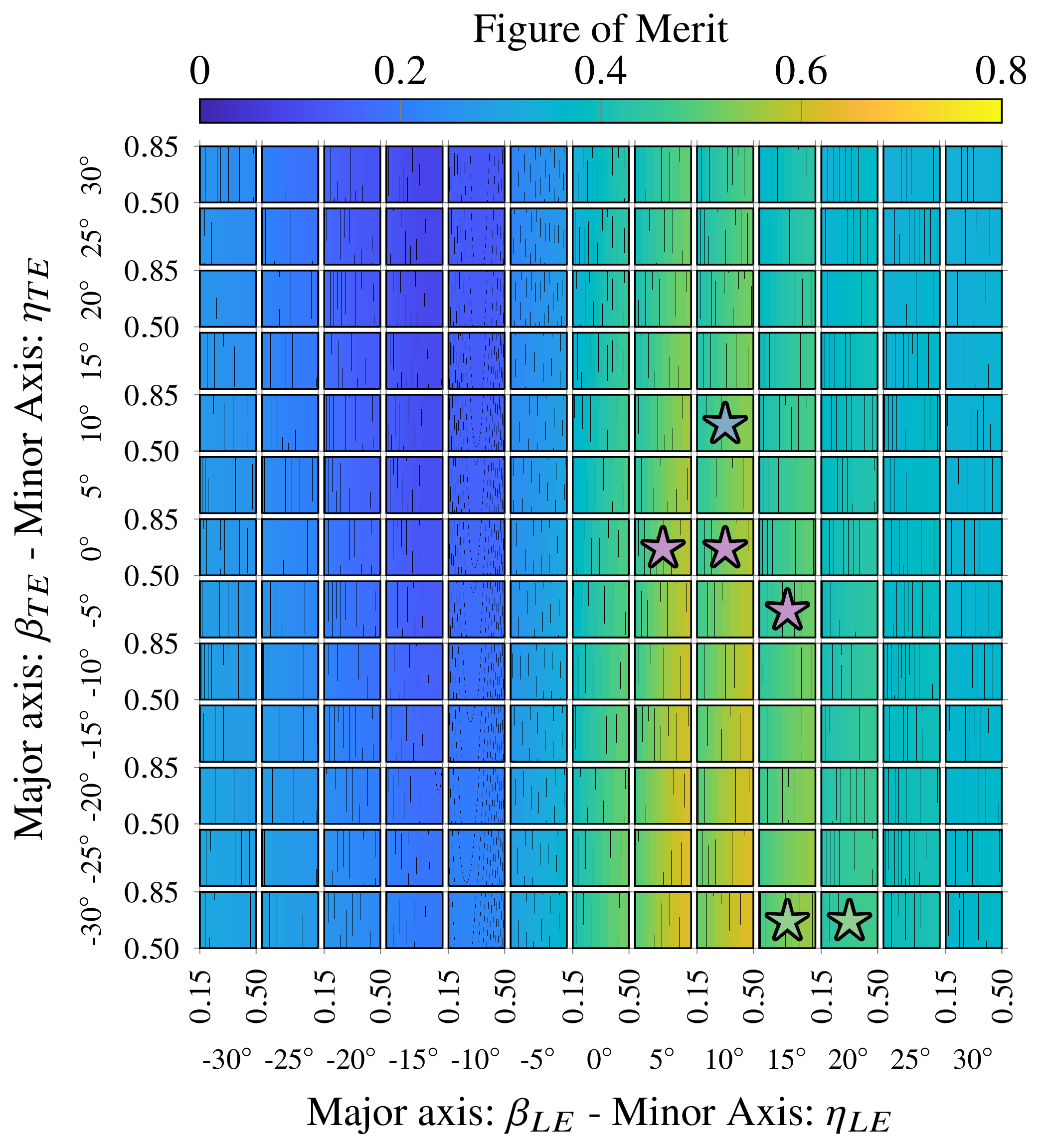}
		\caption{FM trends for a 2-bladed cyclorotor with doubled chord-length with variations in leading edge (x axis) and trailing edge (y axis) shape. Full parameter space mapped out with Kriging surrogate model fitted to URANS results. Stars indicate location of low-fidelity optimised aerofoil (blue), thrust-maximising aerofoils (purple) and torque minimising aerofoils (green)}
		\label{fig:design_space_2blade}
	\end{figure}
	
	URANS simulations were used to assess thrust across configurations with different blade-counts and chord-lengths for the same baseline solidity ($\sigma$), as indicated in Table \ref{table:constant_solidity}. For blade-counts between 2-6, the predicted thrust remains within $\lesssim17\%$ of the baseline 4-bladed configuration. The single streamtube model used in the low-fidelity model, states that throughflow velocity $U_t \propto \sqrt{F_T}$. This $\lesssim17\%$ variation in thrust thus corresponds to an estimated variation in $U_t$ of $\lesssim8\%$. This variation was considered sufficiently small to treat the throughflow effects as approximately constant. This constant throughflow assumption becomes less reliable at larger $N_b$. Maintaining constant solidity requires a reduction in blade chord as blade count increases, which also reduces the reduced frequency ($k$). With a lower reduced frequency, the flow becomes more steady and the dynamic stall transitions towards conventional static stall \citep{McCroskey1981TheStall}, potentially reducing the effectiveness of aerofoil-shape optimisation.

\begin{figure}[h]
		\centering
		\includegraphics{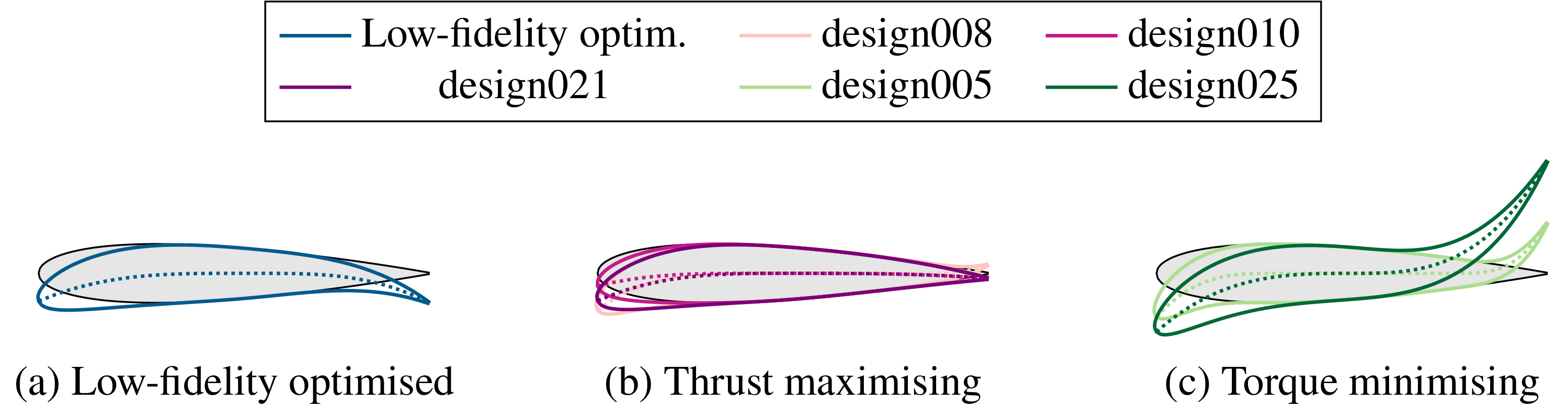}
		\caption{Comparison of optimal aerofoil shapes for a 2-bladed cyclorotor with doubled chord-length. Baseline NACA 0015 shown in grey. }
		\label{fig:opt_shape_anal_2blade}
		\phantomsubcaption\label{fig:opt_shape_anal_2blade_A}
		\phantomsubcaption\label{fig:opt_shape_anal_2blade_B}
		\phantomsubcaption\label{fig:opt_shape_anal_2blade_C}
	\end{figure}

	The variations in $FM$ for the 2-bladed cyclorotor were found to have far less obvious trends in comparison to the 4-bladed configuration, and identified multiple competing regions within the design space (Fig. \ref{fig:design_space_2blade}). To understand these trends, a subset of designs with $FM\geq0.6$ was isolated and compared (Table \ref{table:results_2blade}, Fig. \ref{fig:opt_shape_anal_2blade}). The optimisation revealed two distinct design families for maximising $FM$. The first design family prioritised the maximisation of thrust and comprised aerofoils with small positive LE and TE droops (Fig. \ref{fig:opt_shape_anal_2blade_B}). These aerofoils increased thrust and decreased torque. The shape of thrust-maximising aerofoils align with with the trends observed in \cite{irwin2026roledynamicstallaerofoil} and thus are similar to the low-fidelity optimised shape (Fig. \ref{fig:opt_shape_anal_2blade_A}). For this  configuration, the low-fidelity optimisation method produced the most efficient aerofoil out of all the tested designs.
	
	\begin{table}[h]
		\centering
		\begin{tabular}{r@{\,}l|l|l|l|l|}
			\hhline{~~|----|}
			& & Case & $FM$ & Thrust (\unit{\newton}) & Torque (\unit{\newton\metre}) \\ \hhline{~~|----|}
			& & Baseline & 0.20 & 2.60 & 0.75 \\ \hhline{~~|----|}
			& & Low-fidelity optimisation & 0.68 & 3.92 & 0.41 \\ \hhline{~~|----|}
			\ldelim\{{3}{*}[Thrust-maximising] & & design008 \inlineline{cb_purple_seq2} & 0.60 & 4.08 & 0.50 \\
			& & design010 \inlineline{cb_purple_seq5} & 0.65 & 4.83 & 0.59 \\
			& & design021 \inlineline{cb_purple_seq6} & 0.63 & 4.37 & 0.53 \\ \hhline{~~|----|}
			\ldelim\{{2}{*}[Torque-minimising] & & design005 \inlineline{cb_green_seq3} & 0.65 & 3.33 & 0.34 \\
			& & design025 \inlineline{cb_green_seq6} & 0.65 & 3.20 & 0.32 \\ \hhline{~~|----|}
		\end{tabular}
		\caption{Results for different aerofoils with $FM \ge 0.6$ for a two-bladed cyclorotor with doubled chord length.}
		\label{table:results_2blade}
	\end{table}
	
	The second design family prioritised the minimisation of torque and produced blades with large positive LE droop and large \textit{negative} TE droop (Fig. \ref{fig:opt_shape_anal_2blade_C}). These aerofoils also increased thrust and decreased torque, but produced smaller thrust gains and larger torque reductions than the thrust-maximising designs. The increased chord length places the TE much farther from the centre of rotation during negative pitch, increasing its local velocity and aerodynamic loading. The larger radial distance also increases the moment arm of the TE forces, thereby amplifying their contribution to torque.  Trailing-edge loading is therefore more influential for this configuration, and reducing the local trailing-edge incidence provides a viable mechanism for increasing $FM$ by decreasing torque.

	\begin{figure}[H]
		\centering
		
		\includegraphics{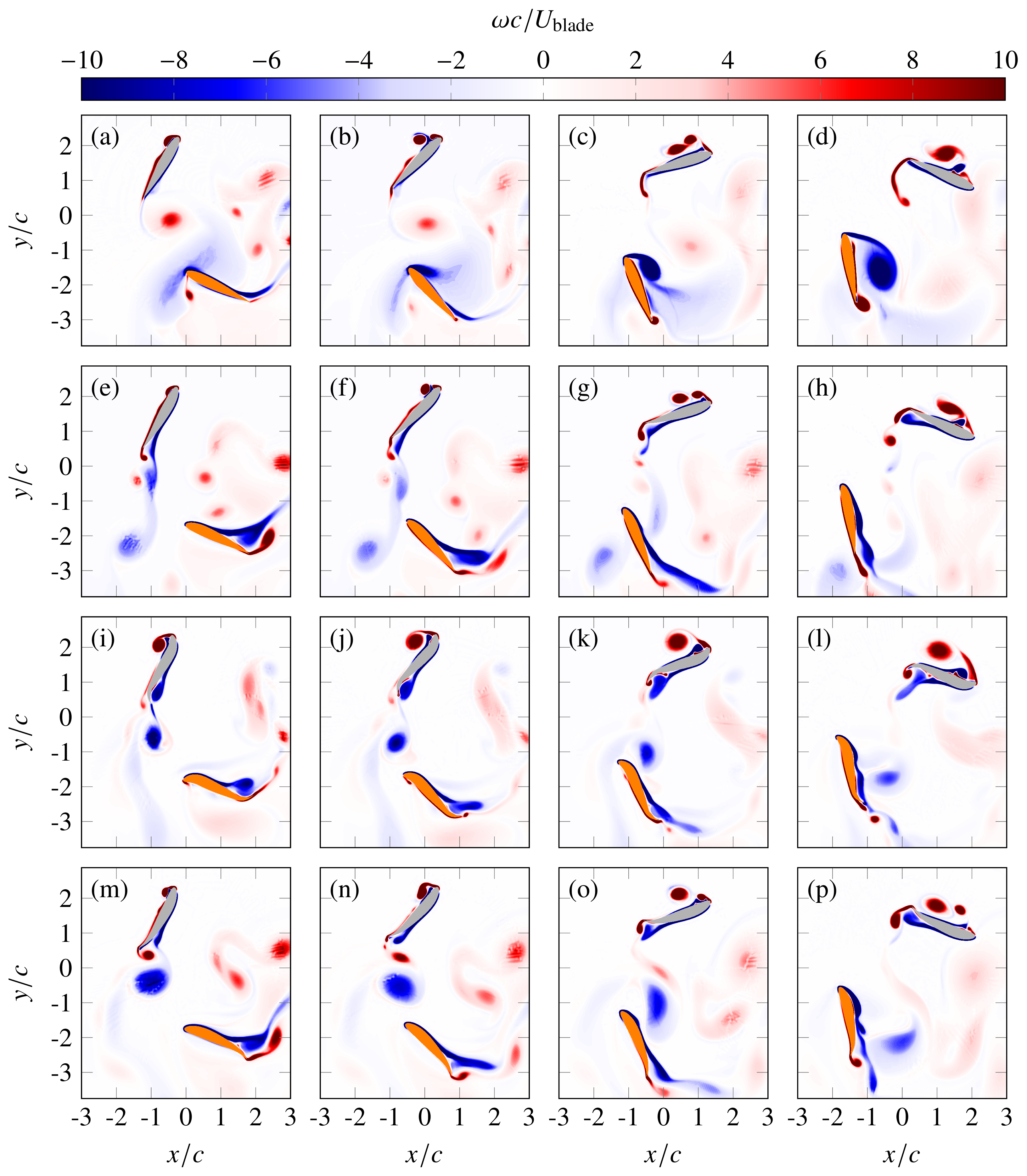}
		\caption{Comparison of URANS vorticity flow fields for a 0.34 solidity 2-bladed cyclorotor using NACA 0015 (a-d),  thrust-maximising aerofoil (e-h), torque minimising aerofoil (i-l) and low-fidelity optimised aerofoil (m-p). Columns from left-to-right show snapshots for $t/T=0.21, 0.26, 0.33, 0.41$. Blade 1 marked in orange.}
		\label{fig:2blade_vorticity}
		\phantomsubcaption\label{fig:2blade_vorticity_a}
		\phantomsubcaption\label{fig:2blade_vorticity_b}
		\phantomsubcaption\label{fig:2blade_vorticity_c}
		\phantomsubcaption\label{fig:2blade_vorticity_d}
		\phantomsubcaption\label{fig:2blade_vorticity_e}
		\phantomsubcaption\label{fig:2blade_vorticity_f}
		\phantomsubcaption\label{fig:2blade_vorticity_g}
		\phantomsubcaption\label{fig:2blade_vorticity_h}
		\phantomsubcaption\label{fig:2blade_vorticity_i}
		\phantomsubcaption\label{fig:2blade_vorticity_j}
		\phantomsubcaption\label{fig:2blade_vorticity_k}
		\phantomsubcaption\label{fig:2blade_vorticity_l}
		\phantomsubcaption\label{fig:2blade_vorticity_m}
		\phantomsubcaption\label{fig:2blade_vorticity_n}
		\phantomsubcaption\label{fig:2blade_vorticity_o}
		\phantomsubcaption\label{fig:2blade_vorticity_p}
	\end{figure}

	This explanation for the two design families is supported by the vorticity flowfields corresponding to each aerofoil shape (Fig. \ref{fig:2blade_vorticity}). The chosen snapshots display blade 1 at 4 different points during the primary thrust peak as it undergoes dynamic stall. For the baseline NACA 0015, the LEV characteristics during the primary thrust peak are similar to the original 4-bladed configuration, with a large coherent LEV separating from the surface of blade 1 (Figs. \ref{fig:2blade_vorticity_a}-\ref{fig:2blade_vorticity_d}). The trailing-edge flow differs more substantially.  The flow in the 4-bladed configuration leaves the TE cleanly, whereas in the 2-bladed configuration the shear layer on the pressure side of the aerofoil wraps around the TE and rolls up into a vortex on the suction side (Figs. \ref{fig:2blade_vorticity_c}-\ref{fig:2blade_vorticity_d}).

	All three optimised aerofoils suppress the negative LEV seen for the baseline aerofoil during the primary thrust peak (Figs. \ref{fig:2blade_vorticity_d},\ref{fig:2blade_vorticity_h},\ref{fig:2blade_vorticity_l},\ref{fig:2blade_vorticity_p}). However, this suppression is accompanied by increased flow activity further downstream on the suction surface, with pronounced shear-layer separation and roll-up between the mid-chord and the trailing edge  (Figs. \ref{fig:2blade_vorticity_e}-\ref{fig:2blade_vorticity_p}). Similar trailing-edge separation was also present, to a lesser extent, in the 4-bladed configuration (Fig. \ref{fig:4blade_vorticity_baseline_opt_anal_opt}). The increased blade-chord of this 2-bladed configuration increases the influence of TE forces on the rotor torque. The TE shape of the torque-minimising aerofoil however, reduces the TE separation and keeps the partially separated shear-layer closer to the blade surface (Fig. \ref{fig:2blade_vorticity_k}). Similarly on the pressure surface, the thrust-maximising and low-fidelity optimised aerofoils shed a strong positive shear-layer that rolls up into a vortex (Figs. \ref{fig:2blade_vorticity_e},\ref{fig:2blade_vorticity_m}), whereas the corresponding shear-layer for the torque-minimising design is much thinner (Fig. \ref{fig:2blade_vorticity_i}). These TE flow features are consistent with reduced aerodynamic loading and, consequently, lower rotor torque.

	While the efficiency of both design families is comparable, the reduced thrust of the torque-minimising family may limit its practical utility for this application. Cyclorotors generally operate more efficiently at low RPMs (Section \ref{section:introduction}). A rotor that use the aerofoils from the torque-minimising design family may require being operated at a higher RPM to achieve a required level of thrust. This may offset the efficiency gains of the torque-minimising design, making the thrust-maximising design family more efficient by comparison. Furthermore, depending on the application, there are metrics other than efficiency that need consideration. Increasing RPM will increase the centrifugal loads upon the blades and thus influence the structural design of the rotor \citep{Benedict2010FundamentalApplications}. Increasing RPM will also increase the angular momentum of the rotor, which in air-vehicle applications can complicate controllability due to gyroscopic precession. From an acoustic perspective, cyclorotor noise is concentrated around the blade-passing frequency. Operating at higher RPM shifts this dominant frequency higher into the range of human ear sensitivity, increasing perceived (A-weighted) noise \citep{Halder2019AeroacousticApproach}. These structural, control and acoustic benefits provide additional motivation to optimise aerofoil geometry for high thrust and efficiency at low rotational speeds.

	\subsection{Kinematics vs aerofoil-shape optimisation}

	In this study, aerofoils with a LE droop have been used to improve cyclorotor efficiency by reducing the local LE angle-of-attack. Optimisation of the blade-pitch kinematics provides an alternative method for controlling this LE angle-of-attack and thereby improve efficiency \citep{Benedict2010FundamentalApplications, Walther2019SymmetricNumbers, Shi2022AnalysisRatio, Benedict2016DevelopmentVehicle}. Consequently,  a direct comparison of the trade-offs between optimisation of aerofoil-shape vs blade-pitch kinematics is required to determine whether aerofoil optimisation offers advantages beyond those achievable through kinematic control. This section applies both approaches to the baseline 4-bladed configuration.

\begin{figure}[H]
		\centering
		\includegraphics{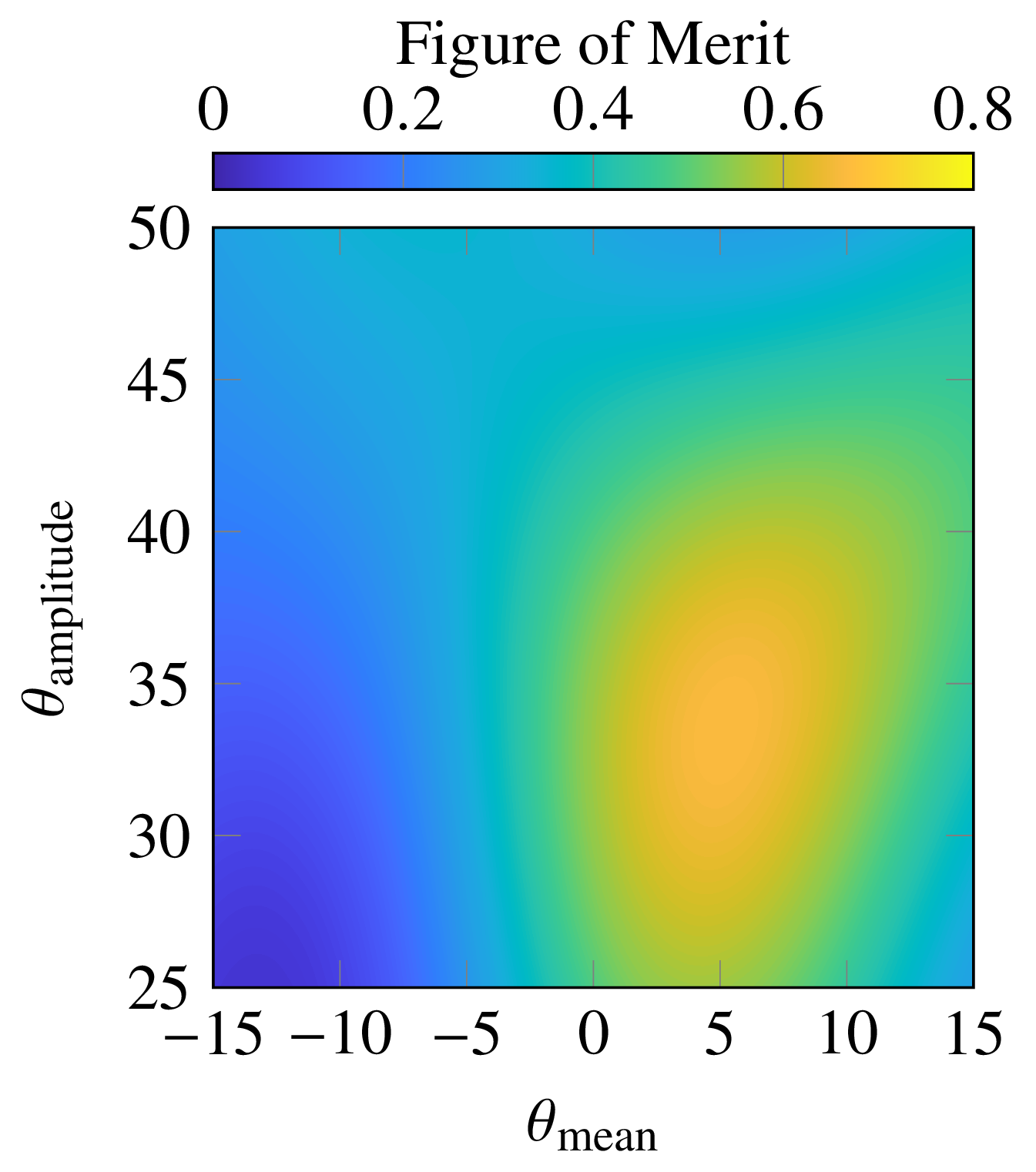}
		\caption{FM trends for a 4-bladed cyclorotor with variations in mean blade pitch ($\theta_{\text{mean}}$) and blade pitch amplitude ($\theta_{\text{amplitude}}$). Optimum at $\theta_{\text{mean}}=\ang{5.2}$ and $\theta_{\text{amplitude}}=\ang{33.4}$. Full parameter space mapped out with Kriging surrogate model fitted to URANS results.}
		\label{fig:design_space_kinematics}
	\end{figure}

\begin{figure}[H]
		\centering
		\includegraphics{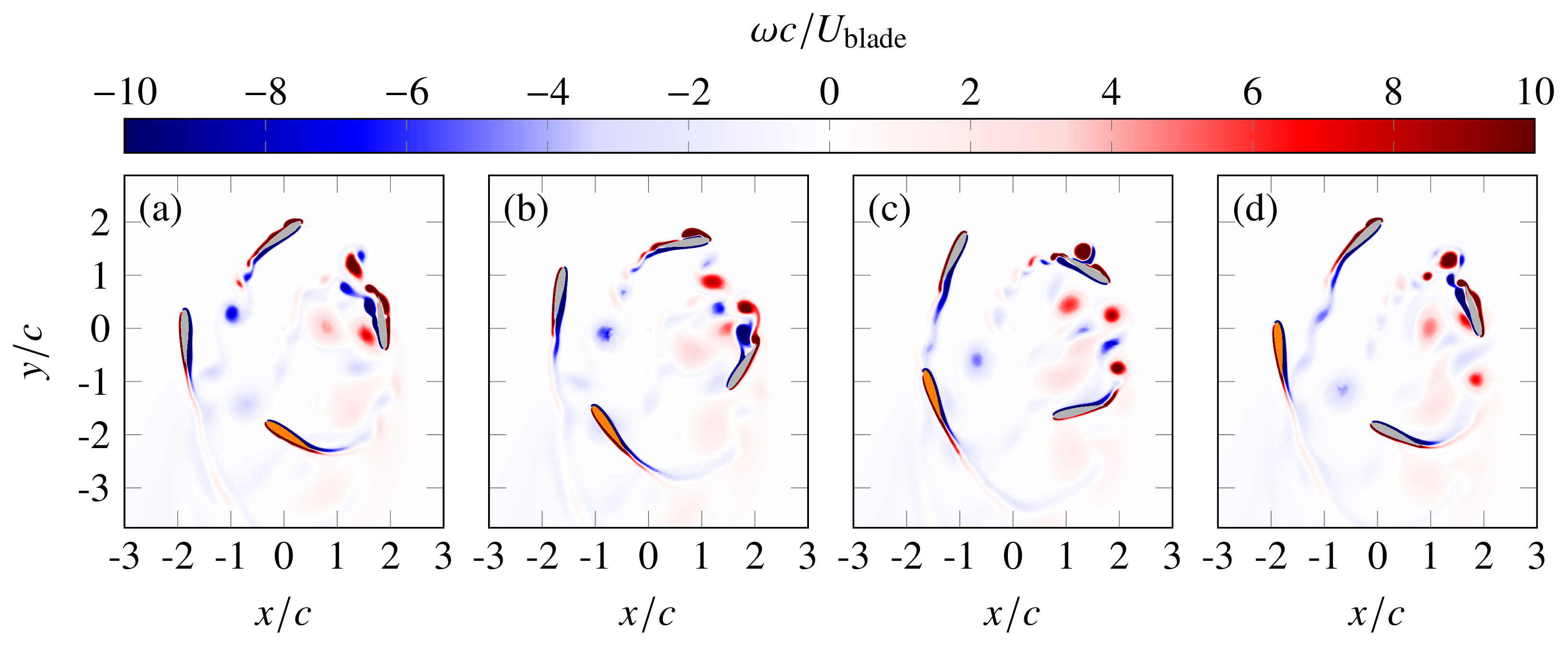}
		\caption{URANS vorticity flow fields for 4-bladed cyclorotor using NACA 0015 and optimised kinematics ($\theta_{\text{mean}}=\ang{5.2}$ and $\theta_{\text{amplitude}}=\ang{33.4}$) . Columns from left-to-right show snapshots for $t/T=0, 0.07, 0.15, 0.25$. Blade 1 marked in orange.}
		\label{fig:4blade_vorticity_opt_kin}
	\end{figure}

	The blade pitch kinematics optimisation produced a clear optimum at $\theta_{\text{mean}}=\ang{5.2}$ and $\theta_{\text{amplitude}}=\ang{33.4}$ (Fig. \ref{fig:design_space_kinematics}). This aligns with previous literature, which also found improved performance using more moderate blade-pitch amplitudes and a small positive bias \cite{Benedict2010FundamentalApplications, Walther2019SymmetricNumbers, Shi2022AnalysisRatio, Benedict2016EffectsCycloidal-rotor}. These kinematics offset the virtual camber effect and reduce blade angle-of-attack during the primary thrust peak, thereby suppressing LEV separation (Fig. \ref{fig:4blade_vorticity_opt_kin}). Consequently, these optimised kinematics follow the same mechanistic principle as the optimised aerofoils.  Therefore, applying aerofoil optimisation to the optimised kinematics is consequently expected to provide limited additional benefit since the targeted leading-edge vortex separation has already been suppressed. 
	
	Cyclorotors with optimised aerofoils and optimised kinematics produced comparable improvements in $FM$ (Table \ref{table:optimised_blade_pitch}). While the optimised aerofoil increased thrust and reduced torque, the optimised kinematics reduced both quantities. Kinematic optimisation therefore exhibited a trade-offs similar to that of the torque-minimising aerofoil family discussed in Section \ref{section:results_optimisation_double_chord}. To meet a required level of thrust, a cyclorotor with optimised kinematics will have to operate at a higher RPM. For this particular configuration, increasing RPM of the optimised kinematics configuration to match the thrust of the baseline configuration does not result in a reduction in $FM$. However, the associated increases in centrifugal loads, gyroscopic precession and noise generation remain relevant practical considerations.

	\begin{table}[H]
		\centering
		\begin{tabular}{|l|l|l|l|}
			\hline
			Case                 & FM   & Thrust & Torque \\ \hline
			Baseline             & 0.39 & 3.00   & 0.49   \\ \hline
			Optimised aerofoil   & 0.65 & 3.52   & 0.37   \\ \hline
			Optimised kinematics & 0.67 & 2.09   & 0.16   \\ \hline
			Optimised kinematics at 28.75 RPM & 0.70 & 3.03   & 0.23   \\ \hline
		\end{tabular}
		\caption{Comparison of URANS predicted performance of a 4-bladed cyclorotor with optimised aerofoils vs optimised blade pitch kinematics.}
		\label{table:optimised_blade_pitch}
	\end{table}

	\section{Conclusion}

	This study reformulated the optimisation principles previously established in \cite{irwin2026roledynamicstallaerofoil} into a generalised low-fidelity method for determining the optimum aerofoil shape for a given cyclorotor configuration. The method suppresses the LEV separation by minimising the effective LE angle-of-attack using throughflow predictions from a single-streamtube model. The method also promotes radial wake ejection by minimising the angle between the TE-camberline-normal and the local curvilinear flow direction.

	The low-fidelity method was first evaluated for the four-bladed configuration previously optimised using a high-fidelity approach \citep{irwin2026roledynamicstallaerofoil}. The resulting aerofoil resembled the high-fidelity optimum, but exhibited larger leading and trailing edges droops. The URANS vorticity flowfields showed that both the low-fidelity and high-fidelity optimised aerofoils successfully suppressed LEV separation in the primary thrust peak in the same way, thus succeeding in improving $FM$. The low-fidelity optimised aerofoil attained 77\% of the improvement in $FM$ attained by the high-fidelity optimised aerofoil, at a fraction of the computational cost.

	The low-fidelity method also identified aerofoils that improved $FM$ for chord-lengths outside the range examined previously \citep{irwin2026roledynamicstallaerofoil}. The increased influence of trailing-edge loading at larger chord lengths extends the optimisation principles established previously \citep{irwin2026roledynamicstallaerofoil} by revealing an additional torque-minimising family. Increasing blade chord increases the contribution of trailing-edge loading to rotor torque, allowing $FM$ to be improved through greater torque reduction rather than increased thrust.
	
	Similarly, optimising cyclorotor kinematics instead of aerofoil shape was found to produce similar improvements in $FM$ but with greatly reduced thrust. The reduced thrust associated with both approaches may limit their practical performance under fixed-thrust requirements as higher rotational speeds are required to recover the target thrust. Although this increase in rotational speed did not reduce $FM$ for the configuration examined, the associated increases in centrifugal loading, rotor angular momentum and acoustic frequency remain important practical considerations. 
	
	In conclusion, this study proposes a computationally efficient tool for preliminary cyclorotor aerofoil design while also identifying the aerodynamic and operational trade-offs associated with different optimisation strategies.

		\clearpage
		\appendix
		
		\section{Experimental validation of URANS flowfields} \label{section:URANS_comparison}
		
		\begin{figure}[b!]
			\centering
			\includegraphics{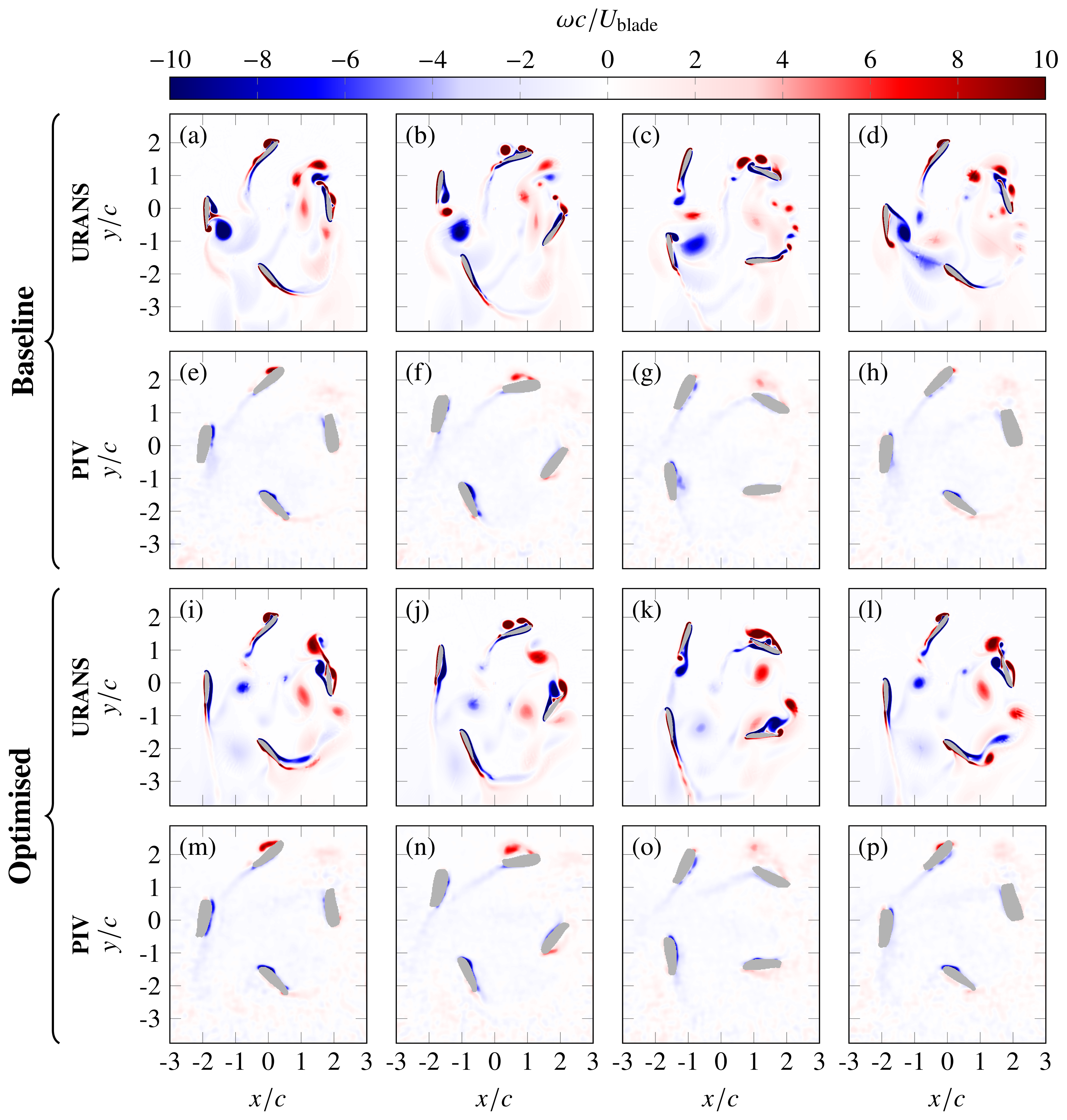}
			\caption{Comparison of vorticity flow fields for 4-bladed cyclorotor using NACA 0015 (a-h) and an optimised aerofoil (i-p). (a-d, i-l) Flowfields from 2D URANS simulations. (e-h,m-p) Flowfields from PIV. Columns from left-to-right show snapshots for $t/T=0, 0.07, 0.15, 0.25$.}
			\label{fig:URANS_PIV_comparison}
		\end{figure}
		
		In the previous study, while URANS was used to determine an optimal aerofoil shape to be tested, all investigation/validation of the fundamental principles behind aerofoil optimisation was done with experimental data \citep{irwin2026roledynamicstallaerofoil}. In contrast, this study predominantly uses URANS data. URANS is a simplified version of the full Navier-Stokes equations using turbulence modelling and may not reproduce all features of complex unsteady separated flows. To ascertain the reliability of the URANS results, the PIV and URANS flowfields are compared for the 4-bladed configuration using both the baseline NACA 0015 and the optimised aerofoil from \cite{irwin2026roledynamicstallaerofoil} (Fig. \ref{fig:URANS_PIV_comparison}). 
		
		A primary difference observed between the URANS and PIV flowfields is the intensity of the vorticity, which is significantly higher in the URANS results. PIV determines flowfields by tracking coherent clusters of particles within an interrogation window between images. Consequently, PIV measurements are spatially low-pass filtered, which can lead to vortices in the flow appearing ``smeared" and lower in vorticity.\citep{Raffel2018ParticleGuide}.

		Within the rotor cage and away from the blade surfaces, the PIV measurements show substantially weaker residual vorticity than the URANS predictions, which retain more coherent and persistent vortical structures after shedding from the blades (Fig. \ref{fig:URANS_PIV_comparison}). This difference can partly be attributed to the spatial filtering inherent in PIV measurements. Vortex decay and breakdown are also inherently 3D phenomena that cannot be fully represented in a 2D simulation \citep{Rivera1995TheTurbulence,Zurman-Nasution2020InfluenceFlapping}. 2D simulations may therefore retain shed vortices for longer and predict greater coherence than observed experimentally \citep{Zurman-Nasution2020InfluenceFlapping,Yu2016Two-dimensionalHover}. Although the mid-span cyclorotor flow is predominantly 2D, the experimental blades have open tips and consequently also generate three-dimensional tip-flow structures. The associated induced velocities may reduce the effective blade incidence and moderate the severity of dynamic stall, contributing to differences in the strength and shape of the vortical structures \citep{Yu2016Two-dimensionalHover}.
		
		The near-wall flow shows closer agreement between the PIV and URANS results than the regions away from the blade surfaces. For the baseline aerofoil, both methods indicate boundary layer separation as the blade in the lower half of the rotor returns to zero pitch. The exact flow features at separation differs between the two methods. In the PIV, this separation originates around the mid-chord and the separated flow dissipates quickly, whereas for the URANS the separation originates at the LE, forming a large coherent vortex. For the optimised aerofoil, both the PIV and URANS show that the shear layer on the aerofoil in lower half of the rotor will remain attached. The presence/absence of this separation is the driving principle behind what allows aerofoil optimisation to improve performance \citep{irwin2026roledynamicstallaerofoil}. As such, despite the differences/limitations that URANS has for modelling these flows, it is still able to correctly rank the performances of these two aerofoils because it does correctly predict the presence of flow separation during dynamic stall. 
		
		In conclusion, the URANS flowfields provide a sufficiently reliable representation of the principle flow structure and indicate whether separation occurs. However, the vortical structures away from the blade surfaces should be interpreted with caution because the 2D simulations predict greater strength, coherence, and persistence than observed experimentally.

		\begin{table}[h]
			\centering
			\begin{tabular}{ccccccc}
				& \multicolumn{3}{c||}{Experimental}                                                      & \multicolumn{3}{c}{URANS}                                                             \\ \hline
				\multicolumn{1}{|c|}{Case}                    & \multicolumn{1}{c|}{FM}   & \multicolumn{1}{c|}{Thrust (\unit{\newton})} & \multicolumn{1}{c||}{Torque (\unit{\newton\metre})} & \multicolumn{1}{c|}{FM}   & \multicolumn{1}{c|}{Thrust (\unit{\newton})} & \multicolumn{1}{c|}{Torque (\unit{\newton\metre})} \\ \hline
				\multicolumn{1}{|l|}{Baseline}                & \multicolumn{1}{c|}{0.49} & \multicolumn{1}{c|}{2.54}   & \multicolumn{1}{c||}{0.30}   & \multicolumn{1}{c|}{0.39} & \multicolumn{1}{c|}{3.00}   & \multicolumn{1}{c|}{0.49}   \\ \hline
				\multicolumn{1}{|l|}{Optimisation} & \multicolumn{1}{c|}{0.56} & \multicolumn{1}{c|}{3.22}   & \multicolumn{1}{c||}{0.37}   & \multicolumn{1}{c|}{0.65} & \multicolumn{1}{c|}{3.52}   & \multicolumn{1}{c|}{0.43}   \\ \hline
			\end{tabular}
			\caption{Comparison of performance metrics for 4-bladed cyclorotor at 24RPM using experimental and URANS results.}
			\label{table:comparison_URANS_experiment}
		\end{table}

		This increased strength of the flowfield features is also reflected in the performance metrics (Table \ref{table:comparison_URANS_experiment}). For both aerofoil shapes, URANS predicts both thrust and torque to be larger than the experimental values. The improvement in $FM$ predicted by URANS (67\%) is also much greater than the improvement in $FM$ found experimentally (14\%). While differences in the instantaneous forces arise from the flawed modelling of vortex decay, the time-averaged forces remain similar to experimentally found values \citep{Yu2016Two-dimensionalHover}. Consequently, URANS can be relied to upon for ranking the relative performance of different aerofoil shapes \citep{Tang2017UnsteadyModel, Zhang2018ThePropellers}.

	\bibliographystyle{apalike}
	\bibliography{references}
	
\end{document}

%% file: colourstuff.tex
\definecolor{symm}{HTML}{b30000}
\definecolor{optim}{HTML}{045a8d}
\definecolor{default_green}{HTML}{4daf4a}
\definecolor{default_orange}{HTML}{ff7f00}
\definecolor{default_purple}{HTML}{984ea3}
\definecolor{linkage_color_1}{HTML}{e41a1c}
\definecolor{linkage_color_2}{HTML}{377eb8}
\definecolor{linkage_color_3}{HTML}{4daf4a}
\definecolor{linkage_color_4}{HTML}{984ea3}
\definecolor{parula-1}{rgb}{0.2422,0.1504,0.6603}
\definecolor{parula-2}{rgb}{0.2444,0.1534,0.6728}
\definecolor{parula-3}{rgb}{0.2464,0.1569,0.6847}
\definecolor{parula-4}{rgb}{0.2484,0.1607,0.6961}
\definecolor{parula-5}{rgb}{0.2503,0.1648,0.7071}
\definecolor{parula-6}{rgb}{0.2522,0.1689,0.7179}
\definecolor{parula-7}{rgb}{0.2540,0.1732,0.7286}
\definecolor{parula-8}{rgb}{0.2558,0.1773,0.7393}
\definecolor{parula-9}{rgb}{0.2576,0.1814,0.7501}
\definecolor{parula-10}{rgb}{0.2594,0.1854,0.7610}
\definecolor{parula-11}{rgb}{0.2611,0.1893,0.7719}
\definecolor{parula-12}{rgb}{0.2628,0.1932,0.7828}
\definecolor{parula-13}{rgb}{0.2645,0.1972,0.7937}
\definecolor{parula-14}{rgb}{0.2661,0.2011,0.8043}
\definecolor{parula-15}{rgb}{0.2676,0.2052,0.8148}
\definecolor{parula-16}{rgb}{0.2691,0.2094,0.8249}
\definecolor{parula-17}{rgb}{0.2704,0.2138,0.8346}
\definecolor{parula-18}{rgb}{0.2717,0.2184,0.8439}
\definecolor{parula-19}{rgb}{0.2729,0.2231,0.8528}
\definecolor{parula-20}{rgb}{0.2740,0.2280,0.8612}
\definecolor{parula-21}{rgb}{0.2749,0.2330,0.8692}
\definecolor{parula-22}{rgb}{0.2758,0.2382,0.8767}
\definecolor{parula-23}{rgb}{0.2766,0.2435,0.8840}
\definecolor{parula-24}{rgb}{0.2774,0.2489,0.8908}
\definecolor{parula-25}{rgb}{0.2781,0.2543,0.8973}
\definecolor{parula-26}{rgb}{0.2788,0.2598,0.9035}
\definecolor{parula-27}{rgb}{0.2794,0.2653,0.9094}
\definecolor{parula-28}{rgb}{0.2798,0.2708,0.9150}
\definecolor{parula-29}{rgb}{0.2802,0.2764,0.9204}
\definecolor{parula-30}{rgb}{0.2806,0.2819,0.9255}
\definecolor{parula-31}{rgb}{0.2809,0.2875,0.9305}
\definecolor{parula-32}{rgb}{0.2811,0.2930,0.9352}
\definecolor{parula-33}{rgb}{0.2813,0.2985,0.9397}
\definecolor{parula-34}{rgb}{0.2814,0.3040,0.9441}
\definecolor{parula-35}{rgb}{0.2814,0.3095,0.9483}
\definecolor{parula-36}{rgb}{0.2813,0.3150,0.9524}
\definecolor{parula-37}{rgb}{0.2811,0.3204,0.9563}
\definecolor{parula-38}{rgb}{0.2809,0.3259,0.9600}
\definecolor{parula-39}{rgb}{0.2807,0.3313,0.9636}
\definecolor{parula-40}{rgb}{0.2803,0.3367,0.9670}
\definecolor{parula-41}{rgb}{0.2798,0.3421,0.9702}
\definecolor{parula-42}{rgb}{0.2791,0.3475,0.9733}
\definecolor{parula-43}{rgb}{0.2784,0.3529,0.9763}
\definecolor{parula-44}{rgb}{0.2776,0.3583,0.9791}
\definecolor{parula-45}{rgb}{0.2766,0.3638,0.9817}
\definecolor{parula-46}{rgb}{0.2754,0.3693,0.9840}
\definecolor{parula-47}{rgb}{0.2741,0.3748,0.9862}
\definecolor{parula-48}{rgb}{0.2726,0.3804,0.9881}
\definecolor{parula-49}{rgb}{0.2710,0.3860,0.9898}
\definecolor{parula-50}{rgb}{0.2691,0.3916,0.9912}
\definecolor{parula-51}{rgb}{0.2670,0.3973,0.9924}
\definecolor{parula-52}{rgb}{0.2647,0.4030,0.9935}
\definecolor{parula-53}{rgb}{0.2621,0.4088,0.9946}
\definecolor{parula-54}{rgb}{0.2591,0.4145,0.9955}
\definecolor{parula-55}{rgb}{0.2556,0.4203,0.9965}
\definecolor{parula-56}{rgb}{0.2517,0.4261,0.9974}
\definecolor{parula-57}{rgb}{0.2473,0.4319,0.9983}
\definecolor{parula-58}{rgb}{0.2424,0.4378,0.9991}
\definecolor{parula-59}{rgb}{0.2369,0.4437,0.9996}
\definecolor{parula-60}{rgb}{0.2311,0.4497,0.9995}
\definecolor{parula-61}{rgb}{0.2250,0.4559,0.9985}
\definecolor{parula-62}{rgb}{0.2189,0.4620,0.9968}
\definecolor{parula-63}{rgb}{0.2128,0.4682,0.9948}
\definecolor{parula-64}{rgb}{0.2066,0.4743,0.9926}
\definecolor{parula-65}{rgb}{0.2006,0.4803,0.9906}
\definecolor{parula-66}{rgb}{0.1950,0.4861,0.9887}
\definecolor{parula-67}{rgb}{0.1903,0.4919,0.9867}
\definecolor{parula-68}{rgb}{0.1869,0.4975,0.9844}
\definecolor{parula-69}{rgb}{0.1847,0.5030,0.9819}
\definecolor{parula-70}{rgb}{0.1831,0.5084,0.9793}
\definecolor{parula-71}{rgb}{0.1818,0.5138,0.9766}
\definecolor{parula-72}{rgb}{0.1806,0.5191,0.9738}
\definecolor{parula-73}{rgb}{0.1795,0.5244,0.9709}
\definecolor{parula-74}{rgb}{0.1785,0.5296,0.9677}
\definecolor{parula-75}{rgb}{0.1778,0.5349,0.9641}
\definecolor{parula-76}{rgb}{0.1773,0.5401,0.9602}
\definecolor{parula-77}{rgb}{0.1768,0.5452,0.9560}
\definecolor{parula-78}{rgb}{0.1764,0.5504,0.9516}
\definecolor{parula-79}{rgb}{0.1755,0.5554,0.9473}
\definecolor{parula-80}{rgb}{0.1740,0.5605,0.9432}
\definecolor{parula-81}{rgb}{0.1716,0.5655,0.9393}
\definecolor{parula-82}{rgb}{0.1686,0.5705,0.9357}
\definecolor{parula-83}{rgb}{0.1649,0.5755,0.9323}
\definecolor{parula-84}{rgb}{0.1610,0.5805,0.9289}
\definecolor{parula-85}{rgb}{0.1573,0.5854,0.9254}
\definecolor{parula-86}{rgb}{0.1540,0.5902,0.9218}
\definecolor{parula-87}{rgb}{0.1513,0.5950,0.9182}
\definecolor{parula-88}{rgb}{0.1492,0.5997,0.9147}
\definecolor{parula-89}{rgb}{0.1475,0.6043,0.9113}
\definecolor{parula-90}{rgb}{0.1461,0.6089,0.9080}
\definecolor{parula-91}{rgb}{0.1446,0.6135,0.9050}
\definecolor{parula-92}{rgb}{0.1429,0.6180,0.9022}
\definecolor{parula-93}{rgb}{0.1408,0.6226,0.8998}
\definecolor{parula-94}{rgb}{0.1383,0.6272,0.8975}
\definecolor{parula-95}{rgb}{0.1354,0.6317,0.8953}
\definecolor{parula-96}{rgb}{0.1321,0.6363,0.8932}
\definecolor{parula-97}{rgb}{0.1288,0.6408,0.8910}
\definecolor{parula-98}{rgb}{0.1253,0.6453,0.8887}
\definecolor{parula-99}{rgb}{0.1219,0.6497,0.8862}
\definecolor{parula-100}{rgb}{0.1185,0.6541,0.8834}
\definecolor{parula-101}{rgb}{0.1152,0.6584,0.8804}
\definecolor{parula-102}{rgb}{0.1119,0.6627,0.8770}
\definecolor{parula-103}{rgb}{0.1085,0.6669,0.8734}
\definecolor{parula-104}{rgb}{0.1048,0.6710,0.8695}
\definecolor{parula-105}{rgb}{0.1009,0.6750,0.8653}
\definecolor{parula-106}{rgb}{0.0964,0.6789,0.8609}
\definecolor{parula-107}{rgb}{0.0914,0.6828,0.8562}
\definecolor{parula-108}{rgb}{0.0855,0.6865,0.8513}
\definecolor{parula-109}{rgb}{0.0789,0.6902,0.8462}
\definecolor{parula-110}{rgb}{0.0713,0.6938,0.8409}
\definecolor{parula-111}{rgb}{0.0628,0.6972,0.8355}
\definecolor{parula-112}{rgb}{0.0535,0.7006,0.8299}
\definecolor{parula-113}{rgb}{0.0433,0.7039,0.8242}
\definecolor{parula-114}{rgb}{0.0328,0.7071,0.8183}
\definecolor{parula-115}{rgb}{0.0234,0.7103,0.8124}
\definecolor{parula-116}{rgb}{0.0155,0.7133,0.8064}
\definecolor{parula-117}{rgb}{0.0091,0.7163,0.8003}
\definecolor{parula-118}{rgb}{0.0046,0.7192,0.7941}
\definecolor{parula-119}{rgb}{0.0019,0.7220,0.7878}
\definecolor{parula-120}{rgb}{0.0009,0.7248,0.7815}
\definecolor{parula-121}{rgb}{0.0018,0.7275,0.7752}
\definecolor{parula-122}{rgb}{0.0046,0.7301,0.7688}
\definecolor{parula-123}{rgb}{0.0094,0.7327,0.7623}
\definecolor{parula-124}{rgb}{0.0162,0.7352,0.7558}
\definecolor{parula-125}{rgb}{0.0253,0.7376,0.7492}
\definecolor{parula-126}{rgb}{0.0369,0.7400,0.7426}
\definecolor{parula-127}{rgb}{0.0504,0.7423,0.7359}
\definecolor{parula-128}{rgb}{0.0638,0.7446,0.7292}
\definecolor{parula-129}{rgb}{0.0770,0.7468,0.7224}
\definecolor{parula-130}{rgb}{0.0899,0.7489,0.7156}
\definecolor{parula-131}{rgb}{0.1023,0.7510,0.7088}
\definecolor{parula-132}{rgb}{0.1141,0.7531,0.7019}
\definecolor{parula-133}{rgb}{0.1252,0.7552,0.6950}
\definecolor{parula-134}{rgb}{0.1354,0.7572,0.6881}
\definecolor{parula-135}{rgb}{0.1448,0.7593,0.6812}
\definecolor{parula-136}{rgb}{0.1532,0.7614,0.6741}
\definecolor{parula-137}{rgb}{0.1609,0.7635,0.6671}
\definecolor{parula-138}{rgb}{0.1678,0.7656,0.6599}
\definecolor{parula-139}{rgb}{0.1741,0.7678,0.6527}
\definecolor{parula-140}{rgb}{0.1799,0.7699,0.6454}
\definecolor{parula-141}{rgb}{0.1853,0.7721,0.6379}
\definecolor{parula-142}{rgb}{0.1905,0.7743,0.6303}
\definecolor{parula-143}{rgb}{0.1954,0.7765,0.6225}
\definecolor{parula-144}{rgb}{0.2003,0.7787,0.6146}
\definecolor{parula-145}{rgb}{0.2061,0.7808,0.6065}
\definecolor{parula-146}{rgb}{0.2118,0.7828,0.5983}
\definecolor{parula-147}{rgb}{0.2178,0.7849,0.5899}
\definecolor{parula-148}{rgb}{0.2244,0.7869,0.5813}
\definecolor{parula-149}{rgb}{0.2318,0.7887,0.5725}
\definecolor{parula-150}{rgb}{0.2401,0.7905,0.5636}
\definecolor{parula-151}{rgb}{0.2491,0.7922,0.5546}
\definecolor{parula-152}{rgb}{0.2589,0.7937,0.5454}
\definecolor{parula-153}{rgb}{0.2695,0.7951,0.5360}
\definecolor{parula-154}{rgb}{0.2809,0.7964,0.5266}
\definecolor{parula-155}{rgb}{0.2929,0.7975,0.5170}
\definecolor{parula-156}{rgb}{0.3052,0.7985,0.5074}
\definecolor{parula-157}{rgb}{0.3176,0.7994,0.4975}
\definecolor{parula-158}{rgb}{0.3301,0.8002,0.4876}
\definecolor{parula-159}{rgb}{0.3424,0.8009,0.4774}
\definecolor{parula-160}{rgb}{0.3548,0.8016,0.4669}
\definecolor{parula-161}{rgb}{0.3671,0.8021,0.4563}
\definecolor{parula-162}{rgb}{0.3795,0.8026,0.4454}
\definecolor{parula-163}{rgb}{0.3921,0.8029,0.4344}
\definecolor{parula-164}{rgb}{0.4050,0.8031,0.4233}
\definecolor{parula-165}{rgb}{0.4184,0.8030,0.4122}
\definecolor{parula-166}{rgb}{0.4322,0.8028,0.4013}
\definecolor{parula-167}{rgb}{0.4463,0.8024,0.3904}
\definecolor{parula-168}{rgb}{0.4608,0.8018,0.3797}
\definecolor{parula-169}{rgb}{0.4753,0.8011,0.3691}
\definecolor{parula-170}{rgb}{0.4899,0.8002,0.3586}
\definecolor{parula-171}{rgb}{0.5044,0.7993,0.3480}
\definecolor{parula-172}{rgb}{0.5187,0.7982,0.3374}
\definecolor{parula-173}{rgb}{0.5329,0.7970,0.3267}
\definecolor{parula-174}{rgb}{0.5470,0.7957,0.3159}
\definecolor{parula-175}{rgb}{0.5609,0.7943,0.3050}
\definecolor{parula-176}{rgb}{0.5748,0.7929,0.2941}
\definecolor{parula-177}{rgb}{0.5886,0.7913,0.2833}
\definecolor{parula-178}{rgb}{0.6024,0.7896,0.2726}
\definecolor{parula-179}{rgb}{0.6161,0.7878,0.2622}
\definecolor{parula-180}{rgb}{0.6297,0.7859,0.2521}
\definecolor{parula-181}{rgb}{0.6433,0.7839,0.2423}
\definecolor{parula-182}{rgb}{0.6567,0.7818,0.2329}
\definecolor{parula-183}{rgb}{0.6701,0.7796,0.2239}
\definecolor{parula-184}{rgb}{0.6833,0.7773,0.2155}
\definecolor{parula-185}{rgb}{0.6963,0.7750,0.2075}
\definecolor{parula-186}{rgb}{0.7091,0.7727,0.1998}
\definecolor{parula-187}{rgb}{0.7218,0.7703,0.1924}
\definecolor{parula-188}{rgb}{0.7344,0.7679,0.1852}
\definecolor{parula-189}{rgb}{0.7468,0.7654,0.1782}
\definecolor{parula-190}{rgb}{0.7590,0.7629,0.1717}
\definecolor{parula-191}{rgb}{0.7710,0.7604,0.1658}
\definecolor{parula-192}{rgb}{0.7829,0.7579,0.1608}
\definecolor{parula-193}{rgb}{0.7945,0.7554,0.1570}
\definecolor{parula-194}{rgb}{0.8060,0.7529,0.1546}
\definecolor{parula-195}{rgb}{0.8172,0.7505,0.1535}
\definecolor{parula-196}{rgb}{0.8281,0.7481,0.1536}
\definecolor{parula-197}{rgb}{0.8389,0.7457,0.1546}
\definecolor{parula-198}{rgb}{0.8495,0.7435,0.1564}
\definecolor{parula-199}{rgb}{0.8600,0.7413,0.1587}
\definecolor{parula-200}{rgb}{0.8703,0.7392,0.1615}
\definecolor{parula-201}{rgb}{0.8804,0.7372,0.1650}
\definecolor{parula-202}{rgb}{0.8903,0.7353,0.1695}
\definecolor{parula-203}{rgb}{0.9000,0.7336,0.1749}
\definecolor{parula-204}{rgb}{0.9093,0.7321,0.1815}
\definecolor{parula-205}{rgb}{0.9184,0.7308,0.1890}
\definecolor{parula-206}{rgb}{0.9272,0.7298,0.1973}
\definecolor{parula-207}{rgb}{0.9357,0.7290,0.2061}
\definecolor{parula-208}{rgb}{0.9440,0.7285,0.2151}
\definecolor{parula-209}{rgb}{0.9523,0.7284,0.2237}
\definecolor{parula-210}{rgb}{0.9606,0.7285,0.2312}
\definecolor{parula-211}{rgb}{0.9689,0.7292,0.2373}
\definecolor{parula-212}{rgb}{0.9770,0.7304,0.2418}
\definecolor{parula-213}{rgb}{0.9842,0.7330,0.2446}
\definecolor{parula-214}{rgb}{0.9900,0.7365,0.2429}
\definecolor{parula-215}{rgb}{0.9946,0.7407,0.2394}
\definecolor{parula-216}{rgb}{0.9966,0.7458,0.2351}
\definecolor{parula-217}{rgb}{0.9971,0.7513,0.2309}
\definecolor{parula-218}{rgb}{0.9972,0.7569,0.2267}
\definecolor{parula-219}{rgb}{0.9971,0.7626,0.2224}
\definecolor{parula-220}{rgb}{0.9969,0.7683,0.2181}
\definecolor{parula-221}{rgb}{0.9966,0.7740,0.2138}
\definecolor{parula-222}{rgb}{0.9962,0.7798,0.2095}
\definecolor{parula-223}{rgb}{0.9957,0.7856,0.2053}
\definecolor{parula-224}{rgb}{0.9949,0.7915,0.2012}
\definecolor{parula-225}{rgb}{0.9938,0.7974,0.1974}
\definecolor{parula-226}{rgb}{0.9923,0.8034,0.1939}
\definecolor{parula-227}{rgb}{0.9906,0.8095,0.1906}
\definecolor{parula-228}{rgb}{0.9885,0.8156,0.1875}
\definecolor{parula-229}{rgb}{0.9861,0.8218,0.1846}
\definecolor{parula-230}{rgb}{0.9835,0.8280,0.1817}
\definecolor{parula-231}{rgb}{0.9807,0.8342,0.1787}
\definecolor{parula-232}{rgb}{0.9778,0.8404,0.1757}
\definecolor{parula-233}{rgb}{0.9748,0.8467,0.1726}
\definecolor{parula-234}{rgb}{0.9720,0.8529,0.1695}
\definecolor{parula-235}{rgb}{0.9694,0.8591,0.1665}
\definecolor{parula-236}{rgb}{0.9671,0.8654,0.1636}
\definecolor{parula-237}{rgb}{0.9651,0.8716,0.1608}
\definecolor{parula-238}{rgb}{0.9634,0.8778,0.1582}
\definecolor{parula-239}{rgb}{0.9619,0.8840,0.1557}
\definecolor{parula-240}{rgb}{0.9608,0.8902,0.1532}
\definecolor{parula-241}{rgb}{0.9601,0.8963,0.1507}
\definecolor{parula-242}{rgb}{0.9596,0.9023,0.1480}
\definecolor{parula-243}{rgb}{0.9595,0.9084,0.1450}
\definecolor{parula-244}{rgb}{0.9597,0.9143,0.1418}
\definecolor{parula-245}{rgb}{0.9601,0.9203,0.1382}
\definecolor{parula-246}{rgb}{0.9608,0.9262,0.1344}
\definecolor{parula-247}{rgb}{0.9618,0.9320,0.1304}
\definecolor{parula-248}{rgb}{0.9629,0.9379,0.1261}
\definecolor{parula-249}{rgb}{0.9642,0.9437,0.1216}
\definecolor{parula-250}{rgb}{0.9657,0.9494,0.1168}
\definecolor{parula-251}{rgb}{0.9674,0.9552,0.1116}
\definecolor{parula-252}{rgb}{0.9692,0.9609,0.1061}
\definecolor{parula-253}{rgb}{0.9711,0.9667,0.1001}
\definecolor{parula-254}{rgb}{0.9730,0.9724,0.0938}
\definecolor{parula-255}{rgb}{0.9749,0.9782,0.0872}
\definecolor{parula-256}{rgb}{0.9769,0.9839,0.0805}
\pgfplotsset{
colormap={parula}{
color={parula-1};
color={parula-2};
color={parula-3};
color={parula-4};
color={parula-5};
color={parula-6};
color={parula-7};
color={parula-8};
color={parula-9};
color={parula-10};
color={parula-11};
color={parula-12};
color={parula-13};
color={parula-14};
color={parula-15};
color={parula-16};
color={parula-17};
color={parula-18};
color={parula-19};
color={parula-20};
color={parula-21};
color={parula-22};
color={parula-23};
color={parula-24};
color={parula-25};
color={parula-26};
color={parula-27};
color={parula-28};
color={parula-29};
color={parula-30};
color={parula-31};
color={parula-32};
color={parula-33};
color={parula-34};
color={parula-35};
color={parula-36};
color={parula-37};
color={parula-38};
color={parula-39};
color={parula-40};
color={parula-41};
color={parula-42};
color={parula-43};
color={parula-44};
color={parula-45};
color={parula-46};
color={parula-47};
color={parula-48};
color={parula-49};
color={parula-50};
color={parula-51};
color={parula-52};
color={parula-53};
color={parula-54};
color={parula-55};
color={parula-56};
color={parula-57};
color={parula-58};
color={parula-59};
color={parula-60};
color={parula-61};
color={parula-62};
color={parula-63};
color={parula-64};
color={parula-65};
color={parula-66};
color={parula-67};
color={parula-68};
color={parula-69};
color={parula-70};
color={parula-71};
color={parula-72};
color={parula-73};
color={parula-74};
color={parula-75};
color={parula-76};
color={parula-77};
color={parula-78};
color={parula-79};
color={parula-80};
color={parula-81};
color={parula-82};
color={parula-83};
color={parula-84};
color={parula-85};
color={parula-86};
color={parula-87};
color={parula-88};
color={parula-89};
color={parula-90};
color={parula-91};
color={parula-92};
color={parula-93};
color={parula-94};
color={parula-95};
color={parula-96};
color={parula-97};
color={parula-98};
color={parula-99};
color={parula-100};
color={parula-101};
color={parula-102};
color={parula-103};
color={parula-104};
color={parula-105};
color={parula-106};
color={parula-107};
color={parula-108};
color={parula-109};
color={parula-110};
color={parula-111};
color={parula-112};
color={parula-113};
color={parula-114};
color={parula-115};
color={parula-116};
color={parula-117};
color={parula-118};
color={parula-119};
color={parula-120};
color={parula-121};
color={parula-122};
color={parula-123};
color={parula-124};
color={parula-125};
color={parula-126};
color={parula-127};
color={parula-128};
color={parula-129};
color={parula-130};
color={parula-131};
color={parula-132};
color={parula-133};
color={parula-134};
color={parula-135};
color={parula-136};
color={parula-137};
color={parula-138};
color={parula-139};
color={parula-140};
color={parula-141};
color={parula-142};
color={parula-143};
color={parula-144};
color={parula-145};
color={parula-146};
color={parula-147};
color={parula-148};
color={parula-149};
color={parula-150};
color={parula-151};
color={parula-152};
color={parula-153};
color={parula-154};
color={parula-155};
color={parula-156};
color={parula-157};
color={parula-158};
color={parula-159};
color={parula-160};
color={parula-161};
color={parula-162};
color={parula-163};
color={parula-164};
color={parula-165};
color={parula-166};
color={parula-167};
color={parula-168};
color={parula-169};
color={parula-170};
color={parula-171};
color={parula-172};
color={parula-173};
color={parula-174};
color={parula-175};
color={parula-176};
color={parula-177};
color={parula-178};
color={parula-179};
color={parula-180};
color={parula-181};
color={parula-182};
color={parula-183};
color={parula-184};
color={parula-185};
color={parula-186};
color={parula-187};
color={parula-188};
color={parula-189};
color={parula-190};
color={parula-191};
color={parula-192};
color={parula-193};
color={parula-194};
color={parula-195};
color={parula-196};
color={parula-197};
color={parula-198};
color={parula-199};
color={parula-200};
color={parula-201};
color={parula-202};
color={parula-203};
color={parula-204};
color={parula-205};
color={parula-206};
color={parula-207};
color={parula-208};
color={parula-209};
color={parula-210};
color={parula-211};
color={parula-212};
color={parula-213};
color={parula-214};
color={parula-215};
color={parula-216};
color={parula-217};
color={parula-218};
color={parula-219};
color={parula-220};
color={parula-221};
color={parula-222};
color={parula-223};
color={parula-224};
color={parula-225};
color={parula-226};
color={parula-227};
color={parula-228};
color={parula-229};
color={parula-230};
color={parula-231};
color={parula-232};
color={parula-233};
color={parula-234};
color={parula-235};
color={parula-236};
color={parula-237};
color={parula-238};
color={parula-239};
color={parula-240};
color={parula-241};
color={parula-242};
color={parula-243};
color={parula-244};
color={parula-245};
color={parula-246};
color={parula-247};
color={parula-248};
color={parula-249};
color={parula-250};
color={parula-251};
color={parula-252};
color={parula-253};
color={parula-254};
color={parula-255};
color={parula-256};
}
}
\definecolor{bluewhitered-1}{rgb}{0.0000,0.0000,0.4000}
\definecolor{bluewhitered-2}{rgb}{0.0000,0.0000,0.4072}
\definecolor{bluewhitered-3}{rgb}{0.0000,0.0000,0.4144}
\definecolor{bluewhitered-4}{rgb}{0.0000,0.0000,0.4216}
\definecolor{bluewhitered-5}{rgb}{0.0000,0.0000,0.4289}
\definecolor{bluewhitered-6}{rgb}{0.0000,0.0000,0.4361}
\definecolor{bluewhitered-7}{rgb}{0.0000,0.0000,0.4433}
\definecolor{bluewhitered-8}{rgb}{0.0000,0.0000,0.4505}
\definecolor{bluewhitered-9}{rgb}{0.0000,0.0000,0.4577}
\definecolor{bluewhitered-10}{rgb}{0.0000,0.0000,0.4649}
\definecolor{bluewhitered-11}{rgb}{0.0000,0.0000,0.4721}
\definecolor{bluewhitered-12}{rgb}{0.0000,0.0000,0.4794}
\definecolor{bluewhitered-13}{rgb}{0.0000,0.0000,0.4866}
\definecolor{bluewhitered-14}{rgb}{0.0000,0.0000,0.4938}
\definecolor{bluewhitered-15}{rgb}{0.0000,0.0000,0.5010}
\definecolor{bluewhitered-16}{rgb}{0.0000,0.0000,0.5082}
\definecolor{bluewhitered-17}{rgb}{0.0000,0.0000,0.5154}
\definecolor{bluewhitered-18}{rgb}{0.0000,0.0000,0.5226}
\definecolor{bluewhitered-19}{rgb}{0.0000,0.0000,0.5299}
\definecolor{bluewhitered-20}{rgb}{0.0000,0.0000,0.5371}
\definecolor{bluewhitered-21}{rgb}{0.0000,0.0000,0.5443}
\definecolor{bluewhitered-22}{rgb}{0.0000,0.0000,0.5515}
\definecolor{bluewhitered-23}{rgb}{0.0000,0.0000,0.5587}
\definecolor{bluewhitered-24}{rgb}{0.0000,0.0000,0.5659}
\definecolor{bluewhitered-25}{rgb}{0.0000,0.0000,0.5731}
\definecolor{bluewhitered-26}{rgb}{0.0000,0.0000,0.5804}
\definecolor{bluewhitered-27}{rgb}{0.0000,0.0000,0.5876}
\definecolor{bluewhitered-28}{rgb}{0.0000,0.0000,0.5948}
\definecolor{bluewhitered-29}{rgb}{0.0000,0.0000,0.6020}
\definecolor{bluewhitered-30}{rgb}{0.0000,0.0000,0.6092}
\definecolor{bluewhitered-31}{rgb}{0.0000,0.0000,0.6164}
\definecolor{bluewhitered-32}{rgb}{0.0000,0.0000,0.6236}
\definecolor{bluewhitered-33}{rgb}{0.0000,0.0000,0.6309}
\definecolor{bluewhitered-34}{rgb}{0.0000,0.0000,0.6381}
\definecolor{bluewhitered-35}{rgb}{0.0000,0.0000,0.6453}
\definecolor{bluewhitered-36}{rgb}{0.0000,0.0000,0.6525}
\definecolor{bluewhitered-37}{rgb}{0.0000,0.0000,0.6597}
\definecolor{bluewhitered-38}{rgb}{0.0000,0.0000,0.6669}
\definecolor{bluewhitered-39}{rgb}{0.0000,0.0000,0.6741}
\definecolor{bluewhitered-40}{rgb}{0.0000,0.0000,0.6814}
\definecolor{bluewhitered-41}{rgb}{0.0000,0.0000,0.6886}
\definecolor{bluewhitered-42}{rgb}{0.0000,0.0000,0.6958}
\definecolor{bluewhitered-43}{rgb}{0.0000,0.0000,0.7030}
\definecolor{bluewhitered-44}{rgb}{0.0000,0.0000,0.7102}
\definecolor{bluewhitered-45}{rgb}{0.0000,0.0000,0.7174}
\definecolor{bluewhitered-46}{rgb}{0.0000,0.0000,0.7246}
\definecolor{bluewhitered-47}{rgb}{0.0000,0.0000,0.7319}
\definecolor{bluewhitered-48}{rgb}{0.0000,0.0000,0.7391}
\definecolor{bluewhitered-49}{rgb}{0.0000,0.0000,0.7463}
\definecolor{bluewhitered-50}{rgb}{0.0000,0.0000,0.7535}
\definecolor{bluewhitered-51}{rgb}{0.0000,0.0000,0.7607}
\definecolor{bluewhitered-52}{rgb}{0.0000,0.0000,0.7679}
\definecolor{bluewhitered-53}{rgb}{0.0000,0.0000,0.7752}
\definecolor{bluewhitered-54}{rgb}{0.0000,0.0000,0.7824}
\definecolor{bluewhitered-55}{rgb}{0.0000,0.0000,0.7896}
\definecolor{bluewhitered-56}{rgb}{0.0000,0.0000,0.7968}
\definecolor{bluewhitered-57}{rgb}{0.0000,0.0000,0.8040}
\definecolor{bluewhitered-58}{rgb}{0.0000,0.0000,0.8112}
\definecolor{bluewhitered-59}{rgb}{0.0000,0.0000,0.8184}
\definecolor{bluewhitered-60}{rgb}{0.0000,0.0000,0.8257}
\definecolor{bluewhitered-61}{rgb}{0.0000,0.0000,0.8329}
\definecolor{bluewhitered-62}{rgb}{0.0000,0.0000,0.8401}
\definecolor{bluewhitered-63}{rgb}{0.0000,0.0000,0.8473}
\definecolor{bluewhitered-64}{rgb}{0.0000,0.0000,0.8545}
\definecolor{bluewhitered-65}{rgb}{0.0000,0.0000,0.8617}
\definecolor{bluewhitered-66}{rgb}{0.0000,0.0000,0.8689}
\definecolor{bluewhitered-67}{rgb}{0.0000,0.0000,0.8762}
\definecolor{bluewhitered-68}{rgb}{0.0000,0.0000,0.8834}
\definecolor{bluewhitered-69}{rgb}{0.0000,0.0000,0.8906}
\definecolor{bluewhitered-70}{rgb}{0.0000,0.0000,0.8978}
\definecolor{bluewhitered-71}{rgb}{0.0000,0.0000,0.9050}
\definecolor{bluewhitered-72}{rgb}{0.0000,0.0000,0.9122}
\definecolor{bluewhitered-73}{rgb}{0.0000,0.0000,0.9194}
\definecolor{bluewhitered-74}{rgb}{0.0000,0.0000,0.9267}
\definecolor{bluewhitered-75}{rgb}{0.0000,0.0000,0.9339}
\definecolor{bluewhitered-76}{rgb}{0.0000,0.0000,0.9411}
\definecolor{bluewhitered-77}{rgb}{0.0000,0.0000,0.9483}
\definecolor{bluewhitered-78}{rgb}{0.0000,0.0000,0.9555}
\definecolor{bluewhitered-79}{rgb}{0.0000,0.0000,0.9627}
\definecolor{bluewhitered-80}{rgb}{0.0000,0.0000,0.9699}
\definecolor{bluewhitered-81}{rgb}{0.0000,0.0000,0.9772}
\definecolor{bluewhitered-82}{rgb}{0.0000,0.0000,0.9844}
\definecolor{bluewhitered-83}{rgb}{0.0000,0.0000,0.9916}
\definecolor{bluewhitered-84}{rgb}{0.0000,0.0000,0.9988}
\definecolor{bluewhitered-85}{rgb}{0.0085,0.0085,1.0000}
\definecolor{bluewhitered-86}{rgb}{0.0187,0.0187,1.0000}
\definecolor{bluewhitered-87}{rgb}{0.0290,0.0290,1.0000}
\definecolor{bluewhitered-88}{rgb}{0.0392,0.0392,1.0000}
\definecolor{bluewhitered-89}{rgb}{0.0494,0.0494,1.0000}
\definecolor{bluewhitered-90}{rgb}{0.0596,0.0596,1.0000}
\definecolor{bluewhitered-91}{rgb}{0.0698,0.0698,1.0000}
\definecolor{bluewhitered-92}{rgb}{0.0801,0.0801,1.0000}
\definecolor{bluewhitered-93}{rgb}{0.0903,0.0903,1.0000}
\definecolor{bluewhitered-94}{rgb}{0.1005,0.1005,1.0000}
\definecolor{bluewhitered-95}{rgb}{0.1107,0.1107,1.0000}
\definecolor{bluewhitered-96}{rgb}{0.1209,0.1209,1.0000}
\definecolor{bluewhitered-97}{rgb}{0.1312,0.1312,1.0000}
\definecolor{bluewhitered-98}{rgb}{0.1414,0.1414,1.0000}
\definecolor{bluewhitered-99}{rgb}{0.1516,0.1516,1.0000}
\definecolor{bluewhitered-100}{rgb}{0.1618,0.1618,1.0000}
\definecolor{bluewhitered-101}{rgb}{0.1720,0.1720,1.0000}
\definecolor{bluewhitered-102}{rgb}{0.1823,0.1823,1.0000}
\definecolor{bluewhitered-103}{rgb}{0.1925,0.1925,1.0000}
\definecolor{bluewhitered-104}{rgb}{0.2027,0.2027,1.0000}
\definecolor{bluewhitered-105}{rgb}{0.2129,0.2129,1.0000}
\definecolor{bluewhitered-106}{rgb}{0.2231,0.2231,1.0000}
\definecolor{bluewhitered-107}{rgb}{0.2334,0.2334,1.0000}
\definecolor{bluewhitered-108}{rgb}{0.2436,0.2436,1.0000}
\definecolor{bluewhitered-109}{rgb}{0.2538,0.2538,1.0000}
\definecolor{bluewhitered-110}{rgb}{0.2640,0.2640,1.0000}
\definecolor{bluewhitered-111}{rgb}{0.2742,0.2742,1.0000}
\definecolor{bluewhitered-112}{rgb}{0.2845,0.2845,1.0000}
\definecolor{bluewhitered-113}{rgb}{0.2947,0.2947,1.0000}
\definecolor{bluewhitered-114}{rgb}{0.3049,0.3049,1.0000}
\definecolor{bluewhitered-115}{rgb}{0.3151,0.3151,1.0000}
\definecolor{bluewhitered-116}{rgb}{0.3254,0.3254,1.0000}
\definecolor{bluewhitered-117}{rgb}{0.3356,0.3356,1.0000}
\definecolor{bluewhitered-118}{rgb}{0.3458,0.3458,1.0000}
\definecolor{bluewhitered-119}{rgb}{0.3560,0.3560,1.0000}
\definecolor{bluewhitered-120}{rgb}{0.3662,0.3662,1.0000}
\definecolor{bluewhitered-121}{rgb}{0.3765,0.3765,1.0000}
\definecolor{bluewhitered-122}{rgb}{0.3867,0.3867,1.0000}
\definecolor{bluewhitered-123}{rgb}{0.3969,0.3969,1.0000}
\definecolor{bluewhitered-124}{rgb}{0.4071,0.4071,1.0000}
\definecolor{bluewhitered-125}{rgb}{0.4173,0.4173,1.0000}
\definecolor{bluewhitered-126}{rgb}{0.4276,0.4276,1.0000}
\definecolor{bluewhitered-127}{rgb}{0.4378,0.4378,1.0000}
\definecolor{bluewhitered-128}{rgb}{0.4480,0.4480,1.0000}
\definecolor{bluewhitered-129}{rgb}{0.4582,0.4582,1.0000}
\definecolor{bluewhitered-130}{rgb}{0.4684,0.4684,1.0000}
\definecolor{bluewhitered-131}{rgb}{0.4787,0.4787,1.0000}
\definecolor{bluewhitered-132}{rgb}{0.4889,0.4889,1.0000}
\definecolor{bluewhitered-133}{rgb}{0.4991,0.4991,1.0000}
\definecolor{bluewhitered-134}{rgb}{0.5093,0.5093,1.0000}
\definecolor{bluewhitered-135}{rgb}{0.5195,0.5195,1.0000}
\definecolor{bluewhitered-136}{rgb}{0.5298,0.5298,1.0000}
\definecolor{bluewhitered-137}{rgb}{0.5400,0.5400,1.0000}
\definecolor{bluewhitered-138}{rgb}{0.5502,0.5502,1.0000}
\definecolor{bluewhitered-139}{rgb}{0.5604,0.5604,1.0000}
\definecolor{bluewhitered-140}{rgb}{0.5706,0.5706,1.0000}
\definecolor{bluewhitered-141}{rgb}{0.5809,0.5809,1.0000}
\definecolor{bluewhitered-142}{rgb}{0.5911,0.5911,1.0000}
\definecolor{bluewhitered-143}{rgb}{0.6013,0.6013,1.0000}
\definecolor{bluewhitered-144}{rgb}{0.6115,0.6115,1.0000}
\definecolor{bluewhitered-145}{rgb}{0.6217,0.6217,1.0000}
\definecolor{bluewhitered-146}{rgb}{0.6320,0.6320,1.0000}
\definecolor{bluewhitered-147}{rgb}{0.6422,0.6422,1.0000}
\definecolor{bluewhitered-148}{rgb}{0.6524,0.6524,1.0000}
\definecolor{bluewhitered-149}{rgb}{0.6626,0.6626,1.0000}
\definecolor{bluewhitered-150}{rgb}{0.6728,0.6728,1.0000}
\definecolor{bluewhitered-151}{rgb}{0.6831,0.6831,1.0000}
\definecolor{bluewhitered-152}{rgb}{0.6933,0.6933,1.0000}
\definecolor{bluewhitered-153}{rgb}{0.7035,0.7035,1.0000}
\definecolor{bluewhitered-154}{rgb}{0.7137,0.7137,1.0000}
\definecolor{bluewhitered-155}{rgb}{0.7239,0.7239,1.0000}
\definecolor{bluewhitered-156}{rgb}{0.7342,0.7342,1.0000}
\definecolor{bluewhitered-157}{rgb}{0.7444,0.7444,1.0000}
\definecolor{bluewhitered-158}{rgb}{0.7546,0.7546,1.0000}
\definecolor{bluewhitered-159}{rgb}{0.7648,0.7648,1.0000}
\definecolor{bluewhitered-160}{rgb}{0.7751,0.7751,1.0000}
\definecolor{bluewhitered-161}{rgb}{0.7853,0.7853,1.0000}
\definecolor{bluewhitered-162}{rgb}{0.7955,0.7955,1.0000}
\definecolor{bluewhitered-163}{rgb}{0.8057,0.8057,1.0000}
\definecolor{bluewhitered-164}{rgb}{0.8159,0.8159,1.0000}
\definecolor{bluewhitered-165}{rgb}{0.8262,0.8262,1.0000}
\definecolor{bluewhitered-166}{rgb}{0.8364,0.8364,1.0000}
\definecolor{bluewhitered-167}{rgb}{0.8466,0.8466,1.0000}
\definecolor{bluewhitered-168}{rgb}{0.8512,0.8512,1.0000}
\definecolor{bluewhitered-169}{rgb}{0.8530,0.8530,1.0000}
\definecolor{bluewhitered-170}{rgb}{0.8548,0.8548,1.0000}
\definecolor{bluewhitered-171}{rgb}{0.8566,0.8566,1.0000}
\definecolor{bluewhitered-172}{rgb}{0.8584,0.8584,1.0000}
\definecolor{bluewhitered-173}{rgb}{0.8602,0.8602,1.0000}
\definecolor{bluewhitered-174}{rgb}{0.8620,0.8620,1.0000}
\definecolor{bluewhitered-175}{rgb}{0.8638,0.8638,1.0000}
\definecolor{bluewhitered-176}{rgb}{0.8656,0.8656,1.0000}
\definecolor{bluewhitered-177}{rgb}{0.8674,0.8674,1.0000}
\definecolor{bluewhitered-178}{rgb}{0.8692,0.8692,1.0000}
\definecolor{bluewhitered-179}{rgb}{0.8710,0.8710,1.0000}
\definecolor{bluewhitered-180}{rgb}{0.8728,0.8728,1.0000}
\definecolor{bluewhitered-181}{rgb}{0.8746,0.8746,1.0000}
\definecolor{bluewhitered-182}{rgb}{0.8765,0.8765,1.0000}
\definecolor{bluewhitered-183}{rgb}{0.8783,0.8783,1.0000}
\definecolor{bluewhitered-184}{rgb}{0.8801,0.8801,1.0000}
\definecolor{bluewhitered-185}{rgb}{0.8819,0.8819,1.0000}
\definecolor{bluewhitered-186}{rgb}{0.8837,0.8837,1.0000}
\definecolor{bluewhitered-187}{rgb}{0.8855,0.8855,1.0000}
\definecolor{bluewhitered-188}{rgb}{0.8873,0.8873,1.0000}
\definecolor{bluewhitered-189}{rgb}{0.8891,0.8891,1.0000}
\definecolor{bluewhitered-190}{rgb}{0.8909,0.8909,1.0000}
\definecolor{bluewhitered-191}{rgb}{0.8927,0.8927,1.0000}
\definecolor{bluewhitered-192}{rgb}{0.8945,0.8945,1.0000}
\definecolor{bluewhitered-193}{rgb}{0.8963,0.8963,1.0000}
\definecolor{bluewhitered-194}{rgb}{0.8981,0.8981,1.0000}
\definecolor{bluewhitered-195}{rgb}{0.8999,0.8999,1.0000}
\definecolor{bluewhitered-196}{rgb}{0.9017,0.9017,1.0000}
\definecolor{bluewhitered-197}{rgb}{0.9035,0.9035,1.0000}
\definecolor{bluewhitered-198}{rgb}{0.9053,0.9053,1.0000}
\definecolor{bluewhitered-199}{rgb}{0.9071,0.9071,1.0000}
\definecolor{bluewhitered-200}{rgb}{0.9089,0.9089,1.0000}
\definecolor{bluewhitered-201}{rgb}{0.9107,0.9107,1.0000}
\definecolor{bluewhitered-202}{rgb}{0.9125,0.9125,1.0000}
\definecolor{bluewhitered-203}{rgb}{0.9143,0.9143,1.0000}
\definecolor{bluewhitered-204}{rgb}{0.9161,0.9161,1.0000}
\definecolor{bluewhitered-205}{rgb}{0.9179,0.9179,1.0000}
\definecolor{bluewhitered-206}{rgb}{0.9197,0.9197,1.0000}
\definecolor{bluewhitered-207}{rgb}{0.9215,0.9215,1.0000}
\definecolor{bluewhitered-208}{rgb}{0.9233,0.9233,1.0000}
\definecolor{bluewhitered-209}{rgb}{0.9252,0.9252,1.0000}
\definecolor{bluewhitered-210}{rgb}{0.9270,0.9270,1.0000}
\definecolor{bluewhitered-211}{rgb}{0.9288,0.9288,1.0000}
\definecolor{bluewhitered-212}{rgb}{0.9306,0.9306,1.0000}
\definecolor{bluewhitered-213}{rgb}{0.9324,0.9324,1.0000}
\definecolor{bluewhitered-214}{rgb}{0.9342,0.9342,1.0000}
\definecolor{bluewhitered-215}{rgb}{0.9360,0.9360,1.0000}
\definecolor{bluewhitered-216}{rgb}{0.9378,0.9378,1.0000}
\definecolor{bluewhitered-217}{rgb}{0.9396,0.9396,1.0000}
\definecolor{bluewhitered-218}{rgb}{0.9414,0.9414,1.0000}
\definecolor{bluewhitered-219}{rgb}{0.9432,0.9432,1.0000}
\definecolor{bluewhitered-220}{rgb}{0.9450,0.9450,1.0000}
\definecolor{bluewhitered-221}{rgb}{0.9468,0.9468,1.0000}
\definecolor{bluewhitered-222}{rgb}{0.9486,0.9486,1.0000}
\definecolor{bluewhitered-223}{rgb}{0.9504,0.9504,1.0000}
\definecolor{bluewhitered-224}{rgb}{0.9522,0.9522,1.0000}
\definecolor{bluewhitered-225}{rgb}{0.9540,0.9540,1.0000}
\definecolor{bluewhitered-226}{rgb}{0.9558,0.9558,1.0000}
\definecolor{bluewhitered-227}{rgb}{0.9576,0.9576,1.0000}
\definecolor{bluewhitered-228}{rgb}{0.9594,0.9594,1.0000}
\definecolor{bluewhitered-229}{rgb}{0.9612,0.9612,1.0000}
\definecolor{bluewhitered-230}{rgb}{0.9630,0.9630,1.0000}
\definecolor{bluewhitered-231}{rgb}{0.9648,0.9648,1.0000}
\definecolor{bluewhitered-232}{rgb}{0.9666,0.9666,1.0000}
\definecolor{bluewhitered-233}{rgb}{0.9684,0.9684,1.0000}
\definecolor{bluewhitered-234}{rgb}{0.9702,0.9702,1.0000}
\definecolor{bluewhitered-235}{rgb}{0.9720,0.9720,1.0000}
\definecolor{bluewhitered-236}{rgb}{0.9738,0.9738,1.0000}
\definecolor{bluewhitered-237}{rgb}{0.9757,0.9757,1.0000}
\definecolor{bluewhitered-238}{rgb}{0.9775,0.9775,1.0000}
\definecolor{bluewhitered-239}{rgb}{0.9793,0.9793,1.0000}
\definecolor{bluewhitered-240}{rgb}{0.9811,0.9811,1.0000}
\definecolor{bluewhitered-241}{rgb}{0.9829,0.9829,1.0000}
\definecolor{bluewhitered-242}{rgb}{0.9847,0.9847,1.0000}
\definecolor{bluewhitered-243}{rgb}{0.9865,0.9865,1.0000}
\definecolor{bluewhitered-244}{rgb}{0.9883,0.9883,1.0000}
\definecolor{bluewhitered-245}{rgb}{0.9901,0.9901,1.0000}
\definecolor{bluewhitered-246}{rgb}{0.9919,0.9919,1.0000}
\definecolor{bluewhitered-247}{rgb}{0.9937,0.9937,1.0000}
\definecolor{bluewhitered-248}{rgb}{0.9955,0.9955,1.0000}
\definecolor{bluewhitered-249}{rgb}{0.9973,0.9973,1.0000}
\definecolor{bluewhitered-250}{rgb}{0.9991,0.9991,1.0000}
\definecolor{bluewhitered-251}{rgb}{1.0000,0.9991,0.9991}
\definecolor{bluewhitered-252}{rgb}{1.0000,0.9973,0.9973}
\definecolor{bluewhitered-253}{rgb}{1.0000,0.9955,0.9955}
\definecolor{bluewhitered-254}{rgb}{1.0000,0.9937,0.9937}
\definecolor{bluewhitered-255}{rgb}{1.0000,0.9919,0.9919}
\definecolor{bluewhitered-256}{rgb}{1.0000,0.9901,0.9901}
\definecolor{bluewhitered-257}{rgb}{1.0000,0.9883,0.9883}
\definecolor{bluewhitered-258}{rgb}{1.0000,0.9865,0.9865}
\definecolor{bluewhitered-259}{rgb}{1.0000,0.9847,0.9847}
\definecolor{bluewhitered-260}{rgb}{1.0000,0.9829,0.9829}
\definecolor{bluewhitered-261}{rgb}{1.0000,0.9811,0.9811}
\definecolor{bluewhitered-262}{rgb}{1.0000,0.9793,0.9793}
\definecolor{bluewhitered-263}{rgb}{1.0000,0.9775,0.9775}
\definecolor{bluewhitered-264}{rgb}{1.0000,0.9757,0.9757}
\definecolor{bluewhitered-265}{rgb}{1.0000,0.9738,0.9738}
\definecolor{bluewhitered-266}{rgb}{1.0000,0.9720,0.9720}
\definecolor{bluewhitered-267}{rgb}{1.0000,0.9702,0.9702}
\definecolor{bluewhitered-268}{rgb}{1.0000,0.9684,0.9684}
\definecolor{bluewhitered-269}{rgb}{1.0000,0.9666,0.9666}
\definecolor{bluewhitered-270}{rgb}{1.0000,0.9648,0.9648}
\definecolor{bluewhitered-271}{rgb}{1.0000,0.9630,0.9630}
\definecolor{bluewhitered-272}{rgb}{1.0000,0.9612,0.9612}
\definecolor{bluewhitered-273}{rgb}{1.0000,0.9594,0.9594}
\definecolor{bluewhitered-274}{rgb}{1.0000,0.9576,0.9576}
\definecolor{bluewhitered-275}{rgb}{1.0000,0.9558,0.9558}
\definecolor{bluewhitered-276}{rgb}{1.0000,0.9540,0.9540}
\definecolor{bluewhitered-277}{rgb}{1.0000,0.9522,0.9522}
\definecolor{bluewhitered-278}{rgb}{1.0000,0.9504,0.9504}
\definecolor{bluewhitered-279}{rgb}{1.0000,0.9486,0.9486}
\definecolor{bluewhitered-280}{rgb}{1.0000,0.9468,0.9468}
\definecolor{bluewhitered-281}{rgb}{1.0000,0.9450,0.9450}
\definecolor{bluewhitered-282}{rgb}{1.0000,0.9432,0.9432}
\definecolor{bluewhitered-283}{rgb}{1.0000,0.9414,0.9414}
\definecolor{bluewhitered-284}{rgb}{1.0000,0.9396,0.9396}
\definecolor{bluewhitered-285}{rgb}{1.0000,0.9378,0.9378}
\definecolor{bluewhitered-286}{rgb}{1.0000,0.9360,0.9360}
\definecolor{bluewhitered-287}{rgb}{1.0000,0.9342,0.9342}
\definecolor{bluewhitered-288}{rgb}{1.0000,0.9324,0.9324}
\definecolor{bluewhitered-289}{rgb}{1.0000,0.9306,0.9306}
\definecolor{bluewhitered-290}{rgb}{1.0000,0.9288,0.9288}
\definecolor{bluewhitered-291}{rgb}{1.0000,0.9270,0.9270}
\definecolor{bluewhitered-292}{rgb}{1.0000,0.9252,0.9252}
\definecolor{bluewhitered-293}{rgb}{1.0000,0.9233,0.9233}
\definecolor{bluewhitered-294}{rgb}{1.0000,0.9215,0.9215}
\definecolor{bluewhitered-295}{rgb}{1.0000,0.9197,0.9197}
\definecolor{bluewhitered-296}{rgb}{1.0000,0.9179,0.9179}
\definecolor{bluewhitered-297}{rgb}{1.0000,0.9161,0.9161}
\definecolor{bluewhitered-298}{rgb}{1.0000,0.9143,0.9143}
\definecolor{bluewhitered-299}{rgb}{1.0000,0.9125,0.9125}
\definecolor{bluewhitered-300}{rgb}{1.0000,0.9107,0.9107}
\definecolor{bluewhitered-301}{rgb}{1.0000,0.9089,0.9089}
\definecolor{bluewhitered-302}{rgb}{1.0000,0.9071,0.9071}
\definecolor{bluewhitered-303}{rgb}{1.0000,0.9053,0.9053}
\definecolor{bluewhitered-304}{rgb}{1.0000,0.9035,0.9035}
\definecolor{bluewhitered-305}{rgb}{1.0000,0.9017,0.9017}
\definecolor{bluewhitered-306}{rgb}{1.0000,0.8999,0.8999}
\definecolor{bluewhitered-307}{rgb}{1.0000,0.8981,0.8981}
\definecolor{bluewhitered-308}{rgb}{1.0000,0.8963,0.8963}
\definecolor{bluewhitered-309}{rgb}{1.0000,0.8945,0.8945}
\definecolor{bluewhitered-310}{rgb}{1.0000,0.8927,0.8927}
\definecolor{bluewhitered-311}{rgb}{1.0000,0.8909,0.8909}
\definecolor{bluewhitered-312}{rgb}{1.0000,0.8891,0.8891}
\definecolor{bluewhitered-313}{rgb}{1.0000,0.8873,0.8873}
\definecolor{bluewhitered-314}{rgb}{1.0000,0.8855,0.8855}
\definecolor{bluewhitered-315}{rgb}{1.0000,0.8837,0.8837}
\definecolor{bluewhitered-316}{rgb}{1.0000,0.8819,0.8819}
\definecolor{bluewhitered-317}{rgb}{1.0000,0.8801,0.8801}
\definecolor{bluewhitered-318}{rgb}{1.0000,0.8783,0.8783}
\definecolor{bluewhitered-319}{rgb}{1.0000,0.8765,0.8765}
\definecolor{bluewhitered-320}{rgb}{1.0000,0.8746,0.8746}
\definecolor{bluewhitered-321}{rgb}{1.0000,0.8728,0.8728}
\definecolor{bluewhitered-322}{rgb}{1.0000,0.8710,0.8710}
\definecolor{bluewhitered-323}{rgb}{1.0000,0.8692,0.8692}
\definecolor{bluewhitered-324}{rgb}{1.0000,0.8674,0.8674}
\definecolor{bluewhitered-325}{rgb}{1.0000,0.8656,0.8656}
\definecolor{bluewhitered-326}{rgb}{1.0000,0.8638,0.8638}
\definecolor{bluewhitered-327}{rgb}{1.0000,0.8620,0.8620}
\definecolor{bluewhitered-328}{rgb}{1.0000,0.8602,0.8602}
\definecolor{bluewhitered-329}{rgb}{1.0000,0.8584,0.8584}
\definecolor{bluewhitered-330}{rgb}{1.0000,0.8566,0.8566}
\definecolor{bluewhitered-331}{rgb}{1.0000,0.8548,0.8548}
\definecolor{bluewhitered-332}{rgb}{1.0000,0.8530,0.8530}
\definecolor{bluewhitered-333}{rgb}{1.0000,0.8512,0.8512}
\definecolor{bluewhitered-334}{rgb}{1.0000,0.8466,0.8466}
\definecolor{bluewhitered-335}{rgb}{1.0000,0.8364,0.8364}
\definecolor{bluewhitered-336}{rgb}{1.0000,0.8262,0.8262}
\definecolor{bluewhitered-337}{rgb}{1.0000,0.8159,0.8159}
\definecolor{bluewhitered-338}{rgb}{1.0000,0.8057,0.8057}
\definecolor{bluewhitered-339}{rgb}{1.0000,0.7955,0.7955}
\definecolor{bluewhitered-340}{rgb}{1.0000,0.7853,0.7853}
\definecolor{bluewhitered-341}{rgb}{1.0000,0.7751,0.7751}
\definecolor{bluewhitered-342}{rgb}{1.0000,0.7648,0.7648}
\definecolor{bluewhitered-343}{rgb}{1.0000,0.7546,0.7546}
\definecolor{bluewhitered-344}{rgb}{1.0000,0.7444,0.7444}
\definecolor{bluewhitered-345}{rgb}{1.0000,0.7342,0.7342}
\definecolor{bluewhitered-346}{rgb}{1.0000,0.7239,0.7239}
\definecolor{bluewhitered-347}{rgb}{1.0000,0.7137,0.7137}
\definecolor{bluewhitered-348}{rgb}{1.0000,0.7035,0.7035}
\definecolor{bluewhitered-349}{rgb}{1.0000,0.6933,0.6933}
\definecolor{bluewhitered-350}{rgb}{1.0000,0.6831,0.6831}
\definecolor{bluewhitered-351}{rgb}{1.0000,0.6728,0.6728}
\definecolor{bluewhitered-352}{rgb}{1.0000,0.6626,0.6626}
\definecolor{bluewhitered-353}{rgb}{1.0000,0.6524,0.6524}
\definecolor{bluewhitered-354}{rgb}{1.0000,0.6422,0.6422}
\definecolor{bluewhitered-355}{rgb}{1.0000,0.6320,0.6320}
\definecolor{bluewhitered-356}{rgb}{1.0000,0.6217,0.6217}
\definecolor{bluewhitered-357}{rgb}{1.0000,0.6115,0.6115}
\definecolor{bluewhitered-358}{rgb}{1.0000,0.6013,0.6013}
\definecolor{bluewhitered-359}{rgb}{1.0000,0.5911,0.5911}
\definecolor{bluewhitered-360}{rgb}{1.0000,0.5809,0.5809}
\definecolor{bluewhitered-361}{rgb}{1.0000,0.5706,0.5706}
\definecolor{bluewhitered-362}{rgb}{1.0000,0.5604,0.5604}
\definecolor{bluewhitered-363}{rgb}{1.0000,0.5502,0.5502}
\definecolor{bluewhitered-364}{rgb}{1.0000,0.5400,0.5400}
\definecolor{bluewhitered-365}{rgb}{1.0000,0.5298,0.5298}
\definecolor{bluewhitered-366}{rgb}{1.0000,0.5195,0.5195}
\definecolor{bluewhitered-367}{rgb}{1.0000,0.5093,0.5093}
\definecolor{bluewhitered-368}{rgb}{1.0000,0.4991,0.4991}
\definecolor{bluewhitered-369}{rgb}{1.0000,0.4889,0.4889}
\definecolor{bluewhitered-370}{rgb}{1.0000,0.4787,0.4787}
\definecolor{bluewhitered-371}{rgb}{1.0000,0.4684,0.4684}
\definecolor{bluewhitered-372}{rgb}{1.0000,0.4582,0.4582}
\definecolor{bluewhitered-373}{rgb}{1.0000,0.4480,0.4480}
\definecolor{bluewhitered-374}{rgb}{1.0000,0.4378,0.4378}
\definecolor{bluewhitered-375}{rgb}{1.0000,0.4276,0.4276}
\definecolor{bluewhitered-376}{rgb}{1.0000,0.4173,0.4173}
\definecolor{bluewhitered-377}{rgb}{1.0000,0.4071,0.4071}
\definecolor{bluewhitered-378}{rgb}{1.0000,0.3969,0.3969}
\definecolor{bluewhitered-379}{rgb}{1.0000,0.3867,0.3867}
\definecolor{bluewhitered-380}{rgb}{1.0000,0.3765,0.3765}
\definecolor{bluewhitered-381}{rgb}{1.0000,0.3662,0.3662}
\definecolor{bluewhitered-382}{rgb}{1.0000,0.3560,0.3560}
\definecolor{bluewhitered-383}{rgb}{1.0000,0.3458,0.3458}
\definecolor{bluewhitered-384}{rgb}{1.0000,0.3356,0.3356}
\definecolor{bluewhitered-385}{rgb}{1.0000,0.3254,0.3254}
\definecolor{bluewhitered-386}{rgb}{1.0000,0.3151,0.3151}
\definecolor{bluewhitered-387}{rgb}{1.0000,0.3049,0.3049}
\definecolor{bluewhitered-388}{rgb}{1.0000,0.2947,0.2947}
\definecolor{bluewhitered-389}{rgb}{1.0000,0.2845,0.2845}
\definecolor{bluewhitered-390}{rgb}{1.0000,0.2742,0.2742}
\definecolor{bluewhitered-391}{rgb}{1.0000,0.2640,0.2640}
\definecolor{bluewhitered-392}{rgb}{1.0000,0.2538,0.2538}
\definecolor{bluewhitered-393}{rgb}{1.0000,0.2436,0.2436}
\definecolor{bluewhitered-394}{rgb}{1.0000,0.2334,0.2334}
\definecolor{bluewhitered-395}{rgb}{1.0000,0.2231,0.2231}
\definecolor{bluewhitered-396}{rgb}{1.0000,0.2129,0.2129}
\definecolor{bluewhitered-397}{rgb}{1.0000,0.2027,0.2027}
\definecolor{bluewhitered-398}{rgb}{1.0000,0.1925,0.1925}
\definecolor{bluewhitered-399}{rgb}{1.0000,0.1823,0.1823}
\definecolor{bluewhitered-400}{rgb}{1.0000,0.1720,0.1720}
\definecolor{bluewhitered-401}{rgb}{1.0000,0.1618,0.1618}
\definecolor{bluewhitered-402}{rgb}{1.0000,0.1516,0.1516}
\definecolor{bluewhitered-403}{rgb}{1.0000,0.1414,0.1414}
\definecolor{bluewhitered-404}{rgb}{1.0000,0.1312,0.1312}
\definecolor{bluewhitered-405}{rgb}{1.0000,0.1209,0.1209}
\definecolor{bluewhitered-406}{rgb}{1.0000,0.1107,0.1107}
\definecolor{bluewhitered-407}{rgb}{1.0000,0.1005,0.1005}
\definecolor{bluewhitered-408}{rgb}{1.0000,0.0903,0.0903}
\definecolor{bluewhitered-409}{rgb}{1.0000,0.0801,0.0801}
\definecolor{bluewhitered-410}{rgb}{1.0000,0.0698,0.0698}
\definecolor{bluewhitered-411}{rgb}{1.0000,0.0596,0.0596}
\definecolor{bluewhitered-412}{rgb}{1.0000,0.0494,0.0494}
\definecolor{bluewhitered-413}{rgb}{1.0000,0.0392,0.0392}
\definecolor{bluewhitered-414}{rgb}{1.0000,0.0290,0.0290}
\definecolor{bluewhitered-415}{rgb}{1.0000,0.0187,0.0187}
\definecolor{bluewhitered-416}{rgb}{1.0000,0.0085,0.0085}
\definecolor{bluewhitered-417}{rgb}{0.9988,0.0000,0.0000}
\definecolor{bluewhitered-418}{rgb}{0.9916,0.0000,0.0000}
\definecolor{bluewhitered-419}{rgb}{0.9844,0.0000,0.0000}
\definecolor{bluewhitered-420}{rgb}{0.9772,0.0000,0.0000}
\definecolor{bluewhitered-421}{rgb}{0.9699,0.0000,0.0000}
\definecolor{bluewhitered-422}{rgb}{0.9627,0.0000,0.0000}
\definecolor{bluewhitered-423}{rgb}{0.9555,0.0000,0.0000}
\definecolor{bluewhitered-424}{rgb}{0.9483,0.0000,0.0000}
\definecolor{bluewhitered-425}{rgb}{0.9411,0.0000,0.0000}
\definecolor{bluewhitered-426}{rgb}{0.9339,0.0000,0.0000}
\definecolor{bluewhitered-427}{rgb}{0.9267,0.0000,0.0000}
\definecolor{bluewhitered-428}{rgb}{0.9194,0.0000,0.0000}
\definecolor{bluewhitered-429}{rgb}{0.9122,0.0000,0.0000}
\definecolor{bluewhitered-430}{rgb}{0.9050,0.0000,0.0000}
\definecolor{bluewhitered-431}{rgb}{0.8978,0.0000,0.0000}
\definecolor{bluewhitered-432}{rgb}{0.8906,0.0000,0.0000}
\definecolor{bluewhitered-433}{rgb}{0.8834,0.0000,0.0000}
\definecolor{bluewhitered-434}{rgb}{0.8762,0.0000,0.0000}
\definecolor{bluewhitered-435}{rgb}{0.8689,0.0000,0.0000}
\definecolor{bluewhitered-436}{rgb}{0.8617,0.0000,0.0000}
\definecolor{bluewhitered-437}{rgb}{0.8545,0.0000,0.0000}
\definecolor{bluewhitered-438}{rgb}{0.8473,0.0000,0.0000}
\definecolor{bluewhitered-439}{rgb}{0.8401,0.0000,0.0000}
\definecolor{bluewhitered-440}{rgb}{0.8329,0.0000,0.0000}
\definecolor{bluewhitered-441}{rgb}{0.8257,0.0000,0.0000}
\definecolor{bluewhitered-442}{rgb}{0.8184,0.0000,0.0000}
\definecolor{bluewhitered-443}{rgb}{0.8112,0.0000,0.0000}
\definecolor{bluewhitered-444}{rgb}{0.8040,0.0000,0.0000}
\definecolor{bluewhitered-445}{rgb}{0.7968,0.0000,0.0000}
\definecolor{bluewhitered-446}{rgb}{0.7896,0.0000,0.0000}
\definecolor{bluewhitered-447}{rgb}{0.7824,0.0000,0.0000}
\definecolor{bluewhitered-448}{rgb}{0.7752,0.0000,0.0000}
\definecolor{bluewhitered-449}{rgb}{0.7679,0.0000,0.0000}
\definecolor{bluewhitered-450}{rgb}{0.7607,0.0000,0.0000}
\definecolor{bluewhitered-451}{rgb}{0.7535,0.0000,0.0000}
\definecolor{bluewhitered-452}{rgb}{0.7463,0.0000,0.0000}
\definecolor{bluewhitered-453}{rgb}{0.7391,0.0000,0.0000}
\definecolor{bluewhitered-454}{rgb}{0.7319,0.0000,0.0000}
\definecolor{bluewhitered-455}{rgb}{0.7246,0.0000,0.0000}
\definecolor{bluewhitered-456}{rgb}{0.7174,0.0000,0.0000}
\definecolor{bluewhitered-457}{rgb}{0.7102,0.0000,0.0000}
\definecolor{bluewhitered-458}{rgb}{0.7030,0.0000,0.0000}
\definecolor{bluewhitered-459}{rgb}{0.6958,0.0000,0.0000}
\definecolor{bluewhitered-460}{rgb}{0.6886,0.0000,0.0000}
\definecolor{bluewhitered-461}{rgb}{0.6814,0.0000,0.0000}
\definecolor{bluewhitered-462}{rgb}{0.6741,0.0000,0.0000}
\definecolor{bluewhitered-463}{rgb}{0.6669,0.0000,0.0000}
\definecolor{bluewhitered-464}{rgb}{0.6597,0.0000,0.0000}
\definecolor{bluewhitered-465}{rgb}{0.6525,0.0000,0.0000}
\definecolor{bluewhitered-466}{rgb}{0.6453,0.0000,0.0000}
\definecolor{bluewhitered-467}{rgb}{0.6381,0.0000,0.0000}
\definecolor{bluewhitered-468}{rgb}{0.6309,0.0000,0.0000}
\definecolor{bluewhitered-469}{rgb}{0.6236,0.0000,0.0000}
\definecolor{bluewhitered-470}{rgb}{0.6164,0.0000,0.0000}
\definecolor{bluewhitered-471}{rgb}{0.6092,0.0000,0.0000}
\definecolor{bluewhitered-472}{rgb}{0.6020,0.0000,0.0000}
\definecolor{bluewhitered-473}{rgb}{0.5948,0.0000,0.0000}
\definecolor{bluewhitered-474}{rgb}{0.5876,0.0000,0.0000}
\definecolor{bluewhitered-475}{rgb}{0.5804,0.0000,0.0000}
\definecolor{bluewhitered-476}{rgb}{0.5731,0.0000,0.0000}
\definecolor{bluewhitered-477}{rgb}{0.5659,0.0000,0.0000}
\definecolor{bluewhitered-478}{rgb}{0.5587,0.0000,0.0000}
\definecolor{bluewhitered-479}{rgb}{0.5515,0.0000,0.0000}
\definecolor{bluewhitered-480}{rgb}{0.5443,0.0000,0.0000}
\definecolor{bluewhitered-481}{rgb}{0.5371,0.0000,0.0000}
\definecolor{bluewhitered-482}{rgb}{0.5299,0.0000,0.0000}
\definecolor{bluewhitered-483}{rgb}{0.5226,0.0000,0.0000}
\definecolor{bluewhitered-484}{rgb}{0.5154,0.0000,0.0000}
\definecolor{bluewhitered-485}{rgb}{0.5082,0.0000,0.0000}
\definecolor{bluewhitered-486}{rgb}{0.5010,0.0000,0.0000}
\definecolor{bluewhitered-487}{rgb}{0.4938,0.0000,0.0000}
\definecolor{bluewhitered-488}{rgb}{0.4866,0.0000,0.0000}
\definecolor{bluewhitered-489}{rgb}{0.4794,0.0000,0.0000}
\definecolor{bluewhitered-490}{rgb}{0.4721,0.0000,0.0000}
\definecolor{bluewhitered-491}{rgb}{0.4649,0.0000,0.0000}
\definecolor{bluewhitered-492}{rgb}{0.4577,0.0000,0.0000}
\definecolor{bluewhitered-493}{rgb}{0.4505,0.0000,0.0000}
\definecolor{bluewhitered-494}{rgb}{0.4433,0.0000,0.0000}
\definecolor{bluewhitered-495}{rgb}{0.4361,0.0000,0.0000}
\definecolor{bluewhitered-496}{rgb}{0.4289,0.0000,0.0000}
\definecolor{bluewhitered-497}{rgb}{0.4216,0.0000,0.0000}
\definecolor{bluewhitered-498}{rgb}{0.4144,0.0000,0.0000}
\definecolor{bluewhitered-499}{rgb}{0.4072,0.0000,0.0000}
\definecolor{bluewhitered-500}{rgb}{0.4000,0.0000,0.0000}
\pgfplotsset{
colormap={bluewhitered}{
color={bluewhitered-1};
color={bluewhitered-2};
color={bluewhitered-3};
color={bluewhitered-4};
color={bluewhitered-5};
color={bluewhitered-6};
color={bluewhitered-7};
color={bluewhitered-8};
color={bluewhitered-9};
color={bluewhitered-10};
color={bluewhitered-11};
color={bluewhitered-12};
color={bluewhitered-13};
color={bluewhitered-14};
color={bluewhitered-15};
color={bluewhitered-16};
color={bluewhitered-17};
color={bluewhitered-18};
color={bluewhitered-19};
color={bluewhitered-20};
color={bluewhitered-21};
color={bluewhitered-22};
color={bluewhitered-23};
color={bluewhitered-24};
color={bluewhitered-25};
color={bluewhitered-26};
color={bluewhitered-27};
color={bluewhitered-28};
color={bluewhitered-29};
color={bluewhitered-30};
color={bluewhitered-31};
color={bluewhitered-32};
color={bluewhitered-33};
color={bluewhitered-34};
color={bluewhitered-35};
color={bluewhitered-36};
color={bluewhitered-37};
color={bluewhitered-38};
color={bluewhitered-39};
color={bluewhitered-40};
color={bluewhitered-41};
color={bluewhitered-42};
color={bluewhitered-43};
color={bluewhitered-44};
color={bluewhitered-45};
color={bluewhitered-46};
color={bluewhitered-47};
color={bluewhitered-48};
color={bluewhitered-49};
color={bluewhitered-50};
color={bluewhitered-51};
color={bluewhitered-52};
color={bluewhitered-53};
color={bluewhitered-54};
color={bluewhitered-55};
color={bluewhitered-56};
color={bluewhitered-57};
color={bluewhitered-58};
color={bluewhitered-59};
color={bluewhitered-60};
color={bluewhitered-61};
color={bluewhitered-62};
color={bluewhitered-63};
color={bluewhitered-64};
color={bluewhitered-65};
color={bluewhitered-66};
color={bluewhitered-67};
color={bluewhitered-68};
color={bluewhitered-69};
color={bluewhitered-70};
color={bluewhitered-71};
color={bluewhitered-72};
color={bluewhitered-73};
color={bluewhitered-74};
color={bluewhitered-75};
color={bluewhitered-76};
color={bluewhitered-77};
color={bluewhitered-78};
color={bluewhitered-79};
color={bluewhitered-80};
color={bluewhitered-81};
color={bluewhitered-82};
color={bluewhitered-83};
color={bluewhitered-84};
color={bluewhitered-85};
color={bluewhitered-86};
color={bluewhitered-87};
color={bluewhitered-88};
color={bluewhitered-89};
color={bluewhitered-90};
color={bluewhitered-91};
color={bluewhitered-92};
color={bluewhitered-93};
color={bluewhitered-94};
color={bluewhitered-95};
color={bluewhitered-96};
color={bluewhitered-97};
color={bluewhitered-98};
color={bluewhitered-99};
color={bluewhitered-100};
color={bluewhitered-101};
color={bluewhitered-102};
color={bluewhitered-103};
color={bluewhitered-104};
color={bluewhitered-105};
color={bluewhitered-106};
color={bluewhitered-107};
color={bluewhitered-108};
color={bluewhitered-109};
color={bluewhitered-110};
color={bluewhitered-111};
color={bluewhitered-112};
color={bluewhitered-113};
color={bluewhitered-114};
color={bluewhitered-115};
color={bluewhitered-116};
color={bluewhitered-117};
color={bluewhitered-118};
color={bluewhitered-119};
color={bluewhitered-120};
color={bluewhitered-121};
color={bluewhitered-122};
color={bluewhitered-123};
color={bluewhitered-124};
color={bluewhitered-125};
color={bluewhitered-126};
color={bluewhitered-127};
color={bluewhitered-128};
color={bluewhitered-129};
color={bluewhitered-130};
color={bluewhitered-131};
color={bluewhitered-132};
color={bluewhitered-133};
color={bluewhitered-134};
color={bluewhitered-135};
color={bluewhitered-136};
color={bluewhitered-137};
color={bluewhitered-138};
color={bluewhitered-139};
color={bluewhitered-140};
color={bluewhitered-141};
color={bluewhitered-142};
color={bluewhitered-143};
color={bluewhitered-144};
color={bluewhitered-145};
color={bluewhitered-146};
color={bluewhitered-147};
color={bluewhitered-148};
color={bluewhitered-149};
color={bluewhitered-150};
color={bluewhitered-151};
color={bluewhitered-152};
color={bluewhitered-153};
color={bluewhitered-154};
color={bluewhitered-155};
color={bluewhitered-156};
color={bluewhitered-157};
color={bluewhitered-158};
color={bluewhitered-159};
color={bluewhitered-160};
color={bluewhitered-161};
color={bluewhitered-162};
color={bluewhitered-163};
color={bluewhitered-164};
color={bluewhitered-165};
color={bluewhitered-166};
color={bluewhitered-167};
color={bluewhitered-168};
color={bluewhitered-169};
color={bluewhitered-170};
color={bluewhitered-171};
color={bluewhitered-172};
color={bluewhitered-173};
color={bluewhitered-174};
color={bluewhitered-175};
color={bluewhitered-176};
color={bluewhitered-177};
color={bluewhitered-178};
color={bluewhitered-179};
color={bluewhitered-180};
color={bluewhitered-181};
color={bluewhitered-182};
color={bluewhitered-183};
color={bluewhitered-184};
color={bluewhitered-185};
color={bluewhitered-186};
color={bluewhitered-187};
color={bluewhitered-188};
color={bluewhitered-189};
color={bluewhitered-190};
color={bluewhitered-191};
color={bluewhitered-192};
color={bluewhitered-193};
color={bluewhitered-194};
color={bluewhitered-195};
color={bluewhitered-196};
color={bluewhitered-197};
color={bluewhitered-198};
color={bluewhitered-199};
color={bluewhitered-200};
color={bluewhitered-201};
color={bluewhitered-202};
color={bluewhitered-203};
color={bluewhitered-204};
color={bluewhitered-205};
color={bluewhitered-206};
color={bluewhitered-207};
color={bluewhitered-208};
color={bluewhitered-209};
color={bluewhitered-210};
color={bluewhitered-211};
color={bluewhitered-212};
color={bluewhitered-213};
color={bluewhitered-214};
color={bluewhitered-215};
color={bluewhitered-216};
color={bluewhitered-217};
color={bluewhitered-218};
color={bluewhitered-219};
color={bluewhitered-220};
color={bluewhitered-221};
color={bluewhitered-222};
color={bluewhitered-223};
color={bluewhitered-224};
color={bluewhitered-225};
color={bluewhitered-226};
color={bluewhitered-227};
color={bluewhitered-228};
color={bluewhitered-229};
color={bluewhitered-230};
color={bluewhitered-231};
color={bluewhitered-232};
color={bluewhitered-233};
color={bluewhitered-234};
color={bluewhitered-235};
color={bluewhitered-236};
color={bluewhitered-237};
color={bluewhitered-238};
color={bluewhitered-239};
color={bluewhitered-240};
color={bluewhitered-241};
color={bluewhitered-242};
color={bluewhitered-243};
color={bluewhitered-244};
color={bluewhitered-245};
color={bluewhitered-246};
color={bluewhitered-247};
color={bluewhitered-248};
color={bluewhitered-249};
color={bluewhitered-250};
color={bluewhitered-251};
color={bluewhitered-252};
color={bluewhitered-253};
color={bluewhitered-254};
color={bluewhitered-255};
color={bluewhitered-256};
color={bluewhitered-257};
color={bluewhitered-258};
color={bluewhitered-259};
color={bluewhitered-260};
color={bluewhitered-261};
color={bluewhitered-262};
color={bluewhitered-263};
color={bluewhitered-264};
color={bluewhitered-265};
color={bluewhitered-266};
color={bluewhitered-267};
color={bluewhitered-268};
color={bluewhitered-269};
color={bluewhitered-270};
color={bluewhitered-271};
color={bluewhitered-272};
color={bluewhitered-273};
color={bluewhitered-274};
color={bluewhitered-275};
color={bluewhitered-276};
color={bluewhitered-277};
color={bluewhitered-278};
color={bluewhitered-279};
color={bluewhitered-280};
color={bluewhitered-281};
color={bluewhitered-282};
color={bluewhitered-283};
color={bluewhitered-284};
color={bluewhitered-285};
color={bluewhitered-286};
color={bluewhitered-287};
color={bluewhitered-288};
color={bluewhitered-289};
color={bluewhitered-290};
color={bluewhitered-291};
color={bluewhitered-292};
color={bluewhitered-293};
color={bluewhitered-294};
color={bluewhitered-295};
color={bluewhitered-296};
color={bluewhitered-297};
color={bluewhitered-298};
color={bluewhitered-299};
color={bluewhitered-300};
color={bluewhitered-301};
color={bluewhitered-302};
color={bluewhitered-303};
color={bluewhitered-304};
color={bluewhitered-305};
color={bluewhitered-306};
color={bluewhitered-307};
color={bluewhitered-308};
color={bluewhitered-309};
color={bluewhitered-310};
color={bluewhitered-311};
color={bluewhitered-312};
color={bluewhitered-313};
color={bluewhitered-314};
color={bluewhitered-315};
color={bluewhitered-316};
color={bluewhitered-317};
color={bluewhitered-318};
color={bluewhitered-319};
color={bluewhitered-320};
color={bluewhitered-321};
color={bluewhitered-322};
color={bluewhitered-323};
color={bluewhitered-324};
color={bluewhitered-325};
color={bluewhitered-326};
color={bluewhitered-327};
color={bluewhitered-328};
color={bluewhitered-329};
color={bluewhitered-330};
color={bluewhitered-331};
color={bluewhitered-332};
color={bluewhitered-333};
color={bluewhitered-334};
color={bluewhitered-335};
color={bluewhitered-336};
color={bluewhitered-337};
color={bluewhitered-338};
color={bluewhitered-339};
color={bluewhitered-340};
color={bluewhitered-341};
color={bluewhitered-342};
color={bluewhitered-343};
color={bluewhitered-344};
color={bluewhitered-345};
color={bluewhitered-346};
color={bluewhitered-347};
color={bluewhitered-348};
color={bluewhitered-349};
color={bluewhitered-350};
color={bluewhitered-351};
color={bluewhitered-352};
color={bluewhitered-353};
color={bluewhitered-354};
color={bluewhitered-355};
color={bluewhitered-356};
color={bluewhitered-357};
color={bluewhitered-358};
color={bluewhitered-359};
color={bluewhitered-360};
color={bluewhitered-361};
color={bluewhitered-362};
color={bluewhitered-363};
color={bluewhitered-364};
color={bluewhitered-365};
color={bluewhitered-366};
color={bluewhitered-367};
color={bluewhitered-368};
color={bluewhitered-369};
color={bluewhitered-370};
color={bluewhitered-371};
color={bluewhitered-372};
color={bluewhitered-373};
color={bluewhitered-374};
color={bluewhitered-375};
color={bluewhitered-376};
color={bluewhitered-377};
color={bluewhitered-378};
color={bluewhitered-379};
color={bluewhitered-380};
color={bluewhitered-381};
color={bluewhitered-382};
color={bluewhitered-383};
color={bluewhitered-384};
color={bluewhitered-385};
color={bluewhitered-386};
color={bluewhitered-387};
color={bluewhitered-388};
color={bluewhitered-389};
color={bluewhitered-390};
color={bluewhitered-391};
color={bluewhitered-392};
color={bluewhitered-393};
color={bluewhitered-394};
color={bluewhitered-395};
color={bluewhitered-396};
color={bluewhitered-397};
color={bluewhitered-398};
color={bluewhitered-399};
color={bluewhitered-400};
color={bluewhitered-401};
color={bluewhitered-402};
color={bluewhitered-403};
color={bluewhitered-404};
color={bluewhitered-405};
color={bluewhitered-406};
color={bluewhitered-407};
color={bluewhitered-408};
color={bluewhitered-409};
color={bluewhitered-410};
color={bluewhitered-411};
color={bluewhitered-412};
color={bluewhitered-413};
color={bluewhitered-414};
color={bluewhitered-415};
color={bluewhitered-416};
color={bluewhitered-417};
color={bluewhitered-418};
color={bluewhitered-419};
color={bluewhitered-420};
color={bluewhitered-421};
color={bluewhitered-422};
color={bluewhitered-423};
color={bluewhitered-424};
color={bluewhitered-425};
color={bluewhitered-426};
color={bluewhitered-427};
color={bluewhitered-428};
color={bluewhitered-429};
color={bluewhitered-430};
color={bluewhitered-431};
color={bluewhitered-432};
color={bluewhitered-433};
color={bluewhitered-434};
color={bluewhitered-435};
color={bluewhitered-436};
color={bluewhitered-437};
color={bluewhitered-438};
color={bluewhitered-439};
color={bluewhitered-440};
color={bluewhitered-441};
color={bluewhitered-442};
color={bluewhitered-443};
color={bluewhitered-444};
color={bluewhitered-445};
color={bluewhitered-446};
color={bluewhitered-447};
color={bluewhitered-448};
color={bluewhitered-449};
color={bluewhitered-450};
color={bluewhitered-451};
color={bluewhitered-452};
color={bluewhitered-453};
color={bluewhitered-454};
color={bluewhitered-455};
color={bluewhitered-456};
color={bluewhitered-457};
color={bluewhitered-458};
color={bluewhitered-459};
color={bluewhitered-460};
color={bluewhitered-461};
color={bluewhitered-462};
color={bluewhitered-463};
color={bluewhitered-464};
color={bluewhitered-465};
color={bluewhitered-466};
color={bluewhitered-467};
color={bluewhitered-468};
color={bluewhitered-469};
color={bluewhitered-470};
color={bluewhitered-471};
color={bluewhitered-472};
color={bluewhitered-473};
color={bluewhitered-474};
color={bluewhitered-475};
color={bluewhitered-476};
color={bluewhitered-477};
color={bluewhitered-478};
color={bluewhitered-479};
color={bluewhitered-480};
color={bluewhitered-481};
color={bluewhitered-482};
color={bluewhitered-483};
color={bluewhitered-484};
color={bluewhitered-485};
color={bluewhitered-486};
color={bluewhitered-487};
color={bluewhitered-488};
color={bluewhitered-489};
color={bluewhitered-490};
color={bluewhitered-491};
color={bluewhitered-492};
color={bluewhitered-493};
color={bluewhitered-494};
color={bluewhitered-495};
color={bluewhitered-496};
color={bluewhitered-497};
color={bluewhitered-498};
color={bluewhitered-499};
color={bluewhitered-500};
}
}
\definecolor{Umag-1}{rgb}{0.8000,0.8000,0.8000}
\definecolor{Umag-2}{rgb}{0.8006,0.8006,0.7976}
\definecolor{Umag-3}{rgb}{0.8012,0.8012,0.7952}
\definecolor{Umag-4}{rgb}{0.8018,0.8018,0.7928}
\definecolor{Umag-5}{rgb}{0.8024,0.8024,0.7904}
\definecolor{Umag-6}{rgb}{0.8030,0.8030,0.7880}
\definecolor{Umag-7}{rgb}{0.8036,0.8036,0.7856}
\definecolor{Umag-8}{rgb}{0.8042,0.8042,0.7832}
\definecolor{Umag-9}{rgb}{0.8048,0.8048,0.7808}
\definecolor{Umag-10}{rgb}{0.8054,0.8054,0.7784}
\definecolor{Umag-11}{rgb}{0.8060,0.8060,0.7760}
\definecolor{Umag-12}{rgb}{0.8066,0.8066,0.7735}
\definecolor{Umag-13}{rgb}{0.8072,0.8072,0.7711}
\definecolor{Umag-14}{rgb}{0.8078,0.8078,0.7687}
\definecolor{Umag-15}{rgb}{0.8084,0.8084,0.7663}
\definecolor{Umag-16}{rgb}{0.8090,0.8090,0.7639}
\definecolor{Umag-17}{rgb}{0.8096,0.8096,0.7615}
\definecolor{Umag-18}{rgb}{0.8102,0.8102,0.7591}
\definecolor{Umag-19}{rgb}{0.8108,0.8108,0.7567}
\definecolor{Umag-20}{rgb}{0.8114,0.8114,0.7543}
\definecolor{Umag-21}{rgb}{0.8120,0.8120,0.7519}
\definecolor{Umag-22}{rgb}{0.8126,0.8126,0.7495}
\definecolor{Umag-23}{rgb}{0.8132,0.8132,0.7471}
\definecolor{Umag-24}{rgb}{0.8138,0.8138,0.7447}
\definecolor{Umag-25}{rgb}{0.8144,0.8144,0.7423}
\definecolor{Umag-26}{rgb}{0.8150,0.8150,0.7399}
\definecolor{Umag-27}{rgb}{0.8156,0.8156,0.7375}
\definecolor{Umag-28}{rgb}{0.8162,0.8162,0.7351}
\definecolor{Umag-29}{rgb}{0.8168,0.8168,0.7327}
\definecolor{Umag-30}{rgb}{0.8174,0.8174,0.7303}
\definecolor{Umag-31}{rgb}{0.8180,0.8180,0.7279}
\definecolor{Umag-32}{rgb}{0.8186,0.8186,0.7255}
\definecolor{Umag-33}{rgb}{0.8192,0.8192,0.7230}
\definecolor{Umag-34}{rgb}{0.8198,0.8198,0.7206}
\definecolor{Umag-35}{rgb}{0.8204,0.8204,0.7182}
\definecolor{Umag-36}{rgb}{0.8210,0.8210,0.7158}
\definecolor{Umag-37}{rgb}{0.8216,0.8216,0.7134}
\definecolor{Umag-38}{rgb}{0.8222,0.8222,0.7110}
\definecolor{Umag-39}{rgb}{0.8228,0.8228,0.7086}
\definecolor{Umag-40}{rgb}{0.8234,0.8234,0.7062}
\definecolor{Umag-41}{rgb}{0.8240,0.8240,0.7038}
\definecolor{Umag-42}{rgb}{0.8246,0.8246,0.7014}
\definecolor{Umag-43}{rgb}{0.8253,0.8253,0.6990}
\definecolor{Umag-44}{rgb}{0.8259,0.8259,0.6966}
\definecolor{Umag-45}{rgb}{0.8265,0.8265,0.6942}
\definecolor{Umag-46}{rgb}{0.8271,0.8271,0.6918}
\definecolor{Umag-47}{rgb}{0.8277,0.8277,0.6894}
\definecolor{Umag-48}{rgb}{0.8283,0.8283,0.6870}
\definecolor{Umag-49}{rgb}{0.8289,0.8289,0.6846}
\definecolor{Umag-50}{rgb}{0.8295,0.8295,0.6822}
\definecolor{Umag-51}{rgb}{0.8301,0.8301,0.6798}
\definecolor{Umag-52}{rgb}{0.8307,0.8307,0.6774}
\definecolor{Umag-53}{rgb}{0.8313,0.8313,0.6749}
\definecolor{Umag-54}{rgb}{0.8319,0.8319,0.6725}
\definecolor{Umag-55}{rgb}{0.8325,0.8325,0.6701}
\definecolor{Umag-56}{rgb}{0.8331,0.8331,0.6677}
\definecolor{Umag-57}{rgb}{0.8337,0.8337,0.6653}
\definecolor{Umag-58}{rgb}{0.8343,0.8343,0.6629}
\definecolor{Umag-59}{rgb}{0.8349,0.8349,0.6605}
\definecolor{Umag-60}{rgb}{0.8355,0.8355,0.6581}
\definecolor{Umag-61}{rgb}{0.8361,0.8361,0.6557}
\definecolor{Umag-62}{rgb}{0.8367,0.8367,0.6533}
\definecolor{Umag-63}{rgb}{0.8373,0.8373,0.6509}
\definecolor{Umag-64}{rgb}{0.8379,0.8379,0.6485}
\definecolor{Umag-65}{rgb}{0.8385,0.8385,0.6461}
\definecolor{Umag-66}{rgb}{0.8391,0.8391,0.6437}
\definecolor{Umag-67}{rgb}{0.8397,0.8397,0.6413}
\definecolor{Umag-68}{rgb}{0.8403,0.8403,0.6389}
\definecolor{Umag-69}{rgb}{0.8409,0.8409,0.6365}
\definecolor{Umag-70}{rgb}{0.8415,0.8415,0.6341}
\definecolor{Umag-71}{rgb}{0.8421,0.8421,0.6317}
\definecolor{Umag-72}{rgb}{0.8427,0.8427,0.6293}
\definecolor{Umag-73}{rgb}{0.8433,0.8433,0.6269}
\definecolor{Umag-74}{rgb}{0.8439,0.8439,0.6244}
\definecolor{Umag-75}{rgb}{0.8445,0.8445,0.6220}
\definecolor{Umag-76}{rgb}{0.8451,0.8451,0.6196}
\definecolor{Umag-77}{rgb}{0.8457,0.8457,0.6172}
\definecolor{Umag-78}{rgb}{0.8463,0.8463,0.6148}
\definecolor{Umag-79}{rgb}{0.8469,0.8469,0.6124}
\definecolor{Umag-80}{rgb}{0.8475,0.8475,0.6100}
\definecolor{Umag-81}{rgb}{0.8481,0.8481,0.6076}
\definecolor{Umag-82}{rgb}{0.8487,0.8487,0.6052}
\definecolor{Umag-83}{rgb}{0.8493,0.8493,0.6028}
\definecolor{Umag-84}{rgb}{0.8499,0.8499,0.6004}
\definecolor{Umag-85}{rgb}{0.8505,0.8505,0.5980}
\definecolor{Umag-86}{rgb}{0.8511,0.8511,0.5956}
\definecolor{Umag-87}{rgb}{0.8517,0.8517,0.5932}
\definecolor{Umag-88}{rgb}{0.8523,0.8523,0.5908}
\definecolor{Umag-89}{rgb}{0.8529,0.8529,0.5884}
\definecolor{Umag-90}{rgb}{0.8535,0.8535,0.5860}
\definecolor{Umag-91}{rgb}{0.8541,0.8541,0.5836}
\definecolor{Umag-92}{rgb}{0.8547,0.8547,0.5812}
\definecolor{Umag-93}{rgb}{0.8553,0.8553,0.5788}
\definecolor{Umag-94}{rgb}{0.8559,0.8559,0.5764}
\definecolor{Umag-95}{rgb}{0.8565,0.8565,0.5739}
\definecolor{Umag-96}{rgb}{0.8571,0.8571,0.5715}
\definecolor{Umag-97}{rgb}{0.8577,0.8577,0.5691}
\definecolor{Umag-98}{rgb}{0.8583,0.8583,0.5667}
\definecolor{Umag-99}{rgb}{0.8589,0.8589,0.5643}
\definecolor{Umag-100}{rgb}{0.8595,0.8595,0.5619}
\definecolor{Umag-101}{rgb}{0.8601,0.8601,0.5595}
\definecolor{Umag-102}{rgb}{0.8607,0.8607,0.5571}
\definecolor{Umag-103}{rgb}{0.8613,0.8613,0.5547}
\definecolor{Umag-104}{rgb}{0.8619,0.8619,0.5523}
\definecolor{Umag-105}{rgb}{0.8625,0.8625,0.5499}
\definecolor{Umag-106}{rgb}{0.8631,0.8631,0.5475}
\definecolor{Umag-107}{rgb}{0.8637,0.8637,0.5451}
\definecolor{Umag-108}{rgb}{0.8643,0.8643,0.5427}
\definecolor{Umag-109}{rgb}{0.8649,0.8649,0.5403}
\definecolor{Umag-110}{rgb}{0.8655,0.8655,0.5379}
\definecolor{Umag-111}{rgb}{0.8661,0.8661,0.5355}
\definecolor{Umag-112}{rgb}{0.8667,0.8667,0.5331}
\definecolor{Umag-113}{rgb}{0.8673,0.8673,0.5307}
\definecolor{Umag-114}{rgb}{0.8679,0.8679,0.5283}
\definecolor{Umag-115}{rgb}{0.8685,0.8685,0.5259}
\definecolor{Umag-116}{rgb}{0.8691,0.8691,0.5234}
\definecolor{Umag-117}{rgb}{0.8697,0.8697,0.5210}
\definecolor{Umag-118}{rgb}{0.8703,0.8703,0.5186}
\definecolor{Umag-119}{rgb}{0.8709,0.8709,0.5162}
\definecolor{Umag-120}{rgb}{0.8715,0.8715,0.5138}
\definecolor{Umag-121}{rgb}{0.8721,0.8721,0.5114}
\definecolor{Umag-122}{rgb}{0.8727,0.8727,0.5090}
\definecolor{Umag-123}{rgb}{0.8733,0.8733,0.5066}
\definecolor{Umag-124}{rgb}{0.8739,0.8739,0.5042}
\definecolor{Umag-125}{rgb}{0.8745,0.8745,0.5018}
\definecolor{Umag-126}{rgb}{0.8752,0.8752,0.4994}
\definecolor{Umag-127}{rgb}{0.8758,0.8758,0.4970}
\definecolor{Umag-128}{rgb}{0.8764,0.8764,0.4946}
\definecolor{Umag-129}{rgb}{0.8770,0.8770,0.4922}
\definecolor{Umag-130}{rgb}{0.8776,0.8776,0.4898}
\definecolor{Umag-131}{rgb}{0.8782,0.8782,0.4874}
\definecolor{Umag-132}{rgb}{0.8788,0.8788,0.4850}
\definecolor{Umag-133}{rgb}{0.8794,0.8794,0.4826}
\definecolor{Umag-134}{rgb}{0.8800,0.8800,0.4802}
\definecolor{Umag-135}{rgb}{0.8806,0.8806,0.4778}
\definecolor{Umag-136}{rgb}{0.8812,0.8812,0.4754}
\definecolor{Umag-137}{rgb}{0.8818,0.8818,0.4729}
\definecolor{Umag-138}{rgb}{0.8824,0.8824,0.4705}
\definecolor{Umag-139}{rgb}{0.8830,0.8830,0.4681}
\definecolor{Umag-140}{rgb}{0.8836,0.8836,0.4657}
\definecolor{Umag-141}{rgb}{0.8842,0.8842,0.4633}
\definecolor{Umag-142}{rgb}{0.8848,0.8848,0.4609}
\definecolor{Umag-143}{rgb}{0.8854,0.8854,0.4585}
\definecolor{Umag-144}{rgb}{0.8860,0.8860,0.4561}
\definecolor{Umag-145}{rgb}{0.8866,0.8866,0.4537}
\definecolor{Umag-146}{rgb}{0.8872,0.8872,0.4513}
\definecolor{Umag-147}{rgb}{0.8878,0.8878,0.4489}
\definecolor{Umag-148}{rgb}{0.8884,0.8884,0.4465}
\definecolor{Umag-149}{rgb}{0.8890,0.8890,0.4441}
\definecolor{Umag-150}{rgb}{0.8896,0.8896,0.4417}
\definecolor{Umag-151}{rgb}{0.8902,0.8902,0.4393}
\definecolor{Umag-152}{rgb}{0.8908,0.8908,0.4369}
\definecolor{Umag-153}{rgb}{0.8914,0.8914,0.4345}
\definecolor{Umag-154}{rgb}{0.8920,0.8920,0.4321}
\definecolor{Umag-155}{rgb}{0.8926,0.8926,0.4297}
\definecolor{Umag-156}{rgb}{0.8932,0.8932,0.4273}
\definecolor{Umag-157}{rgb}{0.8938,0.8938,0.4248}
\definecolor{Umag-158}{rgb}{0.8944,0.8944,0.4224}
\definecolor{Umag-159}{rgb}{0.8950,0.8950,0.4200}
\definecolor{Umag-160}{rgb}{0.8956,0.8956,0.4176}
\definecolor{Umag-161}{rgb}{0.8962,0.8962,0.4152}
\definecolor{Umag-162}{rgb}{0.8968,0.8968,0.4128}
\definecolor{Umag-163}{rgb}{0.8974,0.8974,0.4104}
\definecolor{Umag-164}{rgb}{0.8980,0.8980,0.4080}
\definecolor{Umag-165}{rgb}{0.8986,0.8986,0.4056}
\definecolor{Umag-166}{rgb}{0.8992,0.8992,0.4032}
\definecolor{Umag-167}{rgb}{0.8998,0.8998,0.4008}
\definecolor{Umag-168}{rgb}{0.9004,0.8988,0.3984}
\definecolor{Umag-169}{rgb}{0.9010,0.8970,0.3960}
\definecolor{Umag-170}{rgb}{0.9016,0.8952,0.3936}
\definecolor{Umag-171}{rgb}{0.9022,0.8934,0.3912}
\definecolor{Umag-172}{rgb}{0.9028,0.8916,0.3888}
\definecolor{Umag-173}{rgb}{0.9034,0.8898,0.3864}
\definecolor{Umag-174}{rgb}{0.9040,0.8880,0.3840}
\definecolor{Umag-175}{rgb}{0.9046,0.8862,0.3816}
\definecolor{Umag-176}{rgb}{0.9052,0.8844,0.3792}
\definecolor{Umag-177}{rgb}{0.9058,0.8826,0.3768}
\definecolor{Umag-178}{rgb}{0.9064,0.8808,0.3743}
\definecolor{Umag-179}{rgb}{0.9070,0.8790,0.3719}
\definecolor{Umag-180}{rgb}{0.9076,0.8772,0.3695}
\definecolor{Umag-181}{rgb}{0.9082,0.8754,0.3671}
\definecolor{Umag-182}{rgb}{0.9088,0.8735,0.3647}
\definecolor{Umag-183}{rgb}{0.9094,0.8717,0.3623}
\definecolor{Umag-184}{rgb}{0.9100,0.8699,0.3599}
\definecolor{Umag-185}{rgb}{0.9106,0.8681,0.3575}
\definecolor{Umag-186}{rgb}{0.9112,0.8663,0.3551}
\definecolor{Umag-187}{rgb}{0.9118,0.8645,0.3527}
\definecolor{Umag-188}{rgb}{0.9124,0.8627,0.3503}
\definecolor{Umag-189}{rgb}{0.9130,0.8609,0.3479}
\definecolor{Umag-190}{rgb}{0.9136,0.8591,0.3455}
\definecolor{Umag-191}{rgb}{0.9142,0.8573,0.3431}
\definecolor{Umag-192}{rgb}{0.9148,0.8555,0.3407}
\definecolor{Umag-193}{rgb}{0.9154,0.8537,0.3383}
\definecolor{Umag-194}{rgb}{0.9160,0.8519,0.3359}
\definecolor{Umag-195}{rgb}{0.9166,0.8501,0.3335}
\definecolor{Umag-196}{rgb}{0.9172,0.8483,0.3311}
\definecolor{Umag-197}{rgb}{0.9178,0.8465,0.3287}
\definecolor{Umag-198}{rgb}{0.9184,0.8447,0.3263}
\definecolor{Umag-199}{rgb}{0.9190,0.8429,0.3238}
\definecolor{Umag-200}{rgb}{0.9196,0.8411,0.3214}
\definecolor{Umag-201}{rgb}{0.9202,0.8393,0.3190}
\definecolor{Umag-202}{rgb}{0.9208,0.8375,0.3166}
\definecolor{Umag-203}{rgb}{0.9214,0.8357,0.3142}
\definecolor{Umag-204}{rgb}{0.9220,0.8339,0.3118}
\definecolor{Umag-205}{rgb}{0.9226,0.8321,0.3094}
\definecolor{Umag-206}{rgb}{0.9232,0.8303,0.3070}
\definecolor{Umag-207}{rgb}{0.9238,0.8285,0.3046}
\definecolor{Umag-208}{rgb}{0.9244,0.8267,0.3022}
\definecolor{Umag-209}{rgb}{0.9251,0.8248,0.2998}
\definecolor{Umag-210}{rgb}{0.9257,0.8230,0.2974}
\definecolor{Umag-211}{rgb}{0.9263,0.8212,0.2950}
\definecolor{Umag-212}{rgb}{0.9269,0.8194,0.2926}
\definecolor{Umag-213}{rgb}{0.9275,0.8176,0.2902}
\definecolor{Umag-214}{rgb}{0.9281,0.8158,0.2878}
\definecolor{Umag-215}{rgb}{0.9287,0.8140,0.2854}
\definecolor{Umag-216}{rgb}{0.9293,0.8122,0.2830}
\definecolor{Umag-217}{rgb}{0.9299,0.8104,0.2806}
\definecolor{Umag-218}{rgb}{0.9305,0.8086,0.2782}
\definecolor{Umag-219}{rgb}{0.9311,0.8068,0.2758}
\definecolor{Umag-220}{rgb}{0.9317,0.8050,0.2733}
\definecolor{Umag-221}{rgb}{0.9323,0.8032,0.2709}
\definecolor{Umag-222}{rgb}{0.9329,0.8014,0.2685}
\definecolor{Umag-223}{rgb}{0.9335,0.7996,0.2661}
\definecolor{Umag-224}{rgb}{0.9341,0.7978,0.2637}
\definecolor{Umag-225}{rgb}{0.9347,0.7960,0.2613}
\definecolor{Umag-226}{rgb}{0.9353,0.7942,0.2589}
\definecolor{Umag-227}{rgb}{0.9359,0.7924,0.2565}
\definecolor{Umag-228}{rgb}{0.9365,0.7906,0.2541}
\definecolor{Umag-229}{rgb}{0.9371,0.7888,0.2517}
\definecolor{Umag-230}{rgb}{0.9377,0.7870,0.2493}
\definecolor{Umag-231}{rgb}{0.9383,0.7852,0.2469}
\definecolor{Umag-232}{rgb}{0.9389,0.7834,0.2445}
\definecolor{Umag-233}{rgb}{0.9395,0.7816,0.2421}
\definecolor{Umag-234}{rgb}{0.9401,0.7798,0.2397}
\definecolor{Umag-235}{rgb}{0.9407,0.7780,0.2373}
\definecolor{Umag-236}{rgb}{0.9413,0.7762,0.2349}
\definecolor{Umag-237}{rgb}{0.9419,0.7743,0.2325}
\definecolor{Umag-238}{rgb}{0.9425,0.7725,0.2301}
\definecolor{Umag-239}{rgb}{0.9431,0.7707,0.2277}
\definecolor{Umag-240}{rgb}{0.9437,0.7689,0.2253}
\definecolor{Umag-241}{rgb}{0.9443,0.7671,0.2228}
\definecolor{Umag-242}{rgb}{0.9449,0.7653,0.2204}
\definecolor{Umag-243}{rgb}{0.9455,0.7635,0.2180}
\definecolor{Umag-244}{rgb}{0.9461,0.7617,0.2156}
\definecolor{Umag-245}{rgb}{0.9467,0.7599,0.2132}
\definecolor{Umag-246}{rgb}{0.9473,0.7581,0.2108}
\definecolor{Umag-247}{rgb}{0.9479,0.7563,0.2084}
\definecolor{Umag-248}{rgb}{0.9485,0.7545,0.2060}
\definecolor{Umag-249}{rgb}{0.9491,0.7527,0.2036}
\definecolor{Umag-250}{rgb}{0.9497,0.7509,0.2012}
\definecolor{Umag-251}{rgb}{0.9503,0.7491,0.1988}
\definecolor{Umag-252}{rgb}{0.9509,0.7473,0.1964}
\definecolor{Umag-253}{rgb}{0.9515,0.7455,0.1940}
\definecolor{Umag-254}{rgb}{0.9521,0.7437,0.1916}
\definecolor{Umag-255}{rgb}{0.9527,0.7419,0.1892}
\definecolor{Umag-256}{rgb}{0.9533,0.7401,0.1868}
\definecolor{Umag-257}{rgb}{0.9539,0.7383,0.1844}
\definecolor{Umag-258}{rgb}{0.9545,0.7365,0.1820}
\definecolor{Umag-259}{rgb}{0.9551,0.7347,0.1796}
\definecolor{Umag-260}{rgb}{0.9557,0.7329,0.1772}
\definecolor{Umag-261}{rgb}{0.9563,0.7311,0.1747}
\definecolor{Umag-262}{rgb}{0.9569,0.7293,0.1723}
\definecolor{Umag-263}{rgb}{0.9575,0.7275,0.1699}
\definecolor{Umag-264}{rgb}{0.9581,0.7257,0.1675}
\definecolor{Umag-265}{rgb}{0.9587,0.7238,0.1651}
\definecolor{Umag-266}{rgb}{0.9593,0.7220,0.1627}
\definecolor{Umag-267}{rgb}{0.9599,0.7202,0.1603}
\definecolor{Umag-268}{rgb}{0.9605,0.7184,0.1579}
\definecolor{Umag-269}{rgb}{0.9611,0.7166,0.1555}
\definecolor{Umag-270}{rgb}{0.9617,0.7148,0.1531}
\definecolor{Umag-271}{rgb}{0.9623,0.7130,0.1507}
\definecolor{Umag-272}{rgb}{0.9629,0.7112,0.1483}
\definecolor{Umag-273}{rgb}{0.9635,0.7094,0.1459}
\definecolor{Umag-274}{rgb}{0.9641,0.7076,0.1435}
\definecolor{Umag-275}{rgb}{0.9647,0.7058,0.1411}
\definecolor{Umag-276}{rgb}{0.9653,0.7040,0.1387}
\definecolor{Umag-277}{rgb}{0.9659,0.7022,0.1363}
\definecolor{Umag-278}{rgb}{0.9665,0.7004,0.1339}
\definecolor{Umag-279}{rgb}{0.9671,0.6986,0.1315}
\definecolor{Umag-280}{rgb}{0.9677,0.6968,0.1291}
\definecolor{Umag-281}{rgb}{0.9683,0.6950,0.1267}
\definecolor{Umag-282}{rgb}{0.9689,0.6932,0.1242}
\definecolor{Umag-283}{rgb}{0.9695,0.6914,0.1218}
\definecolor{Umag-284}{rgb}{0.9701,0.6896,0.1194}
\definecolor{Umag-285}{rgb}{0.9707,0.6878,0.1170}
\definecolor{Umag-286}{rgb}{0.9713,0.6860,0.1146}
\definecolor{Umag-287}{rgb}{0.9719,0.6842,0.1122}
\definecolor{Umag-288}{rgb}{0.9725,0.6824,0.1098}
\definecolor{Umag-289}{rgb}{0.9731,0.6806,0.1074}
\definecolor{Umag-290}{rgb}{0.9737,0.6788,0.1050}
\definecolor{Umag-291}{rgb}{0.9743,0.6770,0.1026}
\definecolor{Umag-292}{rgb}{0.9749,0.6752,0.1002}
\definecolor{Umag-293}{rgb}{0.9756,0.6733,0.0978}
\definecolor{Umag-294}{rgb}{0.9762,0.6715,0.0954}
\definecolor{Umag-295}{rgb}{0.9768,0.6697,0.0930}
\definecolor{Umag-296}{rgb}{0.9774,0.6679,0.0906}
\definecolor{Umag-297}{rgb}{0.9780,0.6661,0.0882}
\definecolor{Umag-298}{rgb}{0.9786,0.6643,0.0858}
\definecolor{Umag-299}{rgb}{0.9792,0.6625,0.0834}
\definecolor{Umag-300}{rgb}{0.9798,0.6607,0.0810}
\definecolor{Umag-301}{rgb}{0.9804,0.6589,0.0786}
\definecolor{Umag-302}{rgb}{0.9810,0.6571,0.0762}
\definecolor{Umag-303}{rgb}{0.9816,0.6553,0.0737}
\definecolor{Umag-304}{rgb}{0.9822,0.6535,0.0713}
\definecolor{Umag-305}{rgb}{0.9828,0.6517,0.0689}
\definecolor{Umag-306}{rgb}{0.9834,0.6499,0.0665}
\definecolor{Umag-307}{rgb}{0.9840,0.6481,0.0641}
\definecolor{Umag-308}{rgb}{0.9846,0.6463,0.0617}
\definecolor{Umag-309}{rgb}{0.9852,0.6445,0.0593}
\definecolor{Umag-310}{rgb}{0.9858,0.6427,0.0569}
\definecolor{Umag-311}{rgb}{0.9864,0.6409,0.0545}
\definecolor{Umag-312}{rgb}{0.9870,0.6391,0.0521}
\definecolor{Umag-313}{rgb}{0.9876,0.6373,0.0497}
\definecolor{Umag-314}{rgb}{0.9882,0.6355,0.0473}
\definecolor{Umag-315}{rgb}{0.9888,0.6337,0.0449}
\definecolor{Umag-316}{rgb}{0.9894,0.6319,0.0425}
\definecolor{Umag-317}{rgb}{0.9900,0.6301,0.0401}
\definecolor{Umag-318}{rgb}{0.9906,0.6283,0.0377}
\definecolor{Umag-319}{rgb}{0.9912,0.6265,0.0353}
\definecolor{Umag-320}{rgb}{0.9918,0.6246,0.0329}
\definecolor{Umag-321}{rgb}{0.9924,0.6228,0.0305}
\definecolor{Umag-322}{rgb}{0.9930,0.6210,0.0281}
\definecolor{Umag-323}{rgb}{0.9936,0.6192,0.0257}
\definecolor{Umag-324}{rgb}{0.9942,0.6174,0.0232}
\definecolor{Umag-325}{rgb}{0.9948,0.6156,0.0208}
\definecolor{Umag-326}{rgb}{0.9954,0.6138,0.0184}
\definecolor{Umag-327}{rgb}{0.9960,0.6120,0.0160}
\definecolor{Umag-328}{rgb}{0.9966,0.6102,0.0136}
\definecolor{Umag-329}{rgb}{0.9972,0.6084,0.0112}
\definecolor{Umag-330}{rgb}{0.9978,0.6066,0.0088}
\definecolor{Umag-331}{rgb}{0.9984,0.6048,0.0064}
\definecolor{Umag-332}{rgb}{0.9990,0.6030,0.0040}
\definecolor{Umag-333}{rgb}{0.9996,0.6012,0.0016}
\definecolor{Umag-334}{rgb}{0.9996,0.5988,0.0000}
\definecolor{Umag-335}{rgb}{0.9984,0.5952,0.0000}
\definecolor{Umag-336}{rgb}{0.9972,0.5916,0.0000}
\definecolor{Umag-337}{rgb}{0.9960,0.5880,0.0000}
\definecolor{Umag-338}{rgb}{0.9948,0.5844,0.0000}
\definecolor{Umag-339}{rgb}{0.9936,0.5808,0.0000}
\definecolor{Umag-340}{rgb}{0.9924,0.5772,0.0000}
\definecolor{Umag-341}{rgb}{0.9912,0.5735,0.0000}
\definecolor{Umag-342}{rgb}{0.9900,0.5699,0.0000}
\definecolor{Umag-343}{rgb}{0.9888,0.5663,0.0000}
\definecolor{Umag-344}{rgb}{0.9876,0.5627,0.0000}
\definecolor{Umag-345}{rgb}{0.9864,0.5591,0.0000}
\definecolor{Umag-346}{rgb}{0.9852,0.5555,0.0000}
\definecolor{Umag-347}{rgb}{0.9840,0.5519,0.0000}
\definecolor{Umag-348}{rgb}{0.9828,0.5483,0.0000}
\definecolor{Umag-349}{rgb}{0.9816,0.5447,0.0000}
\definecolor{Umag-350}{rgb}{0.9804,0.5411,0.0000}
\definecolor{Umag-351}{rgb}{0.9792,0.5375,0.0000}
\definecolor{Umag-352}{rgb}{0.9780,0.5339,0.0000}
\definecolor{Umag-353}{rgb}{0.9768,0.5303,0.0000}
\definecolor{Umag-354}{rgb}{0.9756,0.5267,0.0000}
\definecolor{Umag-355}{rgb}{0.9743,0.5230,0.0000}
\definecolor{Umag-356}{rgb}{0.9731,0.5194,0.0000}
\definecolor{Umag-357}{rgb}{0.9719,0.5158,0.0000}
\definecolor{Umag-358}{rgb}{0.9707,0.5122,0.0000}
\definecolor{Umag-359}{rgb}{0.9695,0.5086,0.0000}
\definecolor{Umag-360}{rgb}{0.9683,0.5050,0.0000}
\definecolor{Umag-361}{rgb}{0.9671,0.5014,0.0000}
\definecolor{Umag-362}{rgb}{0.9659,0.4978,0.0000}
\definecolor{Umag-363}{rgb}{0.9647,0.4942,0.0000}
\definecolor{Umag-364}{rgb}{0.9635,0.4906,0.0000}
\definecolor{Umag-365}{rgb}{0.9623,0.4870,0.0000}
\definecolor{Umag-366}{rgb}{0.9611,0.4834,0.0000}
\definecolor{Umag-367}{rgb}{0.9599,0.4798,0.0000}
\definecolor{Umag-368}{rgb}{0.9587,0.4762,0.0000}
\definecolor{Umag-369}{rgb}{0.9575,0.4725,0.0000}
\definecolor{Umag-370}{rgb}{0.9563,0.4689,0.0000}
\definecolor{Umag-371}{rgb}{0.9551,0.4653,0.0000}
\definecolor{Umag-372}{rgb}{0.9539,0.4617,0.0000}
\definecolor{Umag-373}{rgb}{0.9527,0.4581,0.0000}
\definecolor{Umag-374}{rgb}{0.9515,0.4545,0.0000}
\definecolor{Umag-375}{rgb}{0.9503,0.4509,0.0000}
\definecolor{Umag-376}{rgb}{0.9491,0.4473,0.0000}
\definecolor{Umag-377}{rgb}{0.9479,0.4437,0.0000}
\definecolor{Umag-378}{rgb}{0.9467,0.4401,0.0000}
\definecolor{Umag-379}{rgb}{0.9455,0.4365,0.0000}
\definecolor{Umag-380}{rgb}{0.9443,0.4329,0.0000}
\definecolor{Umag-381}{rgb}{0.9431,0.4293,0.0000}
\definecolor{Umag-382}{rgb}{0.9419,0.4257,0.0000}
\definecolor{Umag-383}{rgb}{0.9407,0.4220,0.0000}
\definecolor{Umag-384}{rgb}{0.9395,0.4184,0.0000}
\definecolor{Umag-385}{rgb}{0.9383,0.4148,0.0000}
\definecolor{Umag-386}{rgb}{0.9371,0.4112,0.0000}
\definecolor{Umag-387}{rgb}{0.9359,0.4076,0.0000}
\definecolor{Umag-388}{rgb}{0.9347,0.4040,0.0000}
\definecolor{Umag-389}{rgb}{0.9335,0.4004,0.0000}
\definecolor{Umag-390}{rgb}{0.9323,0.3968,0.0000}
\definecolor{Umag-391}{rgb}{0.9311,0.3932,0.0000}
\definecolor{Umag-392}{rgb}{0.9299,0.3896,0.0000}
\definecolor{Umag-393}{rgb}{0.9287,0.3860,0.0000}
\definecolor{Umag-394}{rgb}{0.9275,0.3824,0.0000}
\definecolor{Umag-395}{rgb}{0.9263,0.3788,0.0000}
\definecolor{Umag-396}{rgb}{0.9251,0.3752,0.0000}
\definecolor{Umag-397}{rgb}{0.9238,0.3715,0.0000}
\definecolor{Umag-398}{rgb}{0.9226,0.3679,0.0000}
\definecolor{Umag-399}{rgb}{0.9214,0.3643,0.0000}
\definecolor{Umag-400}{rgb}{0.9202,0.3607,0.0000}
\definecolor{Umag-401}{rgb}{0.9190,0.3571,0.0000}
\definecolor{Umag-402}{rgb}{0.9178,0.3535,0.0000}
\definecolor{Umag-403}{rgb}{0.9166,0.3499,0.0000}
\definecolor{Umag-404}{rgb}{0.9154,0.3463,0.0000}
\definecolor{Umag-405}{rgb}{0.9142,0.3427,0.0000}
\definecolor{Umag-406}{rgb}{0.9130,0.3391,0.0000}
\definecolor{Umag-407}{rgb}{0.9118,0.3355,0.0000}
\definecolor{Umag-408}{rgb}{0.9106,0.3319,0.0000}
\definecolor{Umag-409}{rgb}{0.9094,0.3283,0.0000}
\definecolor{Umag-410}{rgb}{0.9082,0.3246,0.0000}
\definecolor{Umag-411}{rgb}{0.9070,0.3210,0.0000}
\definecolor{Umag-412}{rgb}{0.9058,0.3174,0.0000}
\definecolor{Umag-413}{rgb}{0.9046,0.3138,0.0000}
\definecolor{Umag-414}{rgb}{0.9034,0.3102,0.0000}
\definecolor{Umag-415}{rgb}{0.9022,0.3066,0.0000}
\definecolor{Umag-416}{rgb}{0.9010,0.3030,0.0000}
\definecolor{Umag-417}{rgb}{0.8998,0.2994,0.0000}
\definecolor{Umag-418}{rgb}{0.8986,0.2958,0.0000}
\definecolor{Umag-419}{rgb}{0.8974,0.2922,0.0000}
\definecolor{Umag-420}{rgb}{0.8962,0.2886,0.0000}
\definecolor{Umag-421}{rgb}{0.8950,0.2850,0.0000}
\definecolor{Umag-422}{rgb}{0.8938,0.2814,0.0000}
\definecolor{Umag-423}{rgb}{0.8926,0.2778,0.0000}
\definecolor{Umag-424}{rgb}{0.8914,0.2741,0.0000}
\definecolor{Umag-425}{rgb}{0.8902,0.2705,0.0000}
\definecolor{Umag-426}{rgb}{0.8890,0.2669,0.0000}
\definecolor{Umag-427}{rgb}{0.8878,0.2633,0.0000}
\definecolor{Umag-428}{rgb}{0.8866,0.2597,0.0000}
\definecolor{Umag-429}{rgb}{0.8854,0.2561,0.0000}
\definecolor{Umag-430}{rgb}{0.8842,0.2525,0.0000}
\definecolor{Umag-431}{rgb}{0.8830,0.2489,0.0000}
\definecolor{Umag-432}{rgb}{0.8818,0.2453,0.0000}
\definecolor{Umag-433}{rgb}{0.8806,0.2417,0.0000}
\definecolor{Umag-434}{rgb}{0.8794,0.2381,0.0000}
\definecolor{Umag-435}{rgb}{0.8782,0.2345,0.0000}
\definecolor{Umag-436}{rgb}{0.8770,0.2309,0.0000}
\definecolor{Umag-437}{rgb}{0.8758,0.2273,0.0000}
\definecolor{Umag-438}{rgb}{0.8745,0.2236,0.0000}
\definecolor{Umag-439}{rgb}{0.8733,0.2200,0.0000}
\definecolor{Umag-440}{rgb}{0.8721,0.2164,0.0000}
\definecolor{Umag-441}{rgb}{0.8709,0.2128,0.0000}
\definecolor{Umag-442}{rgb}{0.8697,0.2092,0.0000}
\definecolor{Umag-443}{rgb}{0.8685,0.2056,0.0000}
\definecolor{Umag-444}{rgb}{0.8673,0.2020,0.0000}
\definecolor{Umag-445}{rgb}{0.8661,0.1984,0.0000}
\definecolor{Umag-446}{rgb}{0.8649,0.1948,0.0000}
\definecolor{Umag-447}{rgb}{0.8637,0.1912,0.0000}
\definecolor{Umag-448}{rgb}{0.8625,0.1876,0.0000}
\definecolor{Umag-449}{rgb}{0.8613,0.1840,0.0000}
\definecolor{Umag-450}{rgb}{0.8601,0.1804,0.0000}
\definecolor{Umag-451}{rgb}{0.8589,0.1768,0.0000}
\definecolor{Umag-452}{rgb}{0.8577,0.1731,0.0000}
\definecolor{Umag-453}{rgb}{0.8565,0.1695,0.0000}
\definecolor{Umag-454}{rgb}{0.8553,0.1659,0.0000}
\definecolor{Umag-455}{rgb}{0.8541,0.1623,0.0000}
\definecolor{Umag-456}{rgb}{0.8529,0.1587,0.0000}
\definecolor{Umag-457}{rgb}{0.8517,0.1551,0.0000}
\definecolor{Umag-458}{rgb}{0.8505,0.1515,0.0000}
\definecolor{Umag-459}{rgb}{0.8493,0.1479,0.0000}
\definecolor{Umag-460}{rgb}{0.8481,0.1443,0.0000}
\definecolor{Umag-461}{rgb}{0.8469,0.1407,0.0000}
\definecolor{Umag-462}{rgb}{0.8457,0.1371,0.0000}
\definecolor{Umag-463}{rgb}{0.8445,0.1335,0.0000}
\definecolor{Umag-464}{rgb}{0.8433,0.1299,0.0000}
\definecolor{Umag-465}{rgb}{0.8421,0.1263,0.0000}
\definecolor{Umag-466}{rgb}{0.8409,0.1226,0.0000}
\definecolor{Umag-467}{rgb}{0.8397,0.1190,0.0000}
\definecolor{Umag-468}{rgb}{0.8385,0.1154,0.0000}
\definecolor{Umag-469}{rgb}{0.8373,0.1118,0.0000}
\definecolor{Umag-470}{rgb}{0.8361,0.1082,0.0000}
\definecolor{Umag-471}{rgb}{0.8349,0.1046,0.0000}
\definecolor{Umag-472}{rgb}{0.8337,0.1010,0.0000}
\definecolor{Umag-473}{rgb}{0.8325,0.0974,0.0000}
\definecolor{Umag-474}{rgb}{0.8313,0.0938,0.0000}
\definecolor{Umag-475}{rgb}{0.8301,0.0902,0.0000}
\definecolor{Umag-476}{rgb}{0.8289,0.0866,0.0000}
\definecolor{Umag-477}{rgb}{0.8277,0.0830,0.0000}
\definecolor{Umag-478}{rgb}{0.8265,0.0794,0.0000}
\definecolor{Umag-479}{rgb}{0.8253,0.0758,0.0000}
\definecolor{Umag-480}{rgb}{0.8240,0.0721,0.0000}
\definecolor{Umag-481}{rgb}{0.8228,0.0685,0.0000}
\definecolor{Umag-482}{rgb}{0.8216,0.0649,0.0000}
\definecolor{Umag-483}{rgb}{0.8204,0.0613,0.0000}
\definecolor{Umag-484}{rgb}{0.8192,0.0577,0.0000}
\definecolor{Umag-485}{rgb}{0.8180,0.0541,0.0000}
\definecolor{Umag-486}{rgb}{0.8168,0.0505,0.0000}
\definecolor{Umag-487}{rgb}{0.8156,0.0469,0.0000}
\definecolor{Umag-488}{rgb}{0.8144,0.0433,0.0000}
\definecolor{Umag-489}{rgb}{0.8132,0.0397,0.0000}
\definecolor{Umag-490}{rgb}{0.8120,0.0361,0.0000}
\definecolor{Umag-491}{rgb}{0.8108,0.0325,0.0000}
\definecolor{Umag-492}{rgb}{0.8096,0.0289,0.0000}
\definecolor{Umag-493}{rgb}{0.8084,0.0253,0.0000}
\definecolor{Umag-494}{rgb}{0.8072,0.0216,0.0000}
\definecolor{Umag-495}{rgb}{0.8060,0.0180,0.0000}
\definecolor{Umag-496}{rgb}{0.8048,0.0144,0.0000}
\definecolor{Umag-497}{rgb}{0.8036,0.0108,0.0000}
\definecolor{Umag-498}{rgb}{0.8024,0.0072,0.0000}
\definecolor{Umag-499}{rgb}{0.8012,0.0036,0.0000}
\definecolor{Umag-500}{rgb}{0.8000,0.0000,0.0000}
\pgfplotsset{
colormap={Umag}{
color={Umag-1};
color={Umag-2};
color={Umag-3};
color={Umag-4};
color={Umag-5};
color={Umag-6};
color={Umag-7};
color={Umag-8};
color={Umag-9};
color={Umag-10};
color={Umag-11};
color={Umag-12};
color={Umag-13};
color={Umag-14};
color={Umag-15};
color={Umag-16};
color={Umag-17};
color={Umag-18};
color={Umag-19};
color={Umag-20};
color={Umag-21};
color={Umag-22};
color={Umag-23};
color={Umag-24};
color={Umag-25};
color={Umag-26};
color={Umag-27};
color={Umag-28};
color={Umag-29};
color={Umag-30};
color={Umag-31};
color={Umag-32};
color={Umag-33};
color={Umag-34};
color={Umag-35};
color={Umag-36};
color={Umag-37};
color={Umag-38};
color={Umag-39};
color={Umag-40};
color={Umag-41};
color={Umag-42};
color={Umag-43};
color={Umag-44};
color={Umag-45};
color={Umag-46};
color={Umag-47};
color={Umag-48};
color={Umag-49};
color={Umag-50};
color={Umag-51};
color={Umag-52};
color={Umag-53};
color={Umag-54};
color={Umag-55};
color={Umag-56};
color={Umag-57};
color={Umag-58};
color={Umag-59};
color={Umag-60};
color={Umag-61};
color={Umag-62};
color={Umag-63};
color={Umag-64};
color={Umag-65};
color={Umag-66};
color={Umag-67};
color={Umag-68};
color={Umag-69};
color={Umag-70};
color={Umag-71};
color={Umag-72};
color={Umag-73};
color={Umag-74};
color={Umag-75};
color={Umag-76};
color={Umag-77};
color={Umag-78};
color={Umag-79};
color={Umag-80};
color={Umag-81};
color={Umag-82};
color={Umag-83};
color={Umag-84};
color={Umag-85};
color={Umag-86};
color={Umag-87};
color={Umag-88};
color={Umag-89};
color={Umag-90};
color={Umag-91};
color={Umag-92};
color={Umag-93};
color={Umag-94};
color={Umag-95};
color={Umag-96};
color={Umag-97};
color={Umag-98};
color={Umag-99};
color={Umag-100};
color={Umag-101};
color={Umag-102};
color={Umag-103};
color={Umag-104};
color={Umag-105};
color={Umag-106};
color={Umag-107};
color={Umag-108};
color={Umag-109};
color={Umag-110};
color={Umag-111};
color={Umag-112};
color={Umag-113};
color={Umag-114};
color={Umag-115};
color={Umag-116};
color={Umag-117};
color={Umag-118};
color={Umag-119};
color={Umag-120};
color={Umag-121};
color={Umag-122};
color={Umag-123};
color={Umag-124};
color={Umag-125};
color={Umag-126};
color={Umag-127};
color={Umag-128};
color={Umag-129};
color={Umag-130};
color={Umag-131};
color={Umag-132};
color={Umag-133};
color={Umag-134};
color={Umag-135};
color={Umag-136};
color={Umag-137};
color={Umag-138};
color={Umag-139};
color={Umag-140};
color={Umag-141};
color={Umag-142};
color={Umag-143};
color={Umag-144};
color={Umag-145};
color={Umag-146};
color={Umag-147};
color={Umag-148};
color={Umag-149};
color={Umag-150};
color={Umag-151};
color={Umag-152};
color={Umag-153};
color={Umag-154};
color={Umag-155};
color={Umag-156};
color={Umag-157};
color={Umag-158};
color={Umag-159};
color={Umag-160};
color={Umag-161};
color={Umag-162};
color={Umag-163};
color={Umag-164};
color={Umag-165};
color={Umag-166};
color={Umag-167};
color={Umag-168};
color={Umag-169};
color={Umag-170};
color={Umag-171};
color={Umag-172};
color={Umag-173};
color={Umag-174};
color={Umag-175};
color={Umag-176};
color={Umag-177};
color={Umag-178};
color={Umag-179};
color={Umag-180};
color={Umag-181};
color={Umag-182};
color={Umag-183};
color={Umag-184};
color={Umag-185};
color={Umag-186};
color={Umag-187};
color={Umag-188};
color={Umag-189};
color={Umag-190};
color={Umag-191};
color={Umag-192};
color={Umag-193};
color={Umag-194};
color={Umag-195};
color={Umag-196};
color={Umag-197};
color={Umag-198};
color={Umag-199};
color={Umag-200};
color={Umag-201};
color={Umag-202};
color={Umag-203};
color={Umag-204};
color={Umag-205};
color={Umag-206};
color={Umag-207};
color={Umag-208};
color={Umag-209};
color={Umag-210};
color={Umag-211};
color={Umag-212};
color={Umag-213};
color={Umag-214};
color={Umag-215};
color={Umag-216};
color={Umag-217};
color={Umag-218};
color={Umag-219};
color={Umag-220};
color={Umag-221};
color={Umag-222};
color={Umag-223};
color={Umag-224};
color={Umag-225};
color={Umag-226};
color={Umag-227};
color={Umag-228};
color={Umag-229};
color={Umag-230};
color={Umag-231};
color={Umag-232};
color={Umag-233};
color={Umag-234};
color={Umag-235};
color={Umag-236};
color={Umag-237};
color={Umag-238};
color={Umag-239};
color={Umag-240};
color={Umag-241};
color={Umag-242};
color={Umag-243};
color={Umag-244};
color={Umag-245};
color={Umag-246};
color={Umag-247};
color={Umag-248};
color={Umag-249};
color={Umag-250};
color={Umag-251};
color={Umag-252};
color={Umag-253};
color={Umag-254};
color={Umag-255};
color={Umag-256};
color={Umag-257};
color={Umag-258};
color={Umag-259};
color={Umag-260};
color={Umag-261};
color={Umag-262};
color={Umag-263};
color={Umag-264};
color={Umag-265};
color={Umag-266};
color={Umag-267};
color={Umag-268};
color={Umag-269};
color={Umag-270};
color={Umag-271};
color={Umag-272};
color={Umag-273};
color={Umag-274};
color={Umag-275};
color={Umag-276};
color={Umag-277};
color={Umag-278};
color={Umag-279};
color={Umag-280};
color={Umag-281};
color={Umag-282};
color={Umag-283};
color={Umag-284};
color={Umag-285};
color={Umag-286};
color={Umag-287};
color={Umag-288};
color={Umag-289};
color={Umag-290};
color={Umag-291};
color={Umag-292};
color={Umag-293};
color={Umag-294};
color={Umag-295};
color={Umag-296};
color={Umag-297};
color={Umag-298};
color={Umag-299};
color={Umag-300};
color={Umag-301};
color={Umag-302};
color={Umag-303};
color={Umag-304};
color={Umag-305};
color={Umag-306};
color={Umag-307};
color={Umag-308};
color={Umag-309};
color={Umag-310};
color={Umag-311};
color={Umag-312};
color={Umag-313};
color={Umag-314};
color={Umag-315};
color={Umag-316};
color={Umag-317};
color={Umag-318};
color={Umag-319};
color={Umag-320};
color={Umag-321};
color={Umag-322};
color={Umag-323};
color={Umag-324};
color={Umag-325};
color={Umag-326};
color={Umag-327};
color={Umag-328};
color={Umag-329};
color={Umag-330};
color={Umag-331};
color={Umag-332};
color={Umag-333};
color={Umag-334};
color={Umag-335};
color={Umag-336};
color={Umag-337};
color={Umag-338};
color={Umag-339};
color={Umag-340};
color={Umag-341};
color={Umag-342};
color={Umag-343};
color={Umag-344};
color={Umag-345};
color={Umag-346};
color={Umag-347};
color={Umag-348};
color={Umag-349};
color={Umag-350};
color={Umag-351};
color={Umag-352};
color={Umag-353};
color={Umag-354};
color={Umag-355};
color={Umag-356};
color={Umag-357};
color={Umag-358};
color={Umag-359};
color={Umag-360};
color={Umag-361};
color={Umag-362};
color={Umag-363};
color={Umag-364};
color={Umag-365};
color={Umag-366};
color={Umag-367};
color={Umag-368};
color={Umag-369};
color={Umag-370};
color={Umag-371};
color={Umag-372};
color={Umag-373};
color={Umag-374};
color={Umag-375};
color={Umag-376};
color={Umag-377};
color={Umag-378};
color={Umag-379};
color={Umag-380};
color={Umag-381};
color={Umag-382};
color={Umag-383};
color={Umag-384};
color={Umag-385};
color={Umag-386};
color={Umag-387};
color={Umag-388};
color={Umag-389};
color={Umag-390};
color={Umag-391};
color={Umag-392};
color={Umag-393};
color={Umag-394};
color={Umag-395};
color={Umag-396};
color={Umag-397};
color={Umag-398};
color={Umag-399};
color={Umag-400};
color={Umag-401};
color={Umag-402};
color={Umag-403};
color={Umag-404};
color={Umag-405};
color={Umag-406};
color={Umag-407};
color={Umag-408};
color={Umag-409};
color={Umag-410};
color={Umag-411};
color={Umag-412};
color={Umag-413};
color={Umag-414};
color={Umag-415};
color={Umag-416};
color={Umag-417};
color={Umag-418};
color={Umag-419};
color={Umag-420};
color={Umag-421};
color={Umag-422};
color={Umag-423};
color={Umag-424};
color={Umag-425};
color={Umag-426};
color={Umag-427};
color={Umag-428};
color={Umag-429};
color={Umag-430};
color={Umag-431};
color={Umag-432};
color={Umag-433};
color={Umag-434};
color={Umag-435};
color={Umag-436};
color={Umag-437};
color={Umag-438};
color={Umag-439};
color={Umag-440};
color={Umag-441};
color={Umag-442};
color={Umag-443};
color={Umag-444};
color={Umag-445};
color={Umag-446};
color={Umag-447};
color={Umag-448};
color={Umag-449};
color={Umag-450};
color={Umag-451};
color={Umag-452};
color={Umag-453};
color={Umag-454};
color={Umag-455};
color={Umag-456};
color={Umag-457};
color={Umag-458};
color={Umag-459};
color={Umag-460};
color={Umag-461};
color={Umag-462};
color={Umag-463};
color={Umag-464};
color={Umag-465};
color={Umag-466};
color={Umag-467};
color={Umag-468};
color={Umag-469};
color={Umag-470};
color={Umag-471};
color={Umag-472};
color={Umag-473};
color={Umag-474};
color={Umag-475};
color={Umag-476};
color={Umag-477};
color={Umag-478};
color={Umag-479};
color={Umag-480};
color={Umag-481};
color={Umag-482};
color={Umag-483};
color={Umag-484};
color={Umag-485};
color={Umag-486};
color={Umag-487};
color={Umag-488};
color={Umag-489};
color={Umag-490};
color={Umag-491};
color={Umag-492};
color={Umag-493};
color={Umag-494};
color={Umag-495};
color={Umag-496};
color={Umag-497};
color={Umag-498};
color={Umag-499};
color={Umag-500};
}
}
\definecolor{cb_yb_1}{HTML}{a1dab4}
\definecolor{cb_yb_2}{HTML}{41b6c4}
\definecolor{cb_yb_3}{HTML}{2c7fb8}
\definecolor{cb_yb_4}{HTML}{253494}
\definecolor{cb_disc1_1}{HTML}{e41a1c}
\definecolor{cb_disc1_2}{HTML}{377eb8}
\definecolor{cb_disc1_3}{HTML}{4daf4a}
\definecolor{cb_disc1_4}{HTML}{984ea3}
\definecolor{cb_disc1_5}{HTML}{ff7f00}
\definecolor{cb_disc1_6}{HTML}{a65628}
\definecolor{cb2_p1_1}{HTML}{ffffcc}
\definecolor{cb2_p1_2}{HTML}{a1dab4}
\definecolor{cb2_p1_3}{HTML}{41b6c4}
\definecolor{cb2_p1_4}{HTML}{225ea8}
\definecolor{default_green}{HTML}{4daf4a}
\definecolor{default_orange}{HTML}{ff7f00}
\definecolor{default_purple}{HTML}{984ea3}
\definecolor{cb_red_seq1}{HTML}{fef0d9}
\definecolor{cb_red_seq2}{HTML}{fdd49e}
\definecolor{cb_red_seq3}{HTML}{fdbb84}
\definecolor{cb_red_seq4}{HTML}{fc8d59}
\definecolor{cb_red_seq5}{HTML}{e34a33}
\definecolor{cb_red_seq6}{HTML}{b30000}
\definecolor{cb_blue_seq1}{HTML}{f1eef6}
\definecolor{cb_blue_seq2}{HTML}{d0d1e6}
\definecolor{cb_blue_seq3}{HTML}{a6bddb}
\definecolor{cb_blue_seq4}{HTML}{74a9cf}
\definecolor{cb_blue_seq5}{HTML}{2b8cbe}
\definecolor{cb_blue_seq6}{HTML}{045a8d}
\definecolor{cb_green_seq1}{HTML}{ffffcc}
\definecolor{cb_green_seq2}{HTML}{d9f0a3}
\definecolor{cb_green_seq3}{HTML}{addd8e}
\definecolor{cb_green_seq4}{HTML}{78c679}
\definecolor{cb_green_seq5}{HTML}{31a354}
\definecolor{cb_green_seq6}{HTML}{006837}
\definecolor{cb_purple_seq1}{HTML}{feebe2}
\definecolor{cb_purple_seq2}{HTML}{fcc5c0}
\definecolor{cb_purple_seq3}{HTML}{fa9fb5}
\definecolor{cb_purple_seq4}{HTML}{f768a1}
\definecolor{cb_purple_seq5}{HTML}{c51b8a}
\definecolor{cb_purple_seq6}{HTML}{7a0177}